%% file: main.tex
\documentclass[a4paper,11pt]{article}
\pdfoutput=1
\usepackage{jheppub}

\usepackage{amsmath}
\usepackage{mathtools}
\usepackage{multirow}
\usepackage{array}
\usepackage{bm}

\usepackage{graphicx}
\usepackage{lscape}
\usepackage{subfig}

\usepackage{bbm}
\usepackage{hyperref}
\usepackage{booktabs}
\usepackage{float}

\newcommand{\be}{\begin{equation}}
\newcommand{\ee}{\end{equation}}
\newcommand{\bea}{\begin{eqnarray}}
\newcommand{\eea}{\end{eqnarray}}
\newcommand{\bei}{\begin{itemize}}
\newcommand{\eei}{\end{itemize}}
\newcommand{\bean}{\begin{eqnarray*}}
\newcommand{\eean}{\\end{eqnarray*}}
\newcommand{\nn}{\nonumber \\}

\def\eps{\epsilon}

\def\top #1{\mathcal{T}_{#1}}

\allowdisplaybreaks[1]

\newcommand\scalemath[2]{\scalebox{#1}{\mbox{\ensuremath{\displaystyle #2}}}}

\newcommand{\gammaAB}[2]{\ensuremath\gamma}

\newcommand{\dA}{\ensuremath d\AA}

\newcommand{\dAk}[1]{\ensuremath \underbrace{\dA \ldots \dA}_{\text{#1 times}}}

\newcommand{\Den}{\ensuremath D}
\newcommand{\dd}{\ensuremath \mathrm{d}}
\DeclareMathOperator{\dlog}{\mathit{d}log}

\newcommand{\minus}{\ensuremath \scalebox{0.5}[1.0]{\( - \)}}
\newcommand{\pminus}{\hphantom{\minus}}
\newcommand{\FF}{\ensuremath \text{F}}

\newcommand{\GG}{\ensuremath \text{I}}
\newcommand{\GGvec}{\ensuremath \mathbf{I}}
\newcommand{\HH}{\ensuremath \mathbf{H}}

\newcommand{\MM}{\ensuremath \mathbb{M}}
\renewcommand{\AA}{\ensuremath \mathbb{A}}

\newcommand{\unipd}{Dipartimento di Fisica ed Astronomia, Universit\`a di Padova, Via Marzolo 8, 35131 Padova, Italy}
\newcommand{\uzh}{Department of Physics, University of Z{\"u}rich, CH-8057 Z{\"u}rich, Switzerland}
\newcommand{\pdinfn}{INFN, Sezione di Padova, Via Marzolo 8, 35131 Padova, Italy}
\newcommand{\miinfn}{INFN, Sezione di Milano, Via Celoria 16, 20133 Milano, Italy}
\newcommand{\argonne}{High Energy Physics Division, Argonne National Laboratory, Argonne, IL 60439, USA}

\allowdisplaybreaks

\author[a]{Stefano Di Vita,}
\author[b,c]{Stefano Laporta,}
\author[b,c]{Pierpaolo Mastrolia,}
\author[d]{Amedeo Primo,}
\author[e]{Ulrich Schubert}

\affiliation[a]{\miinfn}
\affiliation[b]{\unipd}
\affiliation[c]{\pdinfn}
\affiliation[d]{\uzh}
\affiliation[e]{\argonne}

\emailAdd{stefano.divita@mi.infn.it}
\emailAdd{stefano.laporta@pd.infn.it}
\emailAdd{pierpaolo.mastrolia@pd.infn.it}
\emailAdd{aprimo@physik.uzh.ch}
\emailAdd{schubertmielnik@anl.gov}

\title{Master integrals for the NNLO virtual corrections to $\mu e$ scattering in QED: the non-planar graphs}


\abstract{ We evaluate the master integrals for the two-loop
  non-planar box-diagrams contributing to the elastic scattering of
  muons and electrons at next-to-next-to-leading order in QED. We
  adopt the method of differential equations and the Magnus
  exponential to determine a canonical set of integrals, finally
  expressed as a Taylor series around four space-time dimensions, with
  coefficients written as combination of generalised
  polylogarithms. The electron is treated as massless, while we retain
  full dependence on the muon mass. The considered integrals are also
  relevant for crossing-related processes, such as di-muon production
  at $e^+e^-$ colliders, as well as for the QCD corrections to
  top-pair production at hadron colliders. In particular our results,
  together with the planar master integrals recently computed,
  represent the complete set of functions needed for the evaluation of
  the photonic two-loop virtual next-to-next-to-leading order QED corrections to
  $\mu e \to \mu e$ and $e^+ e^-\to\mu^+\mu^-$.}

\begin{document}

\maketitle
\flushbottom
\input{tex/Introduction}
\input{tex/Notation}
\input{tex/DEQ}
\input{tex/Numerics}
\input{tex/Conclusions}

\section*{Acknowledgments}
We wish to acknowledge stimulating discussions with all members of the
MUonE collaboration, and in particular we wish to thank Giovanni Ossola, Massimo
Passera and William J.~Torres Bobadilla for our constant interaction
and collaboration, and Fedor Ignatov for stimulating discussions.  We
thank Lorenzo Tancredi for suggesting the use of an alternative set of
variables, which simplified the form of the canonical system we
originally got at an earlier stage of the project. We also acknowledge
discussions with Andreas von Manteuffel.  We wish to thank Lance Dixon
for interesting feedback on the project.  We would like to express a
special thanks to the Mainz Institute for Theoretical Physics (MITP)
for its hospitality and support during the workshop ``The evaluation
of the leading hadronic contribution to the muon anomalous magnetic
moment''.  U.~S.\ is supported by the DOE contract DE-AC02-06CH11357.
This research was supported in part by the Swiss National Science
Foundation (SNF) under contract 200020-175595.

\appendix
\input{tex/AuxVert}

\input{tex/dlogforms}

\bibliographystyle{JHEP}
\bibliography{references}

\end{document}

%% file: tex/Introduction.tex
\section{Introduction}

In a previous work \cite{Mastrolia:2017pfy}, 
we began the investigation of the next-to-next-to-leading-order (NNLO) virtual corrections to
the elastic scattering of muons and electrons in Quantum Electrodynamics (QED), by classifying and evaluating the planar two-loop integrals arising from Feynman diagrams at this order in perturbation theory \cite{Mastrolia:2018sso}.

The NNLO QED corrections to the process $\mu e \to \mu e$ are crucial to interpret the high-precision data of future experiments like MUonE, recently proposed at CERN, aiming at measuring the differential cross section of the elastic scattering of high-energy muons on atomic electrons as a function of the spacelike (negative) squared momentum transfer~\cite{Calame:2015fva,Abbiendi:2016xup}. This measurement will provide the running of the effective electromagnetic coupling in the spacelike region and, as a result, a new and independent determination of the leading hadronic contribution to the muon $g$-2~\cite{Calame:2015fva,Abbiendi:2016xup}. In order for this new determination to be competitive with the present dispersive one, which is obtained via timelike data (see \cite{Jegerlehner:2017gek} for a review), the $\mu e$ differential cross section must be measured with statistical and systematic uncertainties of the order of 10ppm. This high experimental precision demands an analogous accuracy in the theoretical prediction. 

Moreover, the NNLO QED corrections for the crossing-related scattering process $e^+ e^-$ $\to \mu^+ \mu^-$ are important for some of the high-precision studies planned at upcoming low-energy $e^+ e^-$ experiments, like Belle-II and VEPP-2000. Two interesting applications would be the following. The forward-backward asymmetry in muon pair production could be exploited to constrain non-standard $ee\mu\mu$ interactions~\cite{Ferber:2016rka}, and the current estimates suggest that the knowledge of the NNLO QED differential cross section is needed, as QED itself produces an asymmetry starting at NLO. The knowledge of the QED radiative corrections to the $e^+e^-\to\mu^+\mu^-$ cross section will also be needed for precise measurements of the ratio $R(s)=\sigma\left(e^+e^-\to\text{hadrons}\right)/\sigma\left(e^+e^-\to\mu^+\mu^-\right)$~\cite{Aleksejevs:2017hyv,Ignatov:2018fss}.

In this work, we complete the task of determining all functions
required by the NNLO QED virtual photonic corrections to $\mu e$ scattering, by
evaluating the two-loop integrals coming from non-planar four-point
Feynman diagrams.  Given the hierarchy between the electron mass $m_e$\!
and the muon mass $m, m_e/m\!\! \sim \! 5\! \cdot\! 10^{-3}$, as in our former
study we consider the approximation $m_e=0$.\footnote{The simplifying
  assumption $m_e=0$ has already been used in the case of the two-loop
  photonic corrections to Bhabha scattering~\cite{Bern:2000ie},
  with which $\mu e$ scattering shares several features. In
  particular, one expects to find logarithmically enhanced terms,
  proportional to $\log(s/m_e^2)$ and $\log^2(s/m_e^2)$, in the
  cross-sections for such processes. Such terms would correspond to
  collinear singularities in the $m_e \to 0$ limit. For Bhabha
  scattering it was possible to precisely recover the coefficients of
  those terms, starting from the calculation in which the collinear
  divergenes are dimensionally regulated, and exploiting the universal
  infrared structure of gauge theories (see~\cite{Actis:2010gg} and
  references therein).} For the non-planar topology,
integration-by-parts identities
(IBPs)~\cite{Tkachov:1981wb,Chetyrkin:1981qh,Laporta:2001dd} yield the
identification of a set of 44 master integrals (MIs), which we compute
analytically by means of the differential equations
method~\cite{Kotikov:1990kg,Remiddi:1997ny,Gehrmann:1999as}.  The
system-solving strategy~\cite{Henn:2013pwa,Argeri:2014qva} is based on
a consolidated procedure, which has been proven to be very effective
in the context of multi-loop integrals involving several
scales~\cite{Argeri:2014qva,DiVita:2014pza,Bonciani:2016ypc,DiVita:2017xlr,Mastrolia:2017pfy}.
Firstly, we identify a set of MIs that obey a system of
first-order differential equations~(DEQs) in the kinematic variables
$s/m^2$ and $t/m^2$ which is linear in the space-time dimension $d$.
Subsequently we employ the Magnus exponential matrix
\cite{Argeri:2014qva} to derive an equivalent system of equations in {\it
  canonical form} \cite{Henn:2013pwa}, where the dependence of the
associated matrices on $(d-4)$ is factorised from the kinematics.  The matrix
associated with the canonical system is a logarithmic differential
form which, in appropriate variables, has a polynomial alphabet. The
canonical MIs can be therefore cast as a Taylor series around $d=4$,
with coefficients written as combinations of generalised
polylogarithms
(GPLs)~\cite{Goncharov:polylog,Remiddi:1999ew,Gehrmann:2001pz,Vollinga:2004sn}.

For certain classes of MIs, like the ones of $\mu e \to \mu e$ and crossing-related processes,
the choice of the boundary conditions may also constitute a challenging problem. Here, we exploit either the regularity conditions at pseudo-thresholds or the expression of integrals which are obtained by solving simpler, auxiliary systems of DEQs. Therefore, we limit the use of direct integration only to a small subset of 
simpler integrals used as input functions. 

The package \texttt{Reduze}~\cite{vonManteuffel:2012np} has been used throughout the calculations, for the IBPs decomposition and for generating the DEQs obeyed by the MIs.
The analytic expressions of the MIs have been numerically evaluated with the help of \texttt{GiNaC}~\cite{Bauer:2000cp} and were successfully tested against the numerical values provided either by the computer code \texttt{SecDec}~\cite{Borowka:2015mxa} or, for the most complicated two-loop non-planar topologies (with 6- and 7-denominators), by an in-house algorithm. For such topologies, we identified an alternative set of {\it quasi-finite} integrals~\cite{vonManteuffel:2014qoa}, more suitable for numerical integration, also with the help of \texttt{Reduze}.

As far as the QED corrections involving two leptons (one massless, one
massive) are concerned, the non-planar four-point functions hereby
presented, together with the planar ones presented earlier
\cite{Mastrolia:2017pfy}, and the three-point functions available in
the literature
\cite{Bonciani:2003te,Mastrolia:2003yz,Bonciani:2003ai}, make the
analytic evaluation of the virtual two-loop amplitudes for the $\mu e$
scattering, as well as for the crossing-related processes, within
reach.
In order to build the corresponding two-loop virtual amplitude, one
would still need to compute the corresponding Feynman diagrams by
reducing them to our set of MIs and carrying out the ultraviolet
renormalization. Finally, the analytic continuation to the $\mu e$
scattering region ($s\geq m^2$, $-(m^2-s)^2\leq s\,t \leq 0$, and
$s\,(2m^2-s) \leq s\,u \leq m^4$) has to be performed. We leave all
these steps to a future publication.

For completeness we mention that, as in the case of Bhabha scattering
(see e.g. ref.~\cite{Actis:2010gg} for a review), further gauge-invariant
sets of NNLO corrections to $\mu e$ scattering will have to be
computed. Those sets consist of vertex and box diagrams with vacuum
polarization insertions of two kinds: $i$) vacuum polarization due to a
heavy-fermion loop (the tau lepton or the top, bottom and charm
quarks), which can be addressed perturbatively at energies much lower
than their masses, the only difference being the presence of a further
scale in the loop integrals; $ii$) vacuum polarization due a
light-quark loop, where the effect of the hadronic interactions cannot be
neglected and a non-perturbative approach, such as the use of
dispersion relations, is required.

In addition, we remark that the MIs of the QED corrections to $e^+ e^-$ $\to \mu^+ \mu^-$ are a subset of those needed for the QCD corrections to the $t{\bar t}$-pair production at hadron colliders. 
The complete two-loop QCD corrections to $pp \to t {\bar t}$ are currently known only numerically \cite{Czakon:2008zk,Czakon:2012pz,Czakon:2012zr,Baernreuther:2012ws,Czakon:2013goa}.
The analytic evaluation of the MIs appearing in the leading-colour corrections to $pp \to t {\bar t}$
were considered in refs.~\cite{Bonciani:2008az,Bonciani:2009nb,Bonciani:2010mn,vonManteuffel:2013uoa,Bonciani:2013ywa}, which also include the simplest non-planar topology, namely the one in which the crossed-loop is fully massless. In our former work \cite{Mastrolia:2017pfy}, we extended the set of available functions for considering also sub-leading colour contributions. 
Very recently, the analytic calculation of the MIs for the planar double-box integral with a closed top loop appeared in~\cite{Adams:2018kez}. 
The analytic result for a non-planar three-point function, which constitutes a sub-diagram of the non-planar double box with closed heavy quark loop, was presented in~\cite{vonManteuffel:2017hms}.
The non-planar graphs hereby considered would also 
contribute to subleading-colour terms to $t{\bar t}$-pair production, and their analytic evaluation was never addressed before. 
The numerical evaluation of (one of the) non-planar integrals computed analytically in this work has been recently considered in \cite{Liu:2017jxz}, in the context of a novel, promising method that aims at the numerical solution of differential equations.

The paper is organised as follows. 
In section~\ref{sec:Notation}, we set our notation and conventions for the four-point topology relevant for $\mu e$ scattering.
In section~\ref{sec:DEQ}, we describe the general features of the systems of DEQs satisfied by the MIs, cast in $\dlog$-form, and we present the results for the non-planar two-loop MIs. 
In section~\ref{sec:Numerics}, we describe the numerical evaluation of the non-planar four-point integrals. 
The information provided in the text is complemented by two appendices: in appendix~\ref{app:A}, we discuss the computation of the auxiliary integrals which have been used to extract some of the boundary constants and, in appendix~\ref{app:C}, we give the matrices associated with the $\dlog$-form.

The analytic expressions of the considered MIs are given in the ancillary files accompanying the \texttt{arXiv} version of this publication.

%%%%%%%%%%%%%%%%%%
%%% Local Variables:
%%% TeX-master: "../main"
%%% End:

%% file: tex/Notation.tex
\section{The non-planar four-point topology}
\label{sec:Notation}
\begin{figure}[t]
\centering
\includegraphics[scale=0.94]{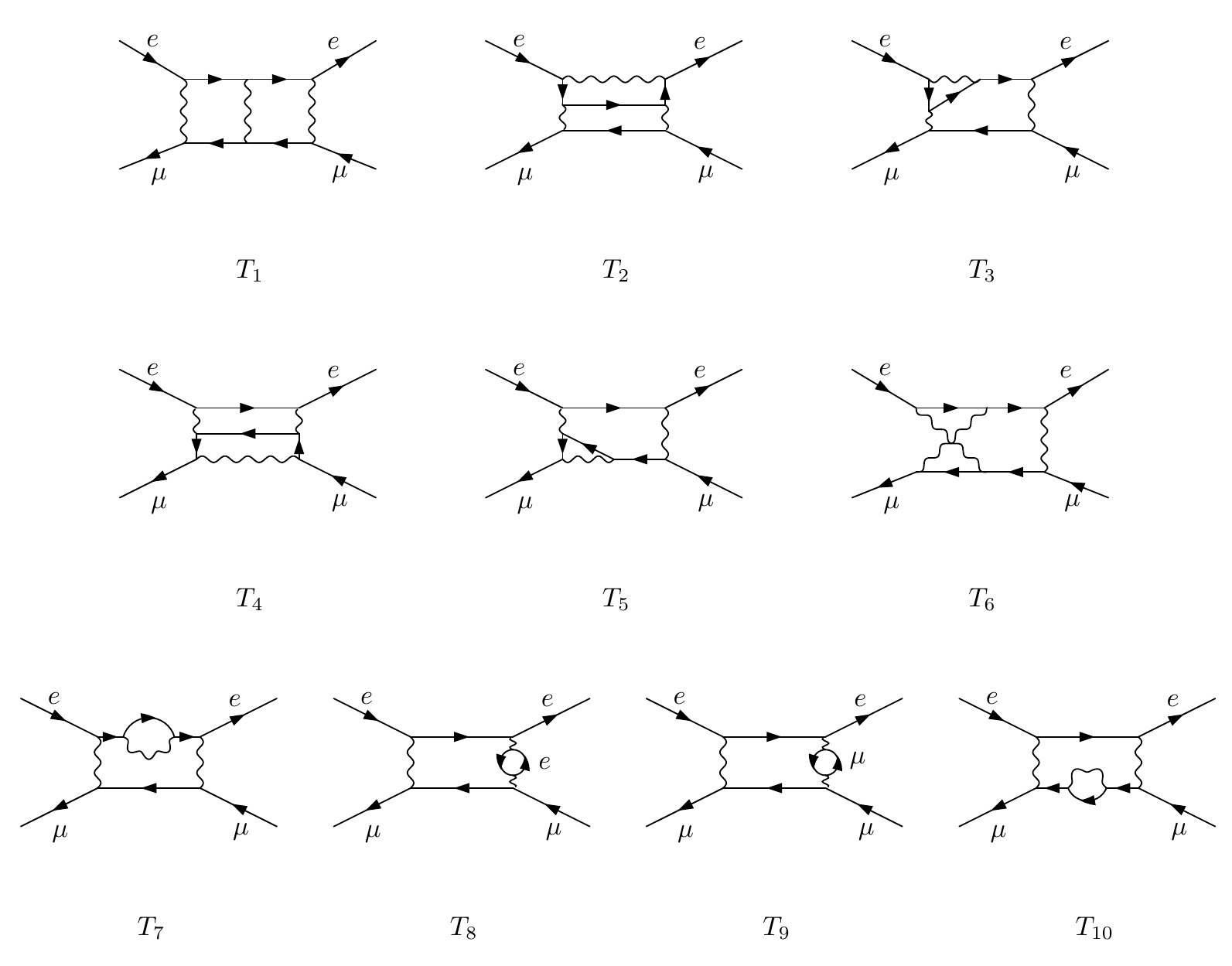}
\caption{Two-loop four-point topologies for $\mu e$ scattering.}
\label{fig:Feyndiag}
\end{figure}

In this paper, we consider the $\mu e$ scattering process
\begin{align}
  \mu^{+}(p_1) + e^{-}(p_2)  \to    e^{-}(p_3) +  \mu^{+}(p_4)\,,
\end{align}
in the approximation of vanishing electron mass, $m_e=0$, i.e. with kinematics specified by
\begin{align}
p_1^2&=p_4^2=m^2\,,\quad p_2^2=p_3^2=0\,,\nn
  s = (p_1+p_2)^2\,,  \quad t&= (p_2-p_3)^2\,, \quad u=(p_1-p_3)^2=2m^2 -t-s  \, ,
\end{align}
where $m$ is the muon mass. Representative Feynman diagrams of the 10 relevant two-loop four-point topologies $T_i$ that contribute to the process are depicted in figure~\ref{fig:Feyndiag}. The computation of the MIs belonging to the topologies $T_{1,2,3,4,5,7,8,9,10}$ has been discussed in~\cite{Mastrolia:2017pfy}. 
In this paper, we complete the evaluation of all MIs required for the two-loop virtual amplitude due to photonic corrections, by determining the analytic expression of the MIs that belong to the non-planar topology $T_6$.

The calculation involves the evaluation, in
$d$ dimensions, of Feynman integrals of the type
\begin{gather}
I^{[d]}(n_1,\ldots,n_9)
\equiv
  \int \widetilde{\dd^d k_1}\widetilde{\dd^d k_2}\,
  \frac{1}{\Den_{1}^{n_1} \ldots \Den_{9}^{n_9}}\,,
\label{eq:def:ourintegrals}
\end{gather}
where $D_{i}$ are inverse scalar propagators.
The analytic calculation described in section~\ref{sec:DEQ} is performed expanding around $d=4$, while the numerical evaluation presented in section~\ref{sec:Numerics} is carried over around $d=6$. We set $\eps\equiv(d_*-d)/2$, where $d_*=4$ and $d_*=6$ according to the case considered, and define  our integration measure
\begin{align}
  \widetilde{\dd^{d}k} = {} & \frac{\dd^{d}k}{i \pi^{d/2}\,\Gamma_\eps } \left(\frac{m^2}{\mu^2}\right)^\epsilon\,,
                                      \label{eq:intmeasure1}
\end{align}
where $\mu$ is the 't Hooft scale of dimensional regularisation and $\Gamma_\eps \equiv \Gamma(1+\eps)$. Notice that our integration measure, when evaluated at $d=4-2\eps$, agrees with eq.(3.2) of~\cite{Mastrolia:2017pfy}.

For the non-planar four-point topology $T_6$, we choose the following set of inverse propagators:
\begin{gather}
\Den_1 = (k_1)^2-m^2,\quad
\Den_2 = (k_2)^2-m^2,\quad
\Den_3 = (k_1+p_1)^2,\quad
\Den_4 = (k_2+p_1)^2, \nn
\Den_5 = (k_1+p_1+p_2)^2,\quad
\Den_6 = (k_2+p_1+p_2)^2,\quad
\Den_7 = (k_1-k_2)^2, \nn
\Den_8 = (k_2+p_1+p_2-p_3)^2,\quad
\Den_9 = (k_1-k_2+p_3)^2\, ,
\label{eq:2Lfamily}
\end{gather}
where  $k_1$ and $k_2$ denote the loop momenta. In particular, with the definition \eqref{eq:intmeasure1}, the tadpole integral $\eps^2 I^{[4-2\eps]}(2,2,0,\ldots,0)$ is normalised to $1$. The Feynman prescription is understood, and it can be recovered by the addition of an arbitrarily small positive imaginary part, $D_i \to D_i + i\omega$.
%%% Local Variables:
%%% TeX-master: "../main"
%%% End:

%% file: tex/DEQ.tex
\section{System of differential equations}
\label{sec:DEQ}
By means of IBPs, the two-loop integrals that belong to $T_6$ can be
reduced to a basis of 44 distinct MIs.  In order to determine the
analytic expression of the latter, we derive their DEQs in the
kinematic variables $s$ and $t$. The evaluation of the MIs can be
further facilitated by parametrising the Mandelstam invariants in term
of two independent dimensionless variables, ${w}$ and ${z}$, which are
defined by
\begin{equation}
% \frac{s-m^2}{m^2} = \frac{(1-w)^2}{w-z^2}, \qquad  \frac{t}{m^2} = -\frac{(1-w)^2}{w}\, .  \quad
\frac{u-m^2}{s-m^2}=-\frac{  z^2}{w}, \qquad  \frac{t}{m^2} = -\frac{(1-w)^2}{w}\, ,
  \label{eq:varswz}
\end{equation}
where the constraint $s+t+u=2m^2$ is understood.
% Due to , one has also
% \begin{align}
% \frac{u-m^2}{s-m^2}=-\frac{  z^2}{w}\,.
%  \end{align}
Such change of variables rationalises the canonical DEQs.\footnote{
At an earlier stage of the project, we found that the variables
$x$ and $y$, defined through,
$$
\frac{s-m^2}{m^2} = 
% - \Big( x - {1 \over 4 x} {(1-y^2)^2 \over y^2} \Big)^2 \, ,
- x^2\bigg( 1 - {(1-y^2)^2 \over 4  x^2 y^2} \bigg)^2 \, ,
\quad 
\frac{t}{m^2}=-\frac{(1-{ y}^2)^2}{{ y}^2} \ ,
$$
remove all irrational terms appearing in the system of DEQs,
individually.
However, as pointed out by  Lorenzo Tancredi - whom we acknowledge for
the suggestion -, it is sufficient to rationalise just those
combinations of irrational terms that appear in the DEQs, by means of $w$ and $z$
defined through in eq.~\eqref{eq:varswz}, to yield a
polynomial alphabet.
}

\begin{figure}[H]
  \centering
  \captionsetup[subfigure]{labelformat=empty}
  \subfloat[$\mathcal{T}_1$]{%
    \includegraphics[width=0.11\textwidth]{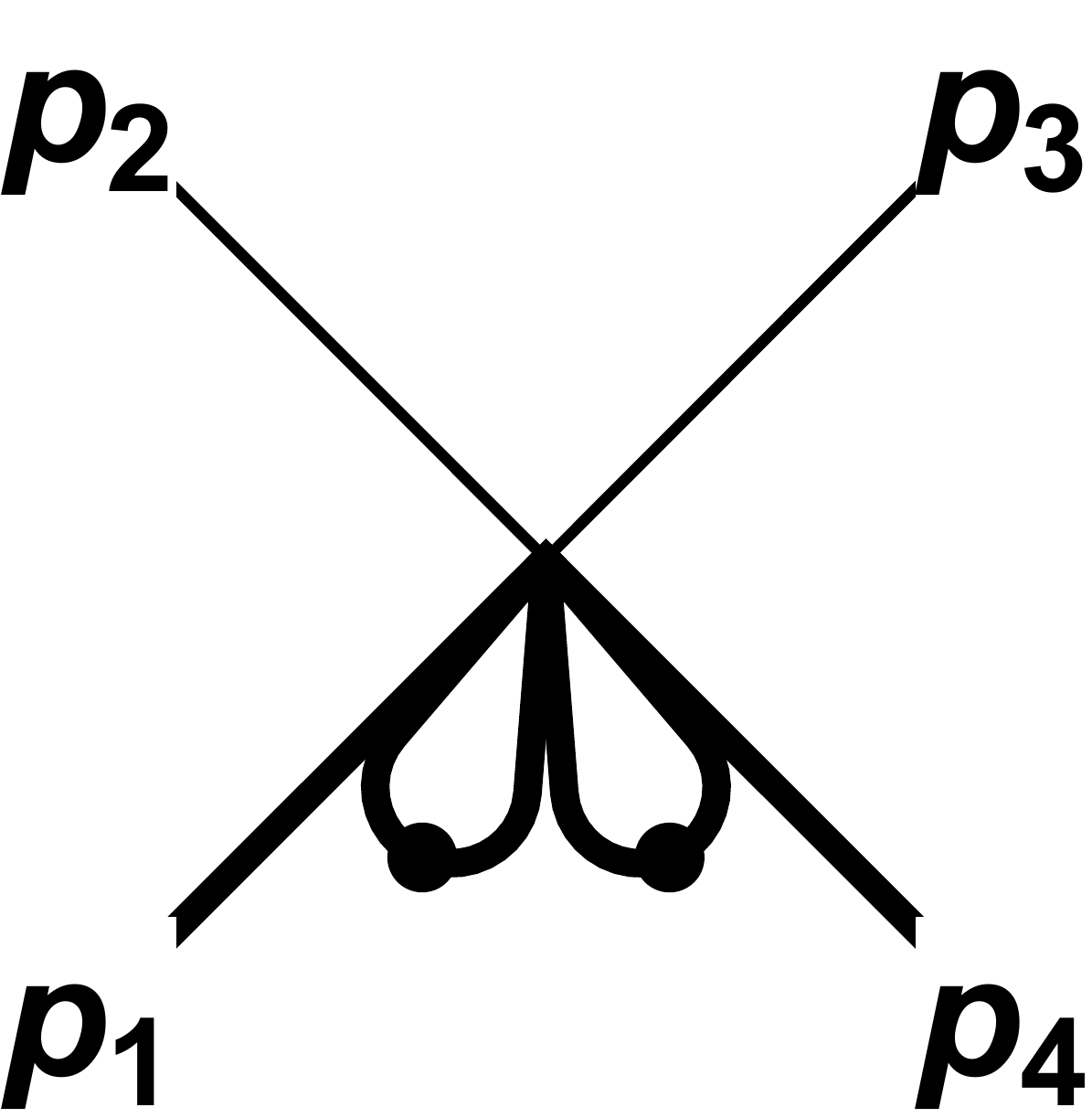}
  }
  \subfloat[$\mathcal{T}_2$]{%
    \includegraphics[width=0.11\textwidth]{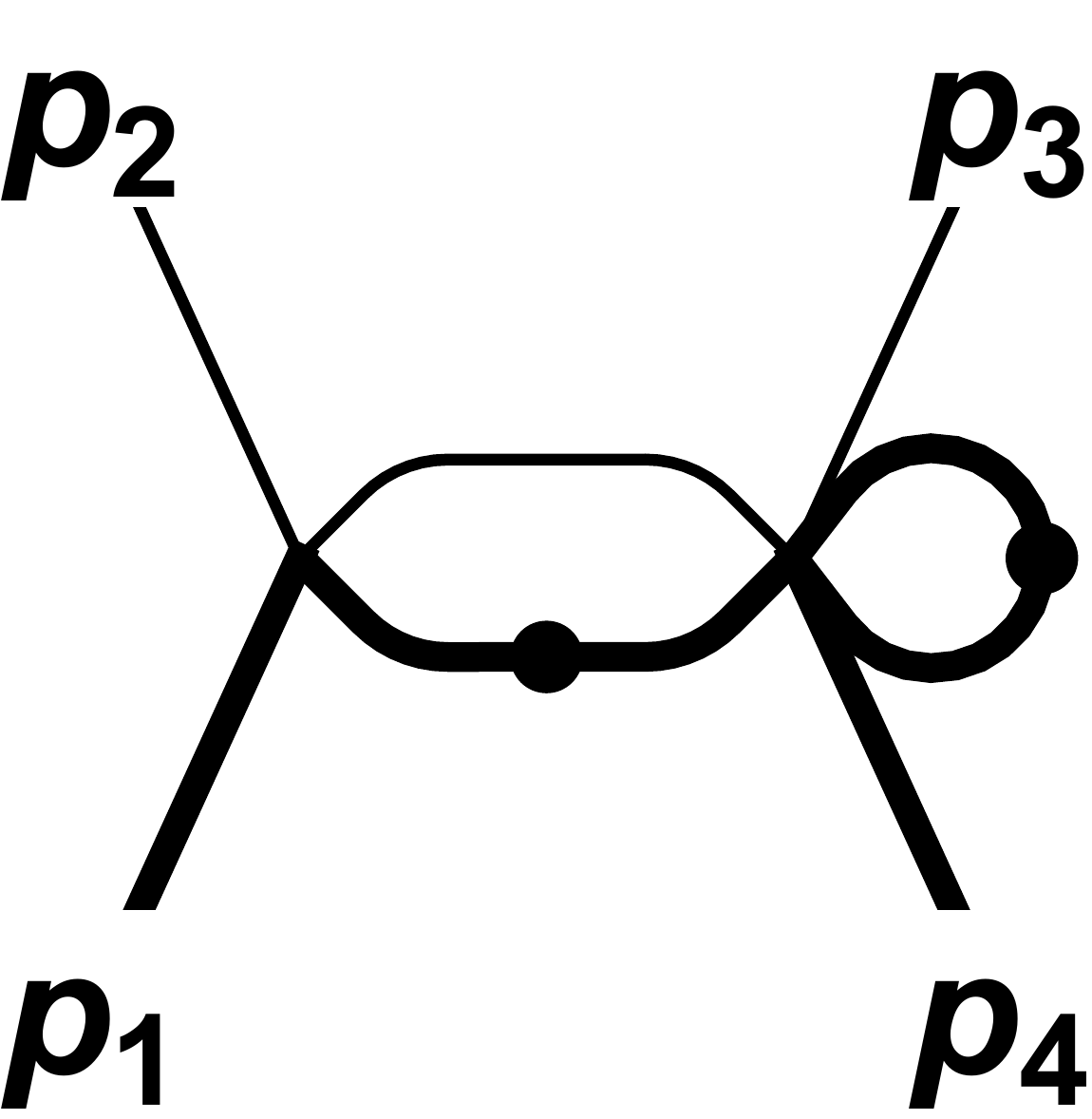}
  }
  \subfloat[$\mathcal{T}_3$]{%
    \includegraphics[width=0.11\textwidth]{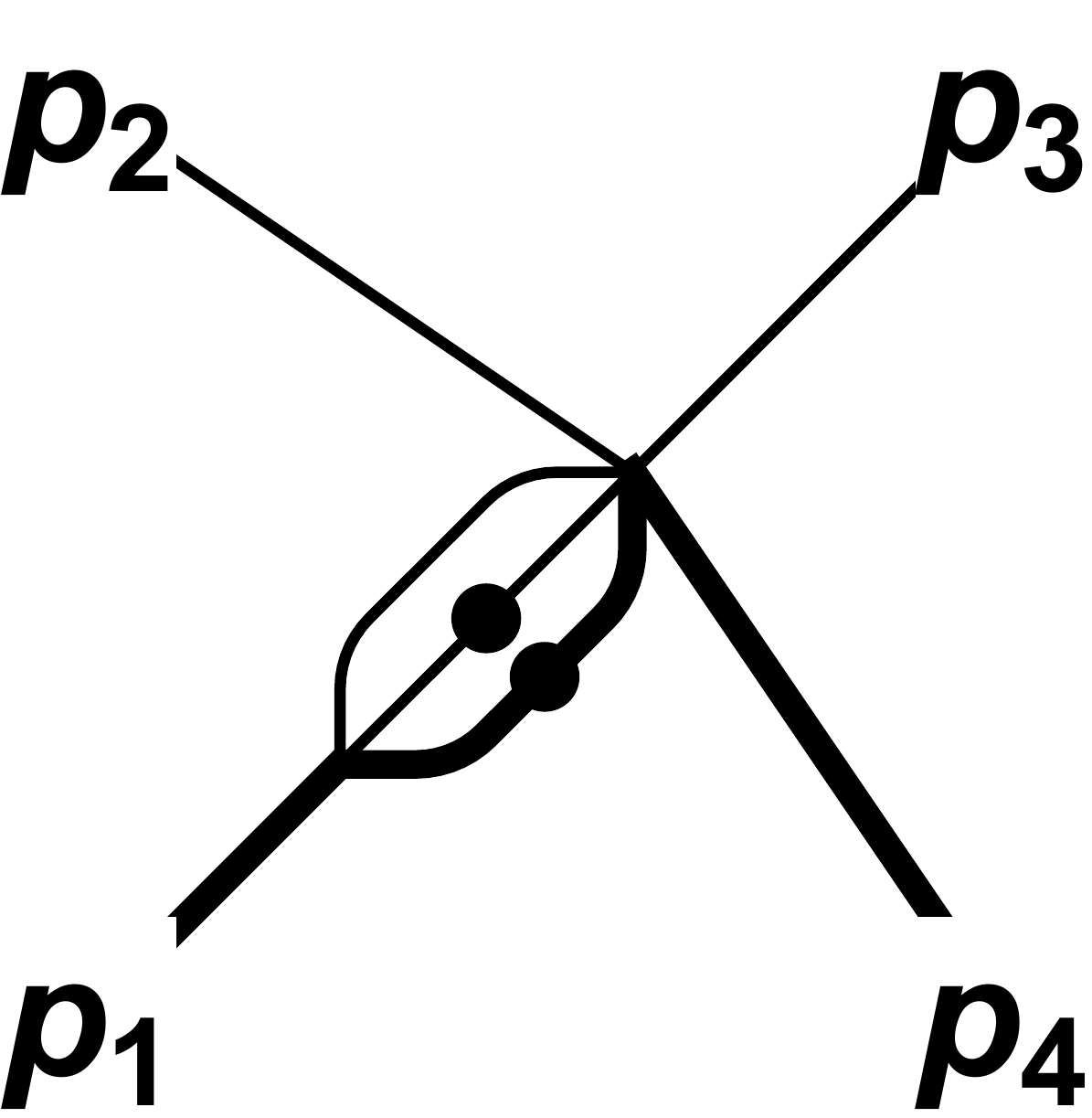}
  }
  \subfloat[$\mathcal{T}_4$]{%
    \includegraphics[width=0.11\textwidth]{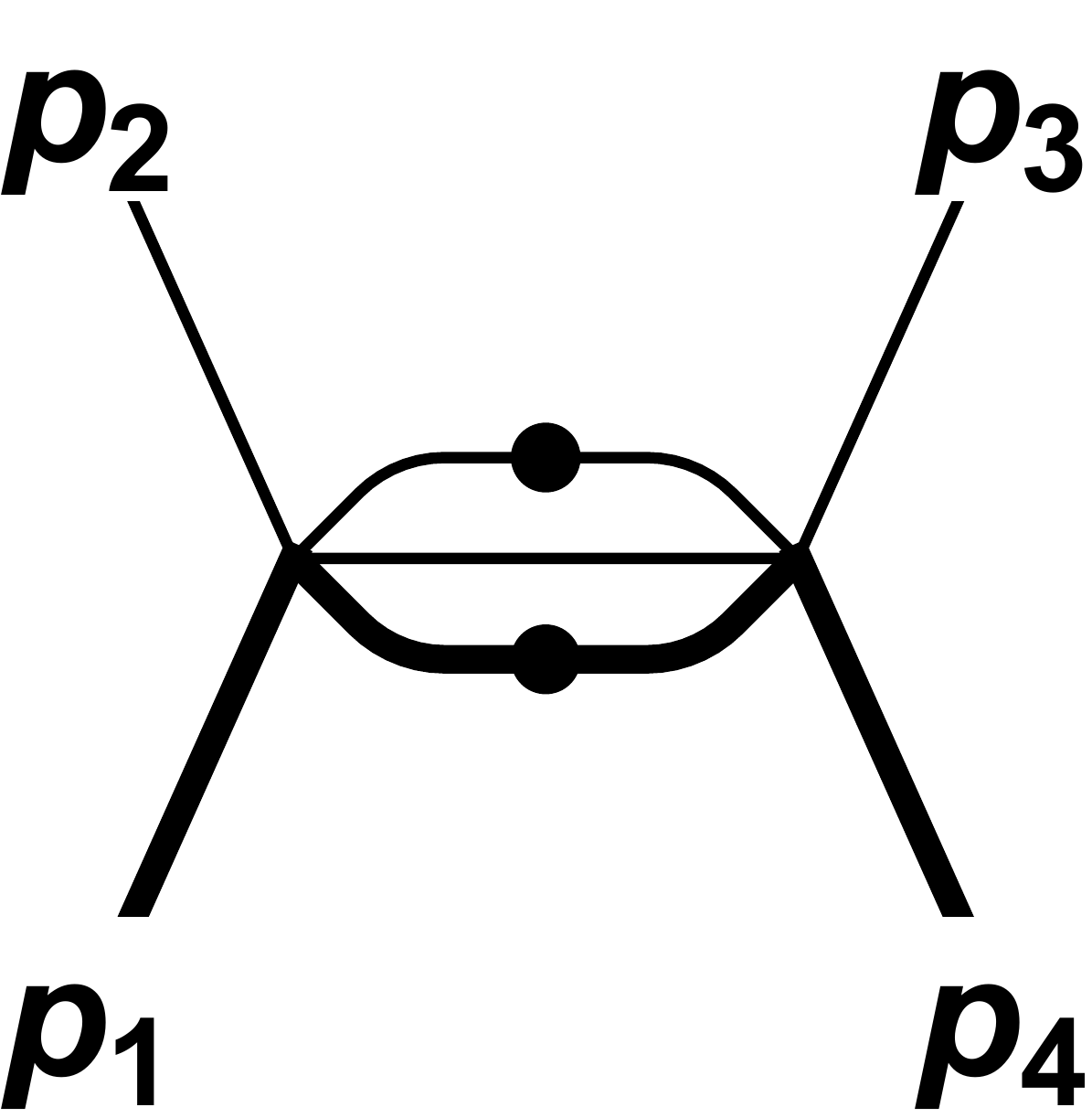}
  }
  \subfloat[$\mathcal{T}_5$]{%
    \includegraphics[width=0.11\textwidth]{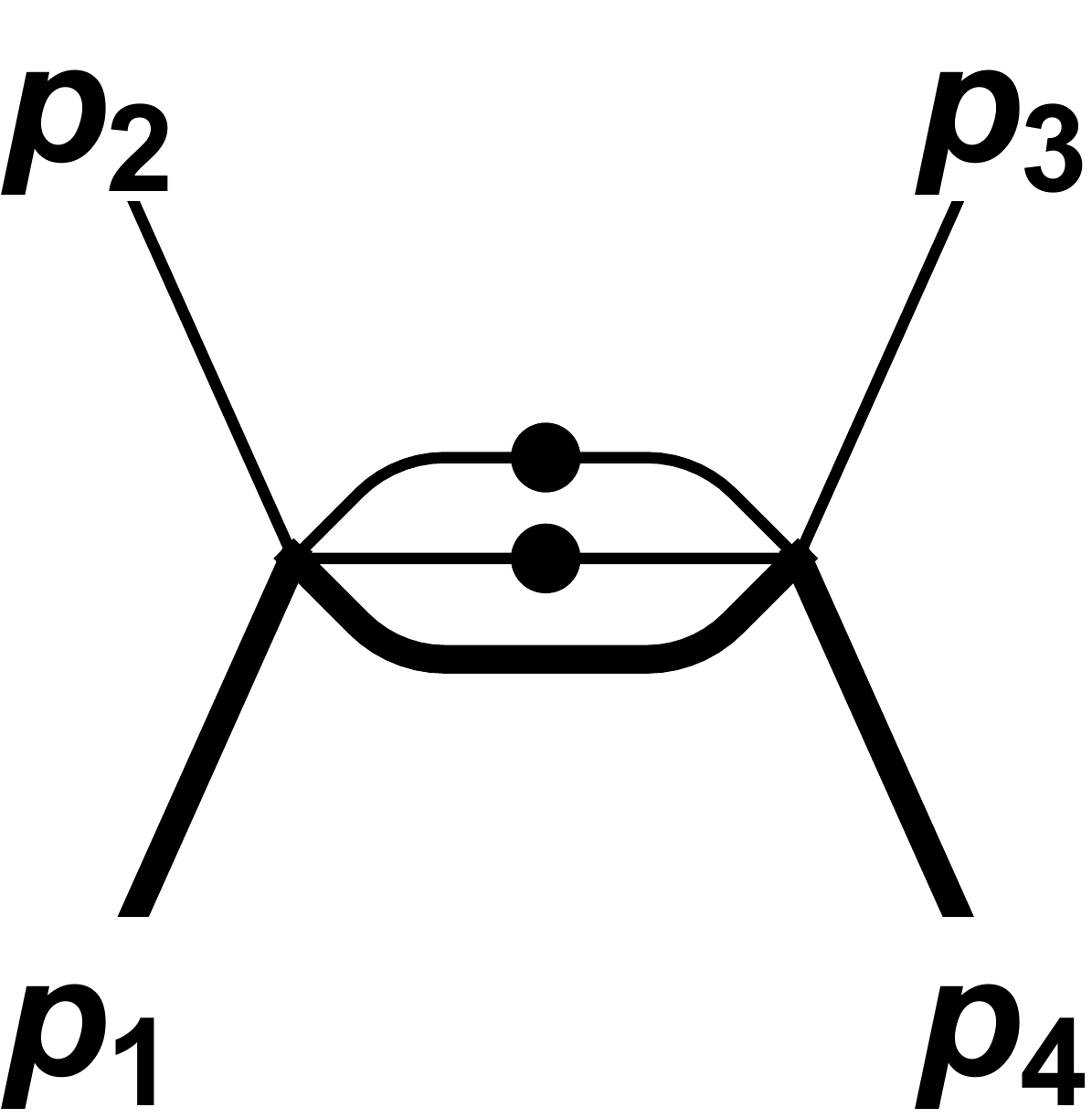}
  }.
  \subfloat[$\mathcal{T}_6$]{%
    \includegraphics[width=0.11\textwidth]{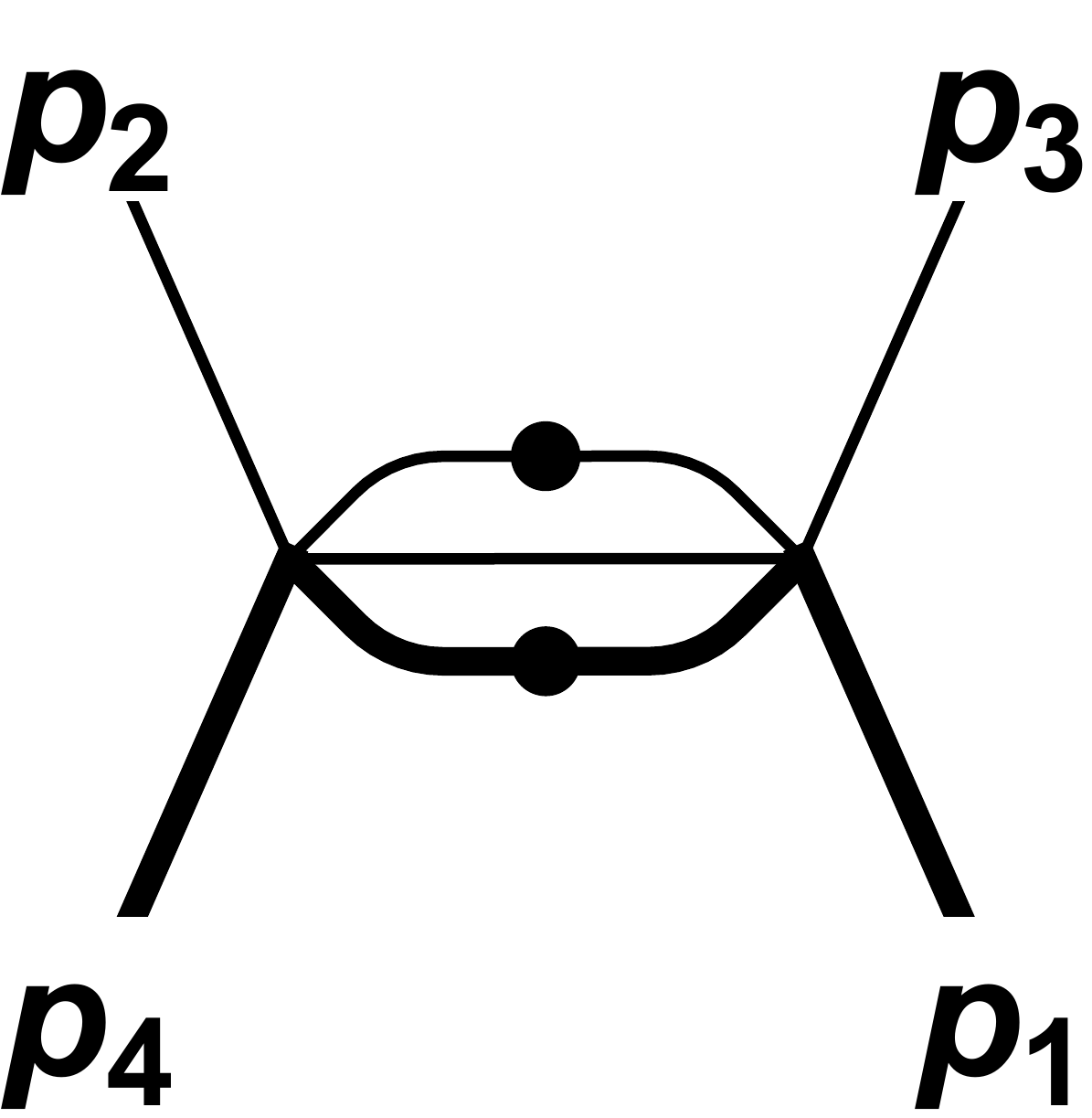}
  } \\
  \subfloat[$\mathcal{T}_7$]{%
    \includegraphics[width=0.11\textwidth]{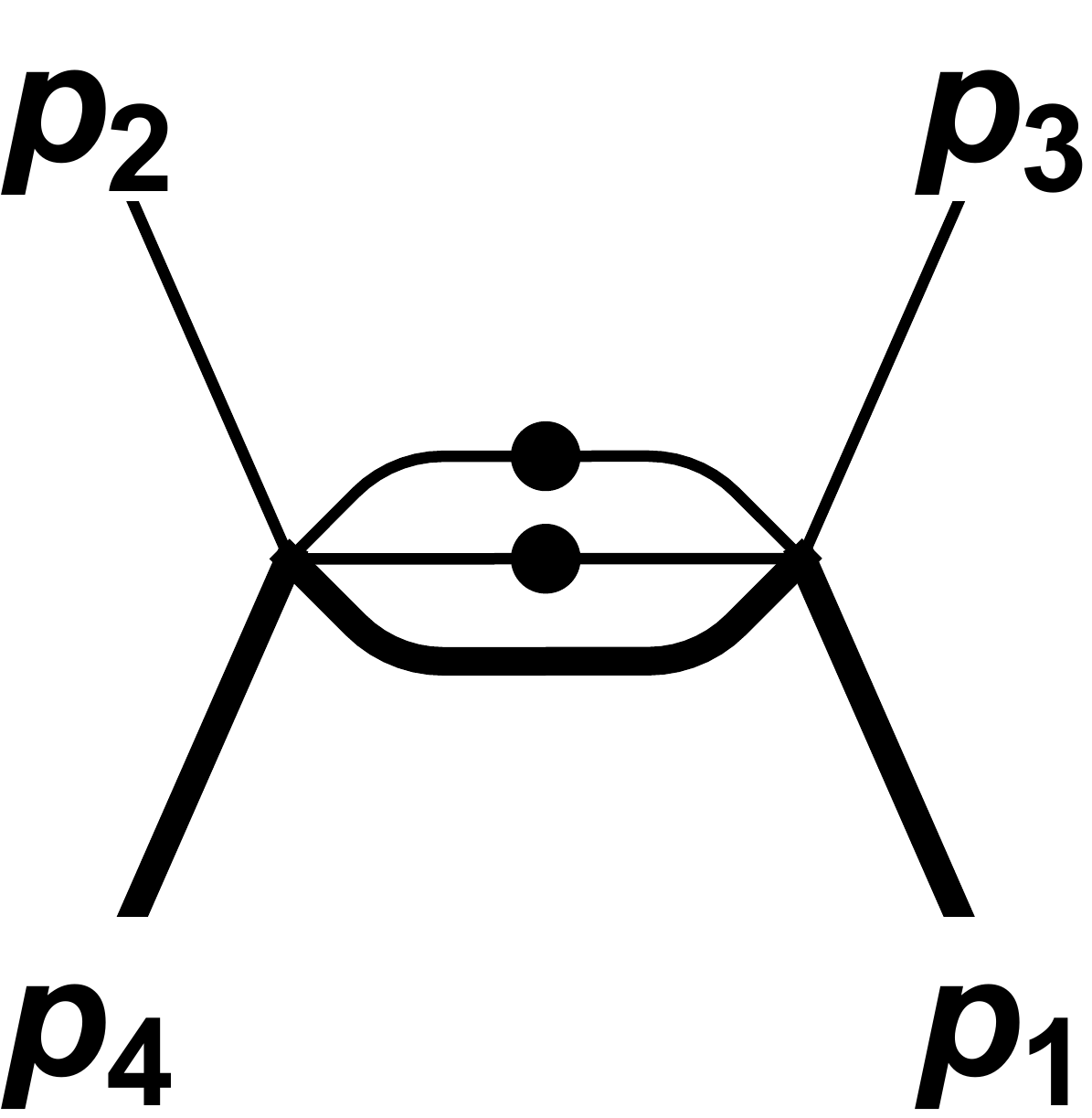}
  }
  \subfloat[$\mathcal{T}_8$]{%
    \includegraphics[width=0.11\textwidth]{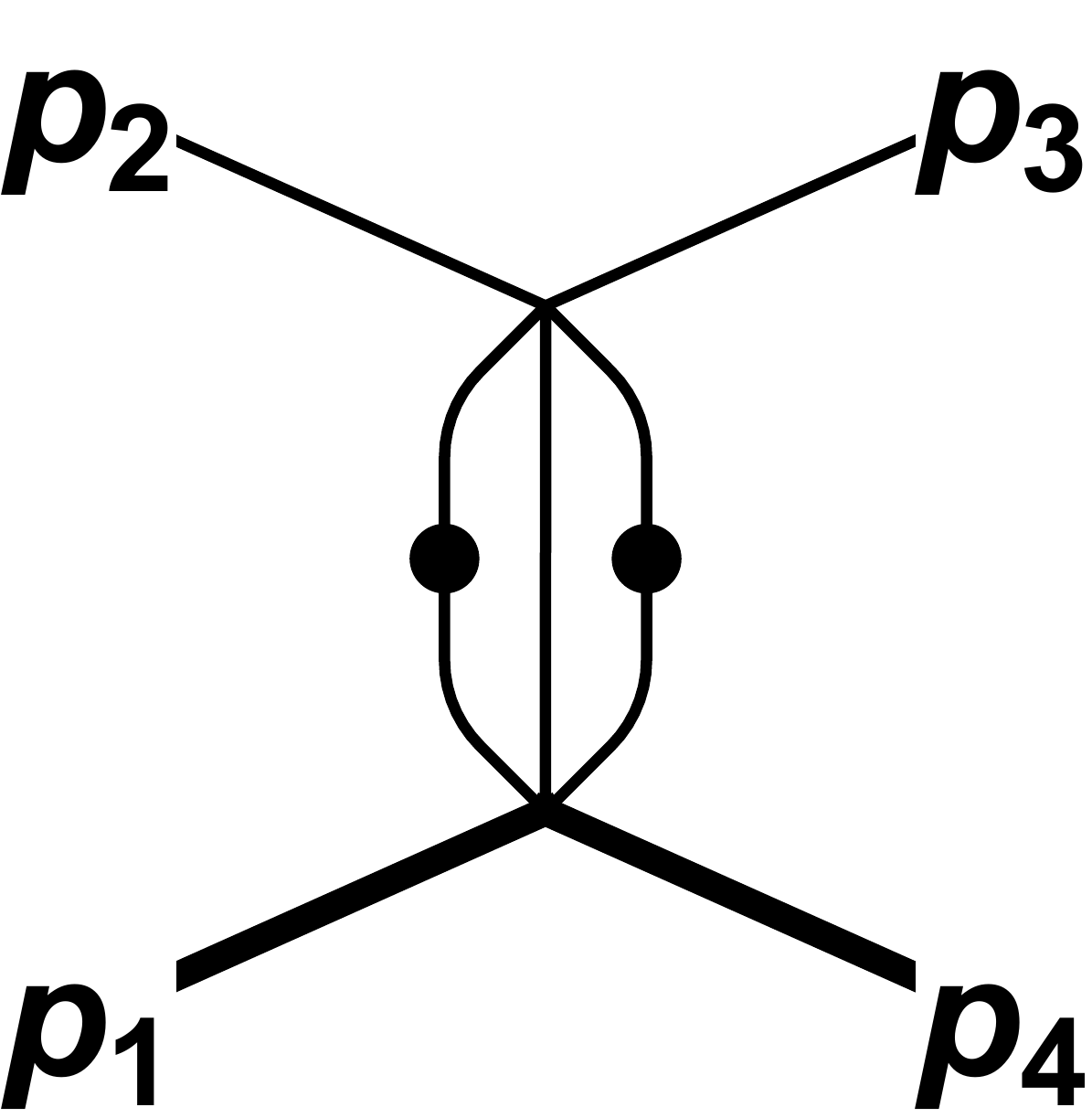}
  }
  \subfloat[$\mathcal{T}_9$]{%
    \includegraphics[width=0.11\textwidth]{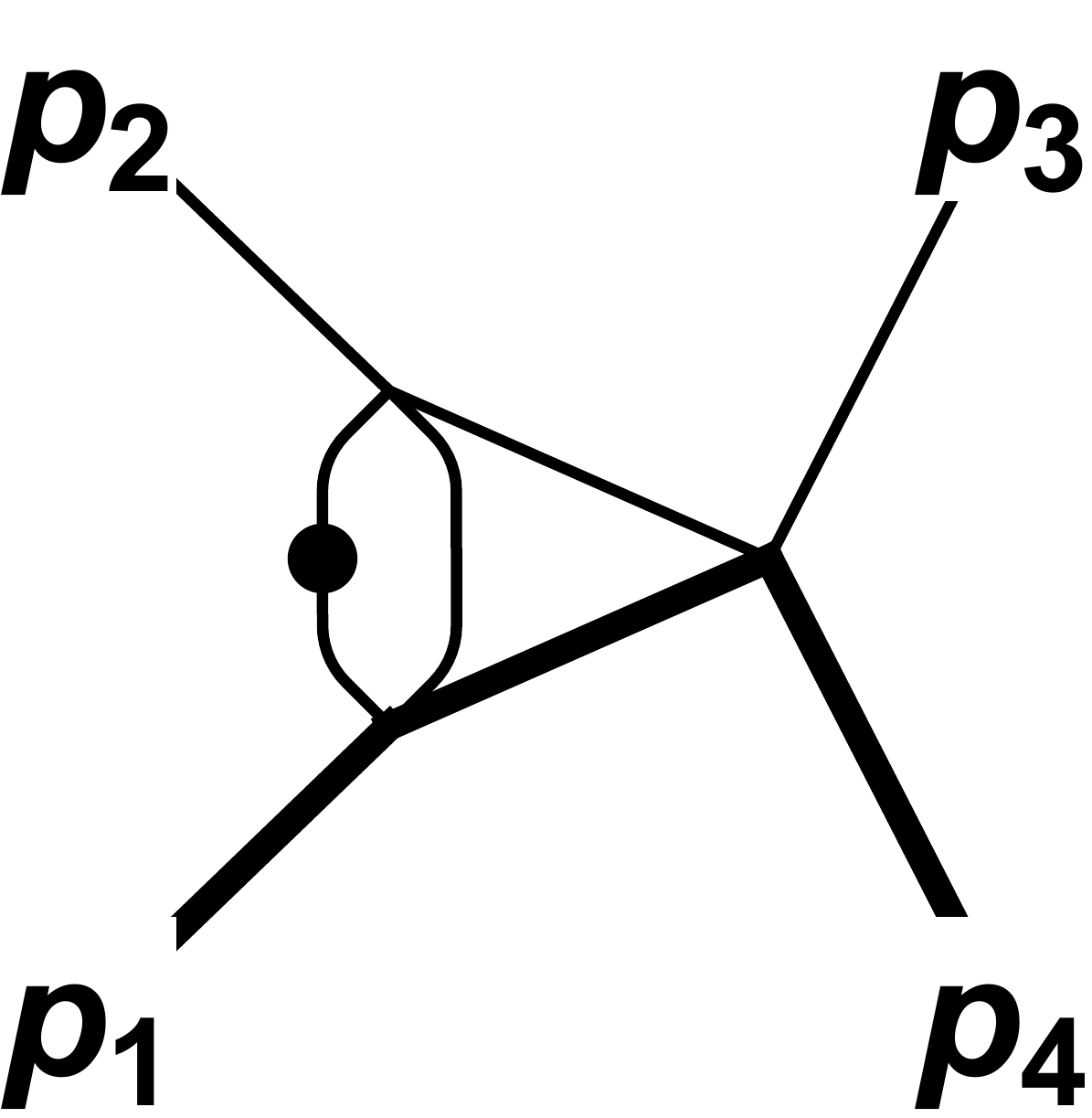}
  }
  \subfloat[$\mathcal{T}_{10}$]{%
    \includegraphics[width=0.11\textwidth]{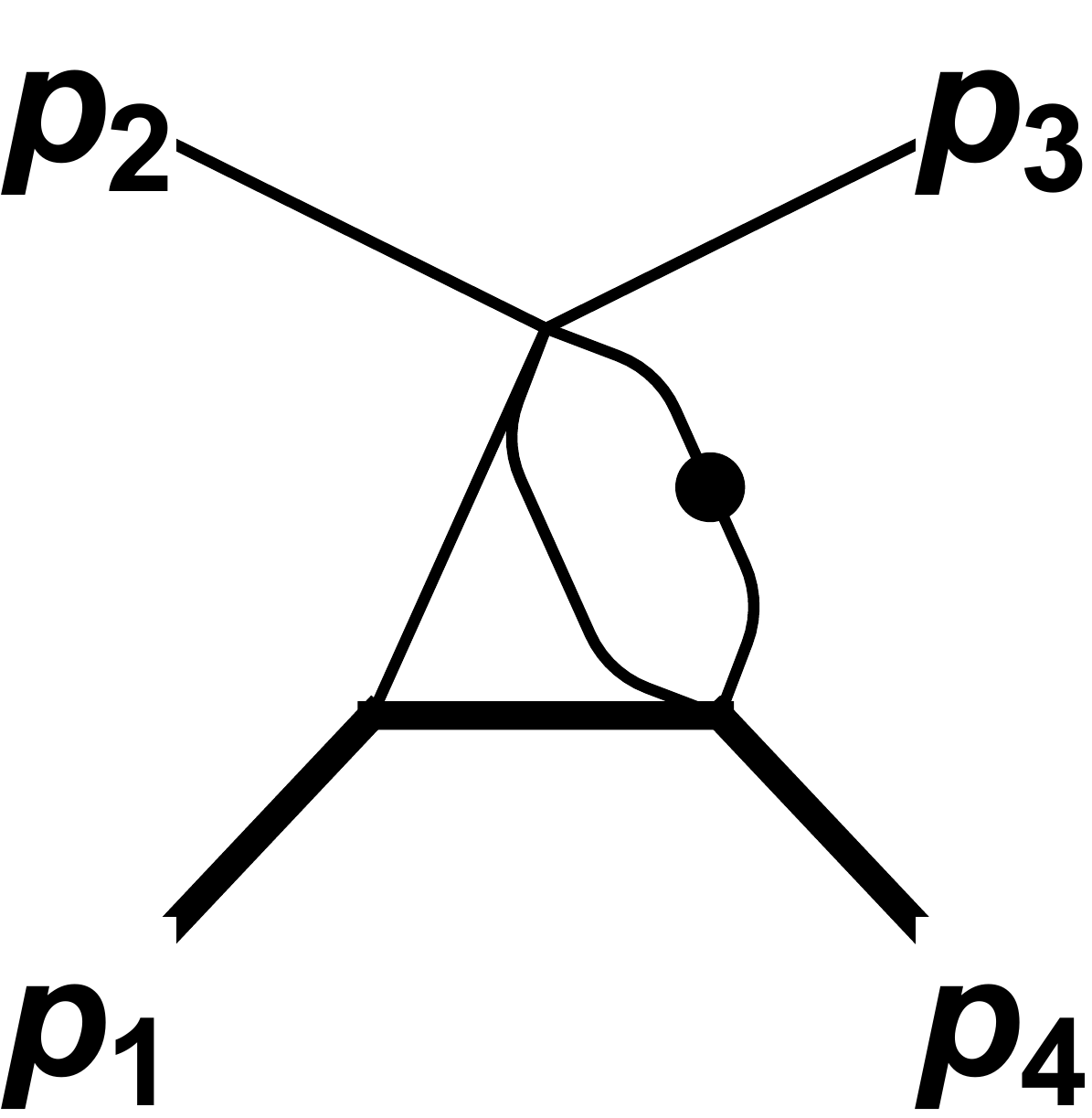}
  }
  \subfloat[$\mathcal{T}_{11}$]{%
    \includegraphics[width=0.11\textwidth]{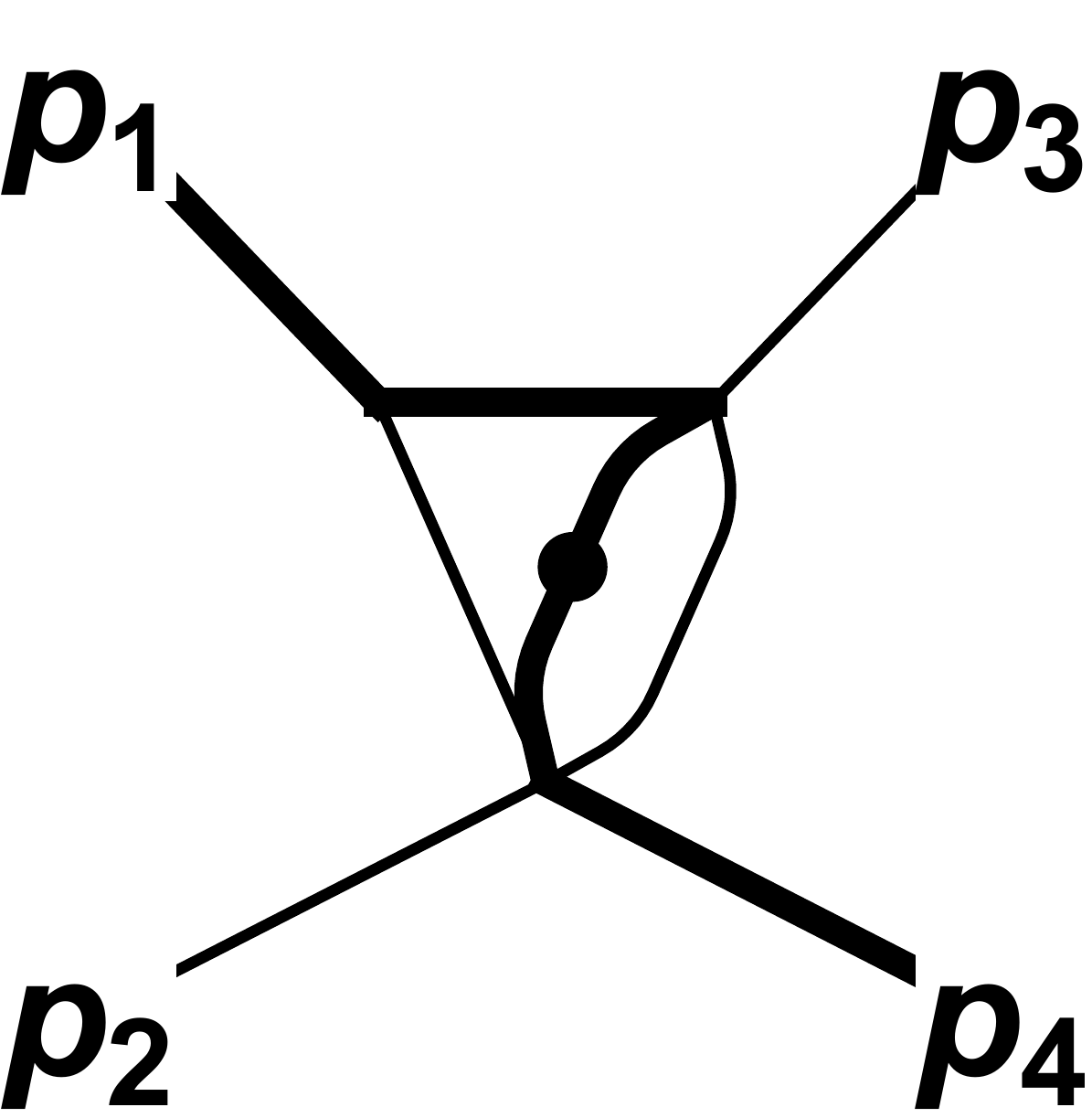}
  }
  \subfloat[$\mathcal{T}_{12}$]{%
    \includegraphics[width=0.11\textwidth]{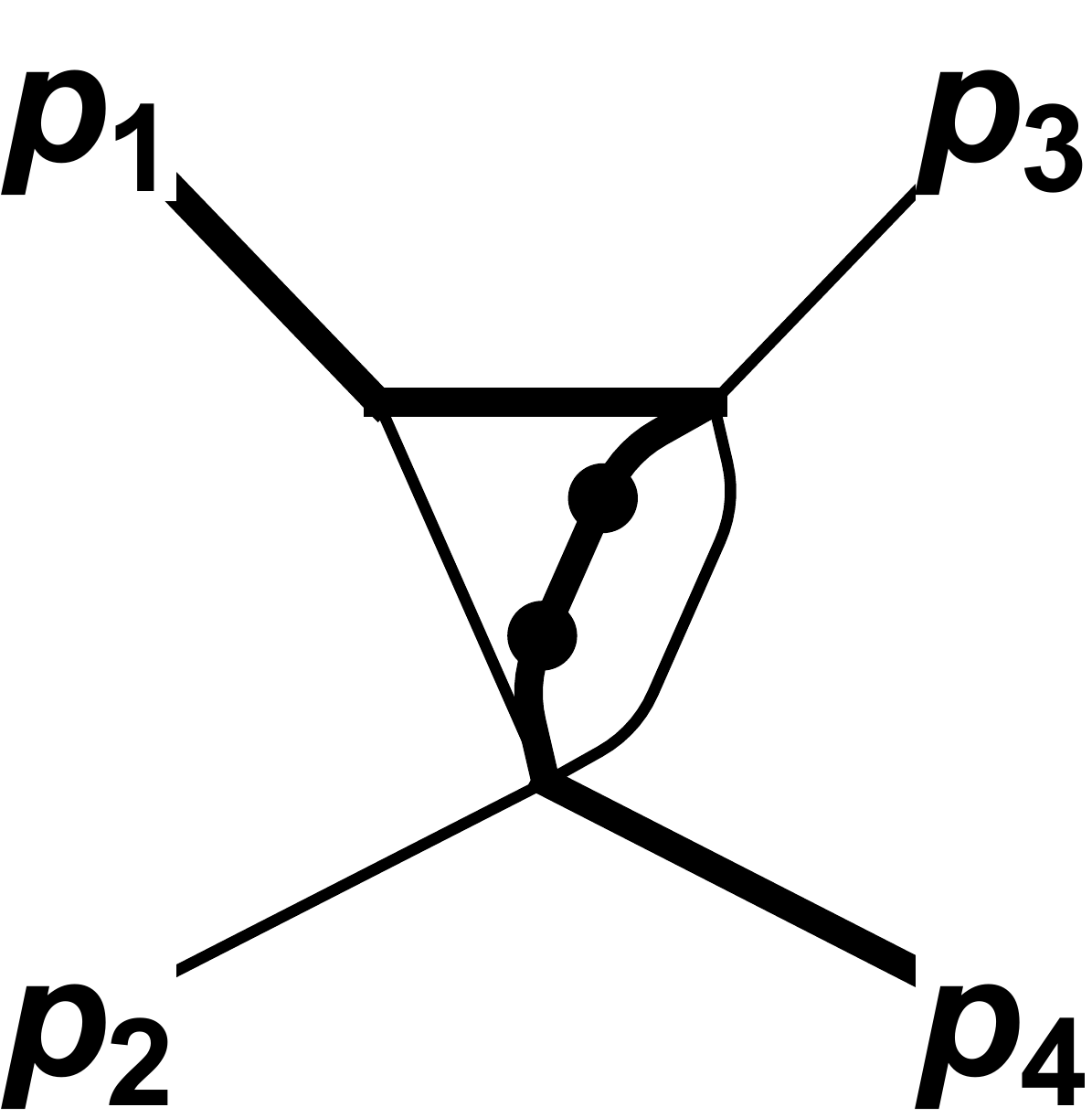}
  } \\
  \subfloat[$\mathcal{T}_{13}$]{%
    \includegraphics[width=0.11\textwidth]{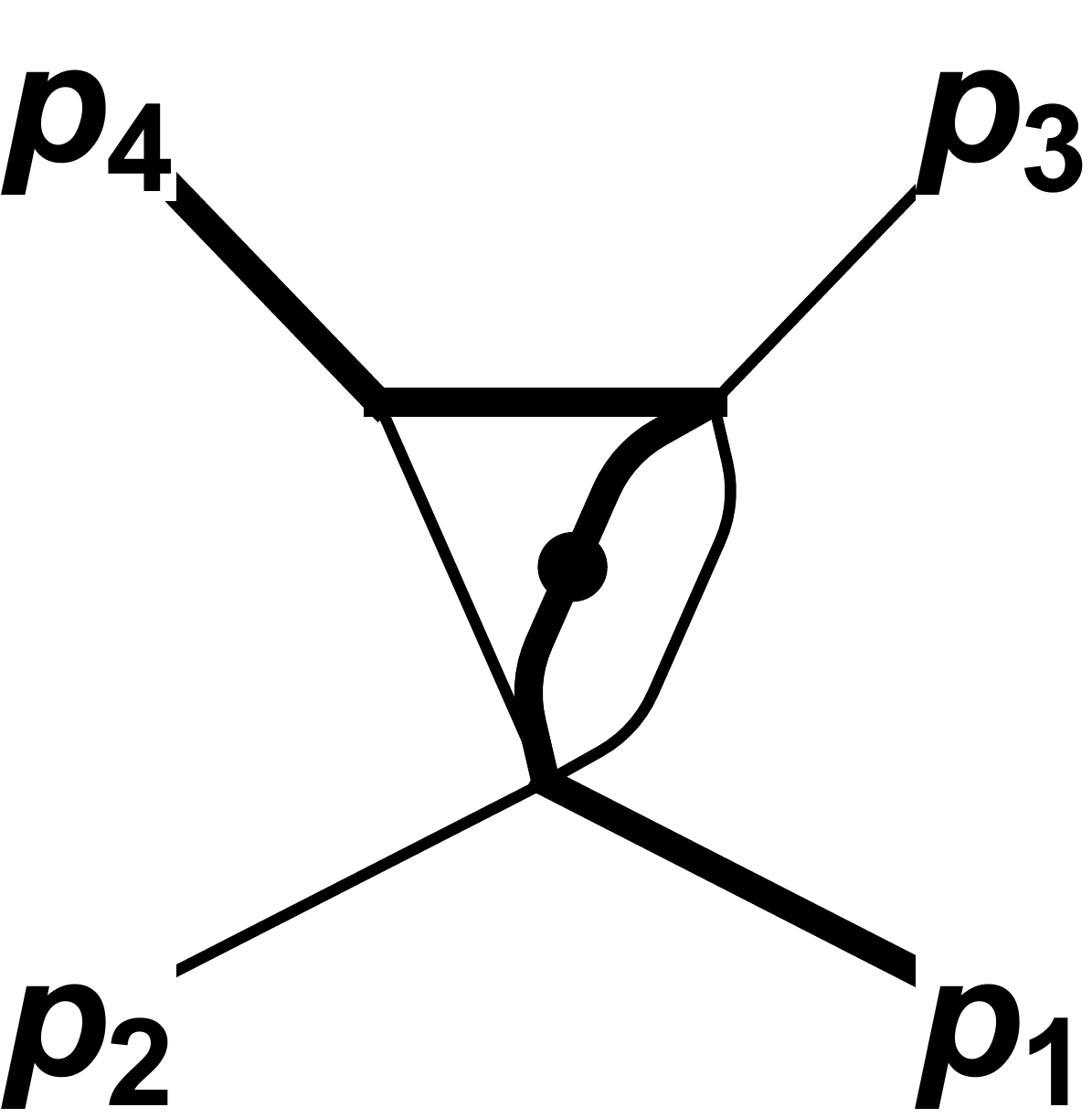}
  }
  \subfloat[$\mathcal{T}_{14}$]{%
    \includegraphics[width=0.11\textwidth]{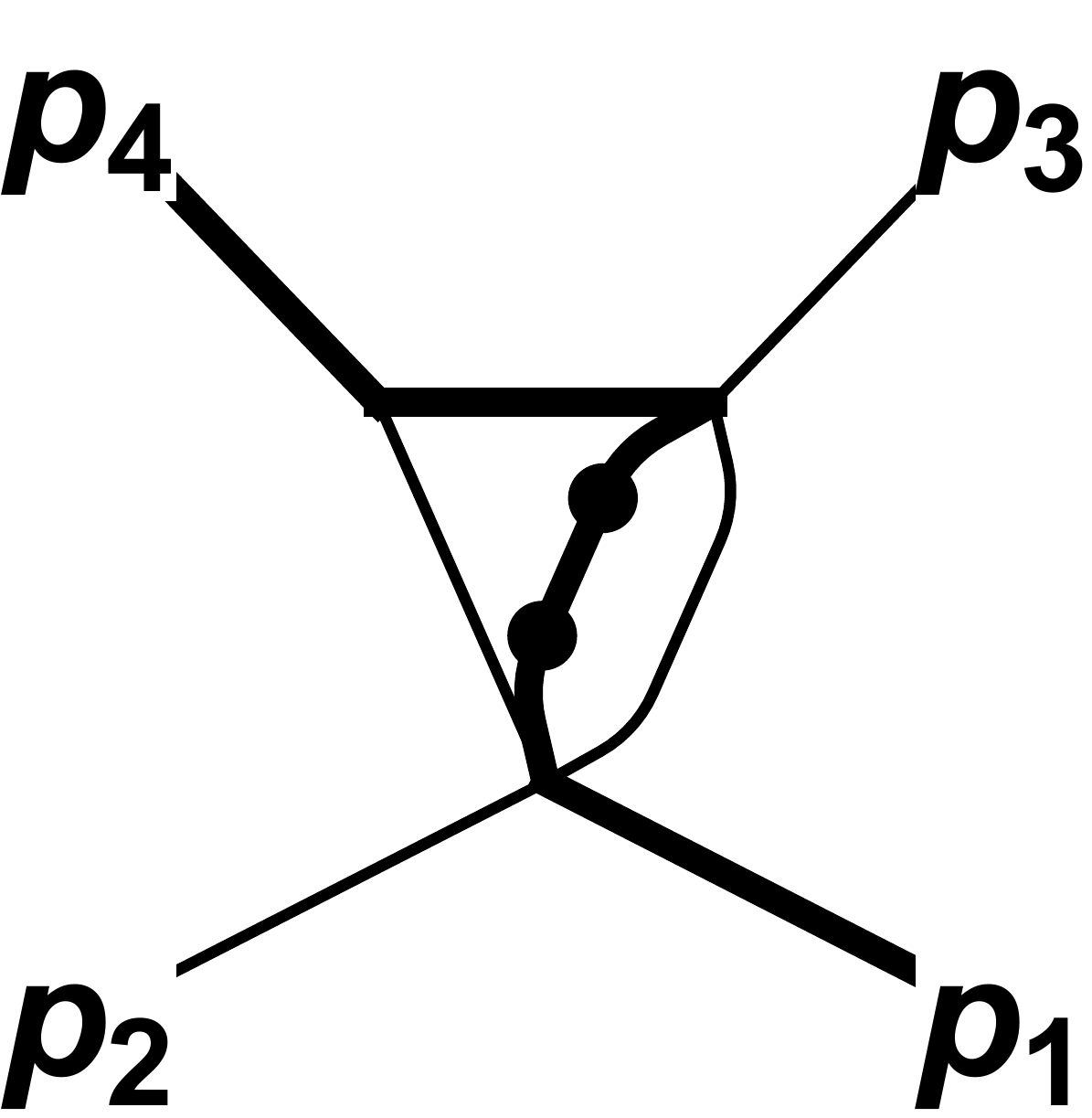}
  }
  \subfloat[$\mathcal{T}_{15}$]{%
    \includegraphics[width=0.11\textwidth]{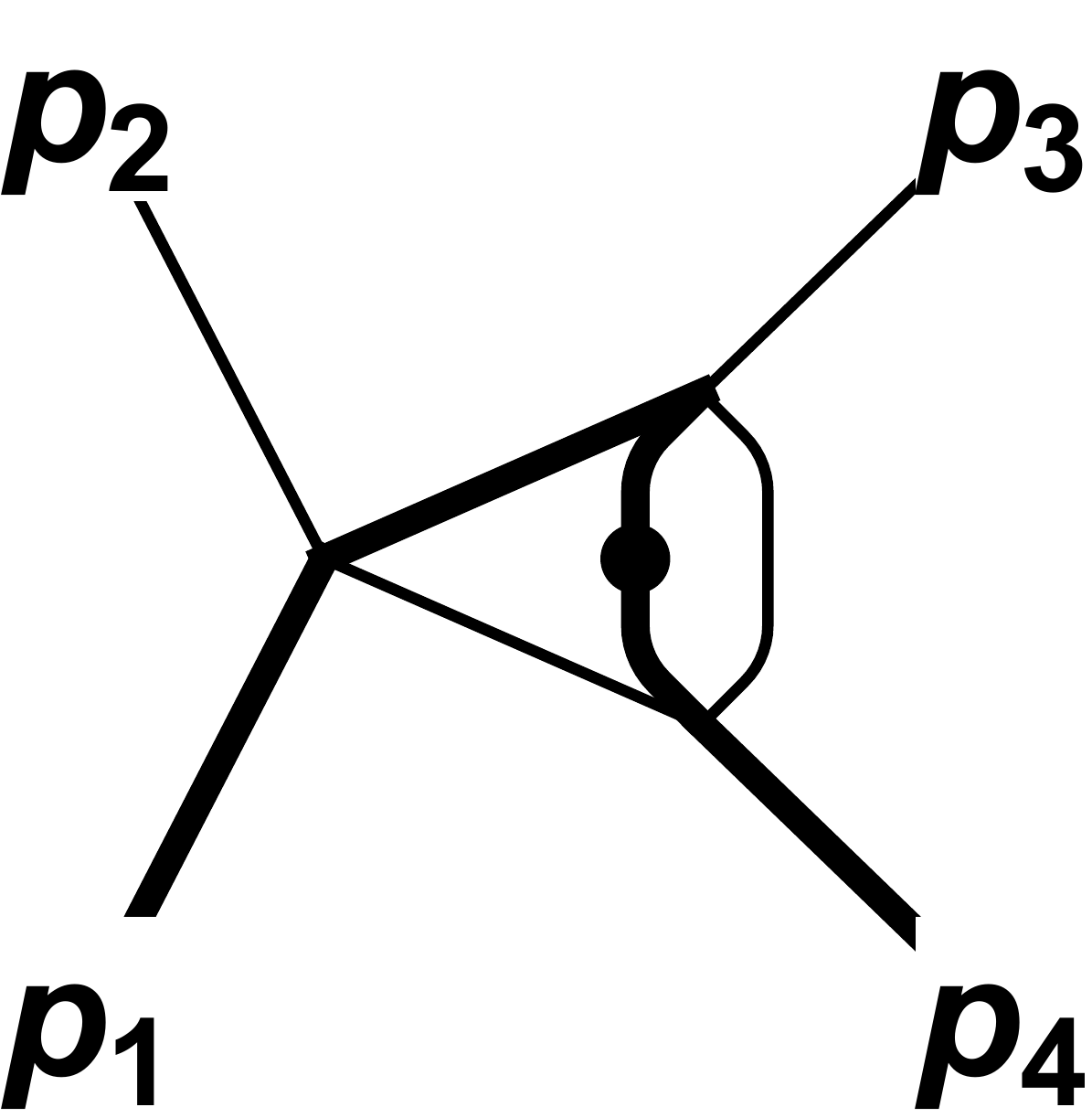}
  }
  \subfloat[$\mathcal{T}_{16}$]{%
    \includegraphics[width=0.11\textwidth]{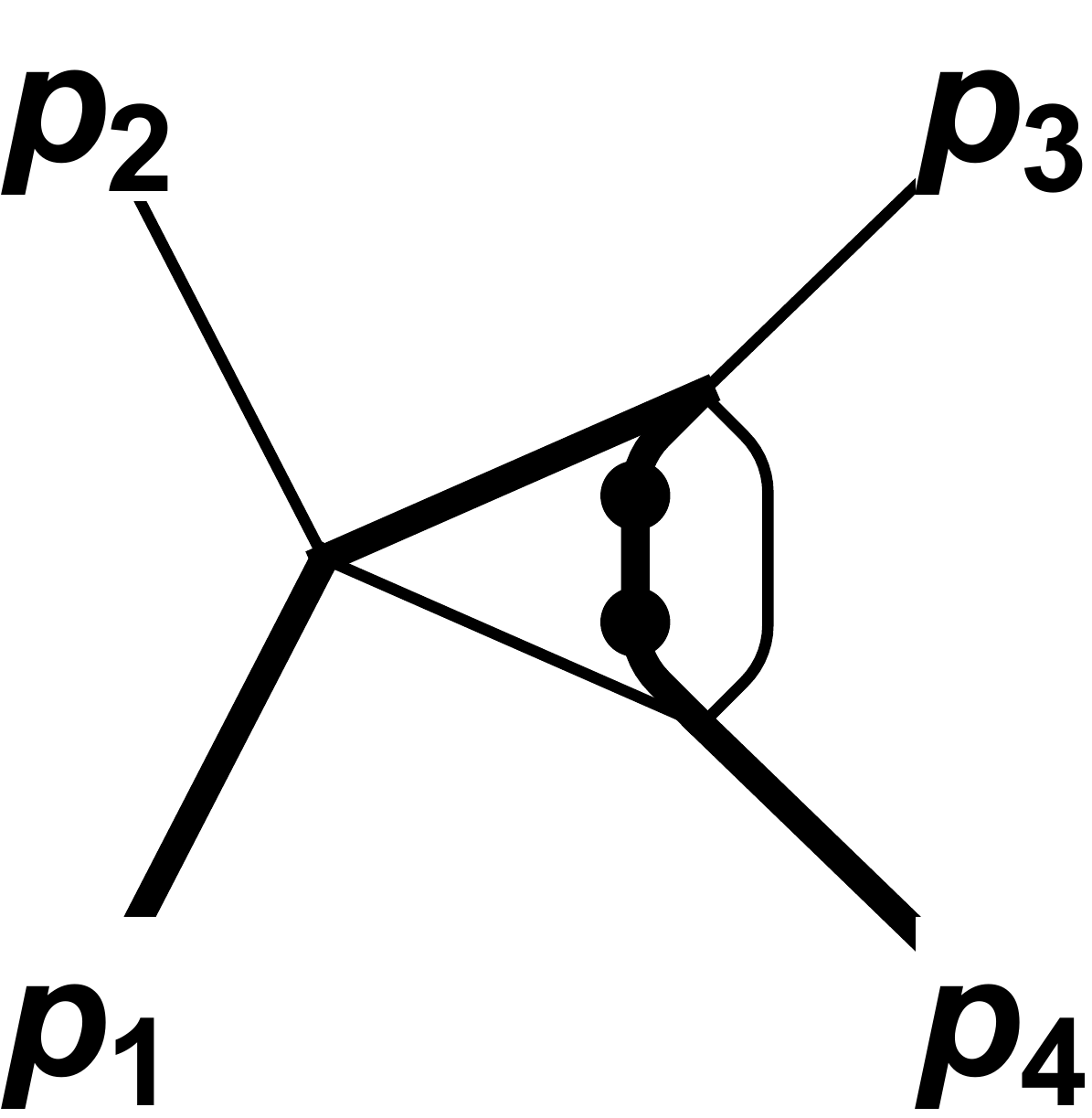}
  }
  \subfloat[$\mathcal{T}_{17}$]{%
    \includegraphics[width=0.11\textwidth]{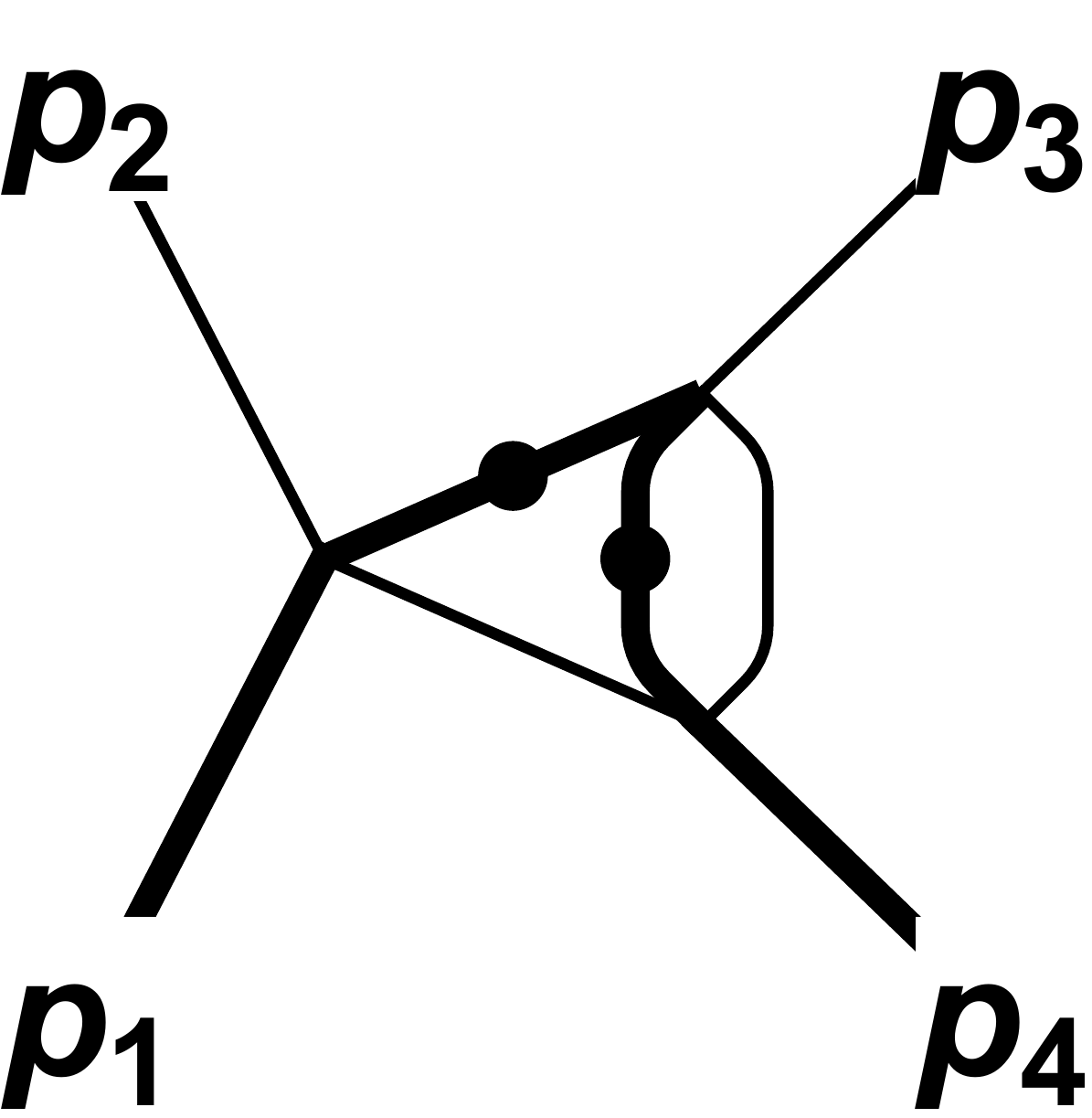}
  }
  \subfloat[$\mathcal{T}_{18}$]{%
    \includegraphics[width=0.11\textwidth]{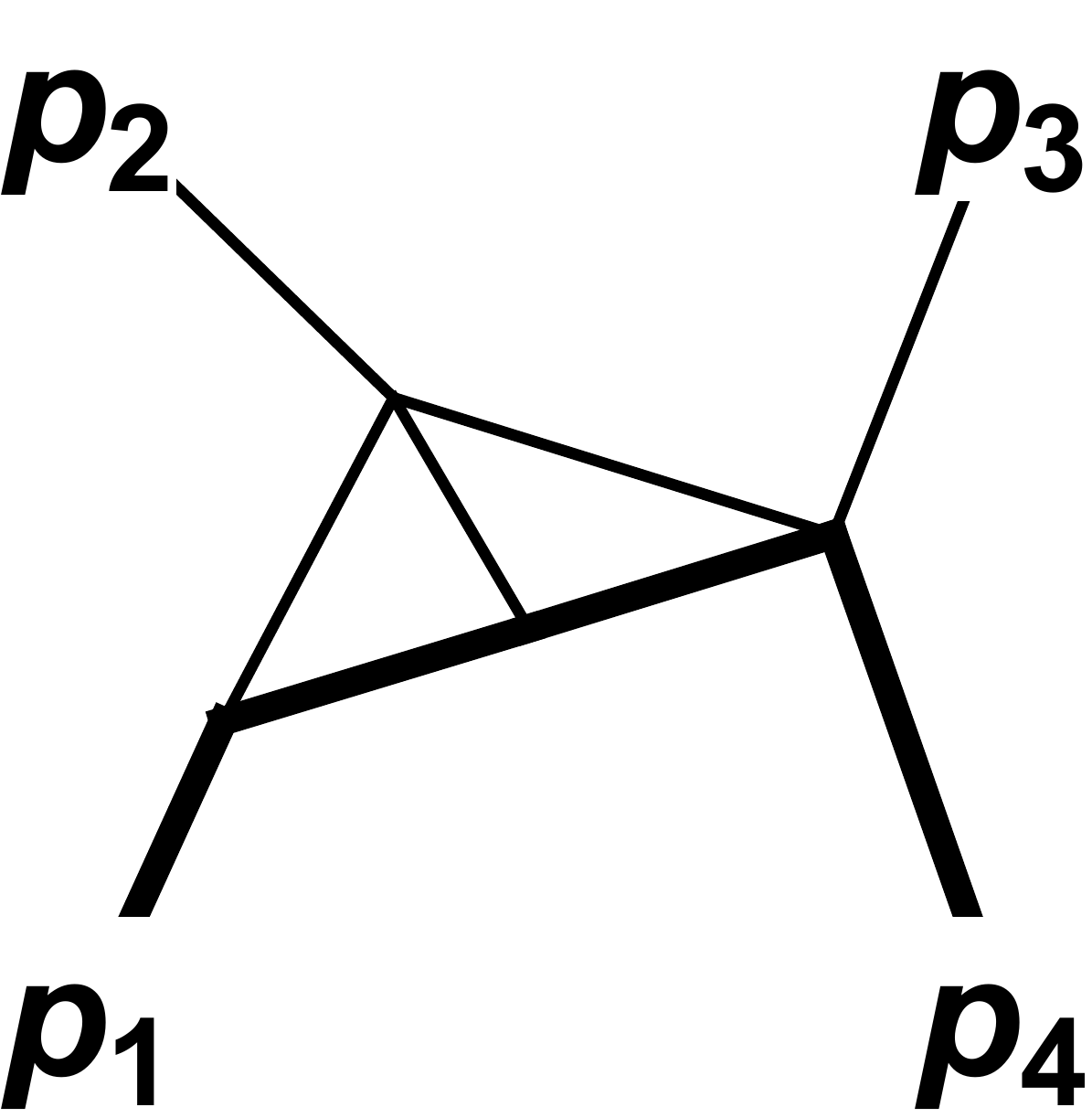}
  } \\
    \subfloat[$\mathcal{T}_{19}$]{%
    \includegraphics[width=0.11\textwidth]{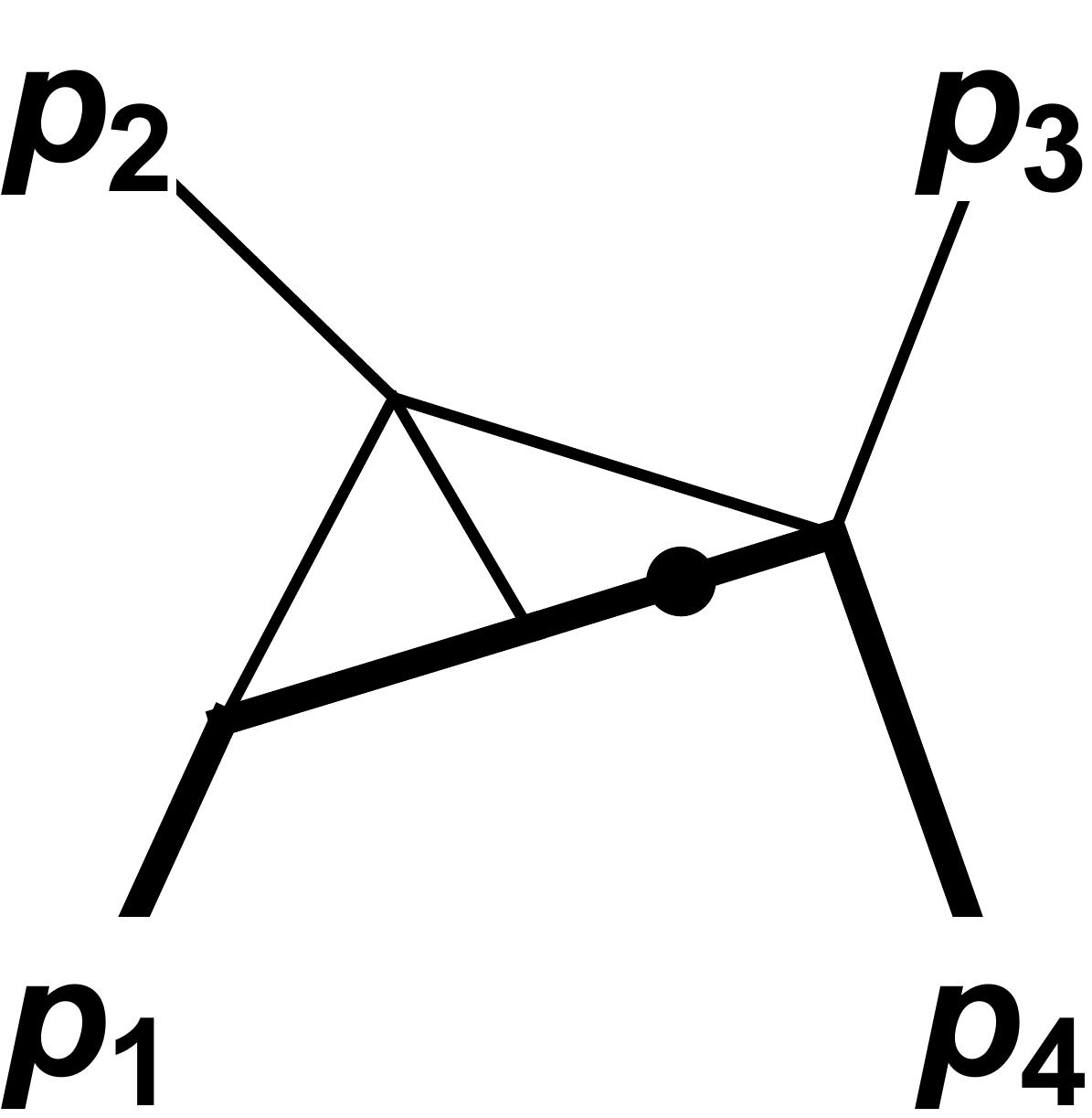}
  }
  \subfloat[$\mathcal{T}_{20}$]{%
    \includegraphics[width=0.11\textwidth]{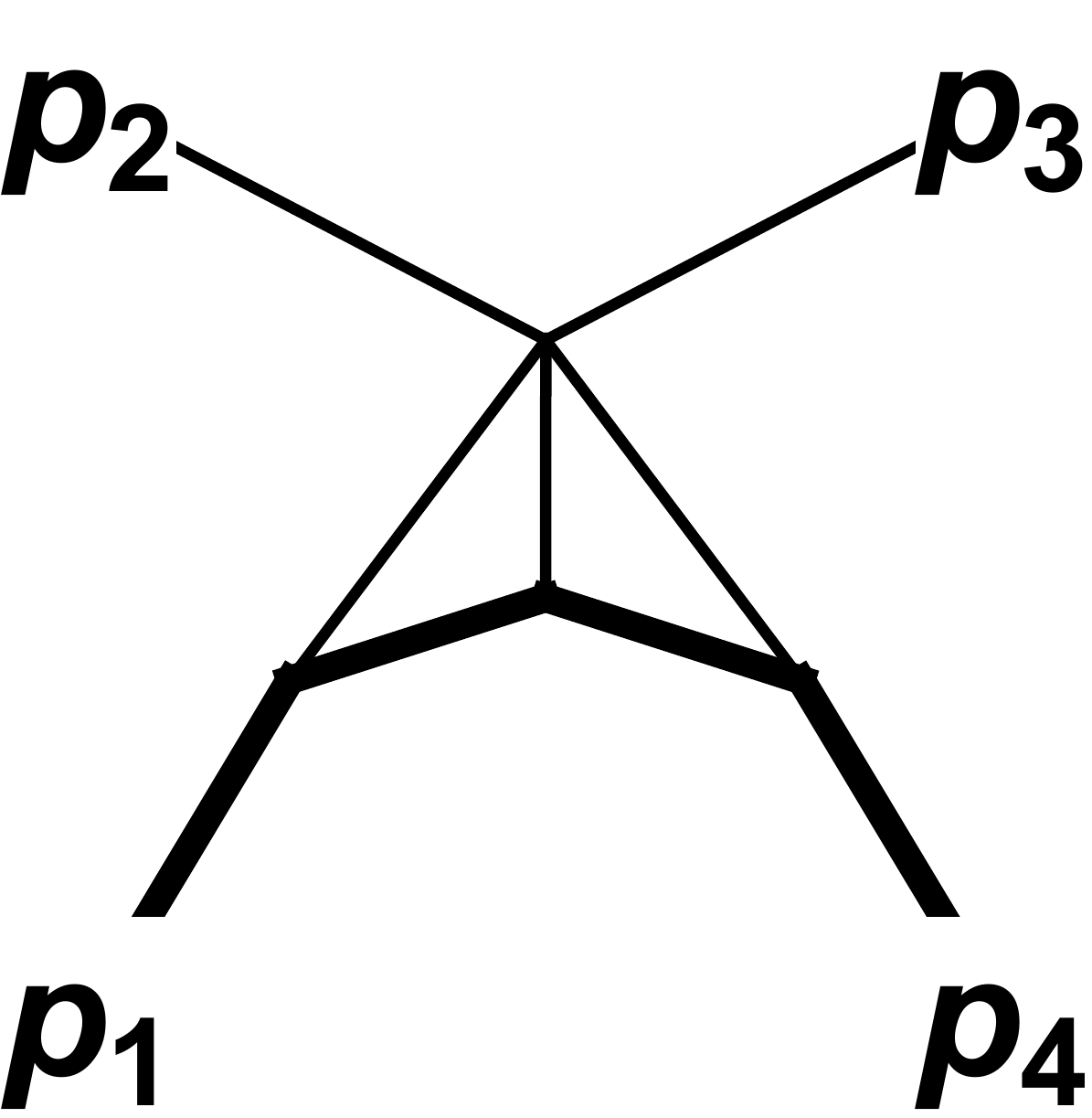}
  }
  \subfloat[$\mathcal{T}_{21}$]{%
    \includegraphics[width=0.11\textwidth]{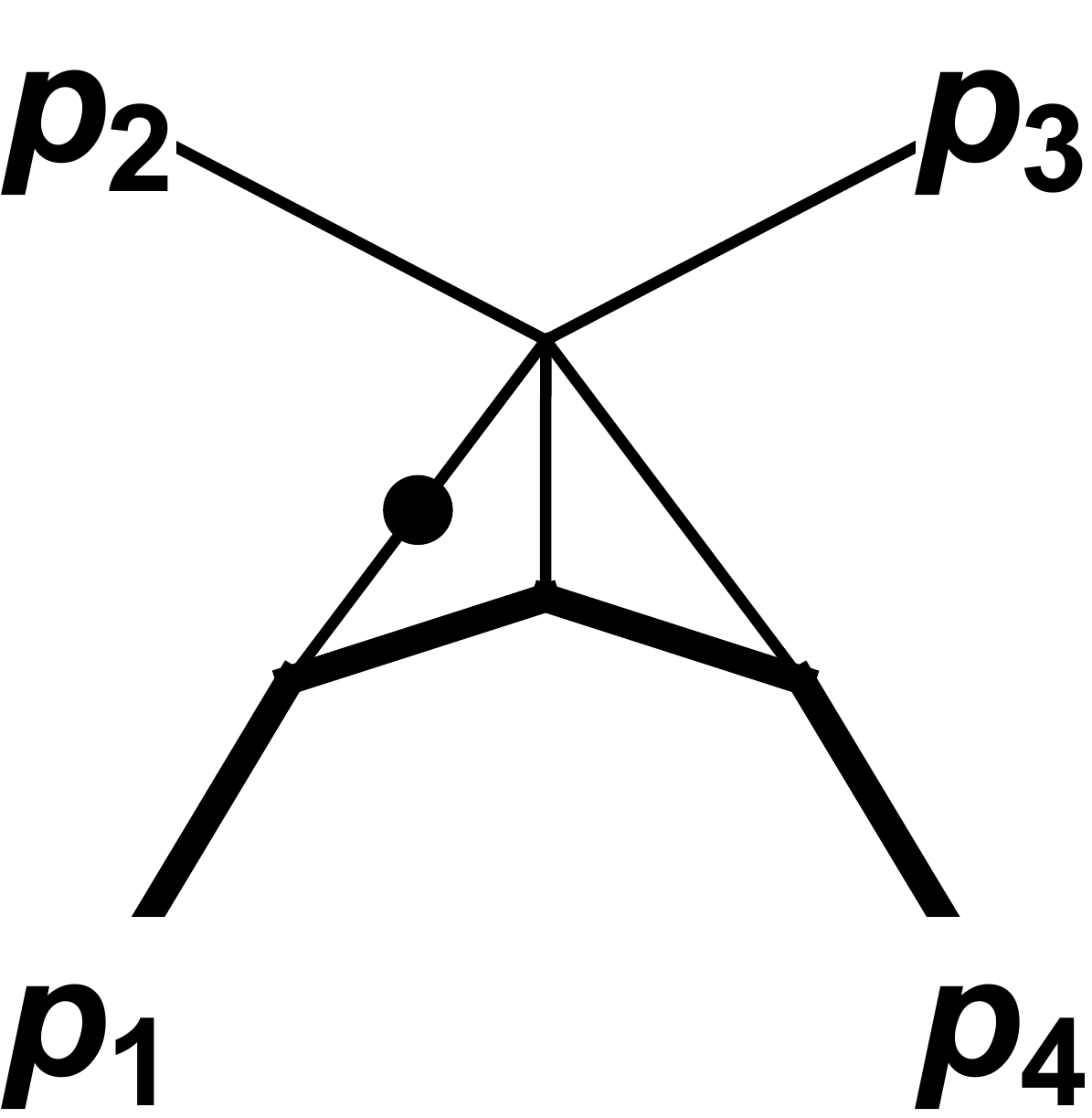}
  }
  \subfloat[$\mathcal{T}_{22}$]{%
    \includegraphics[width=0.11\textwidth]{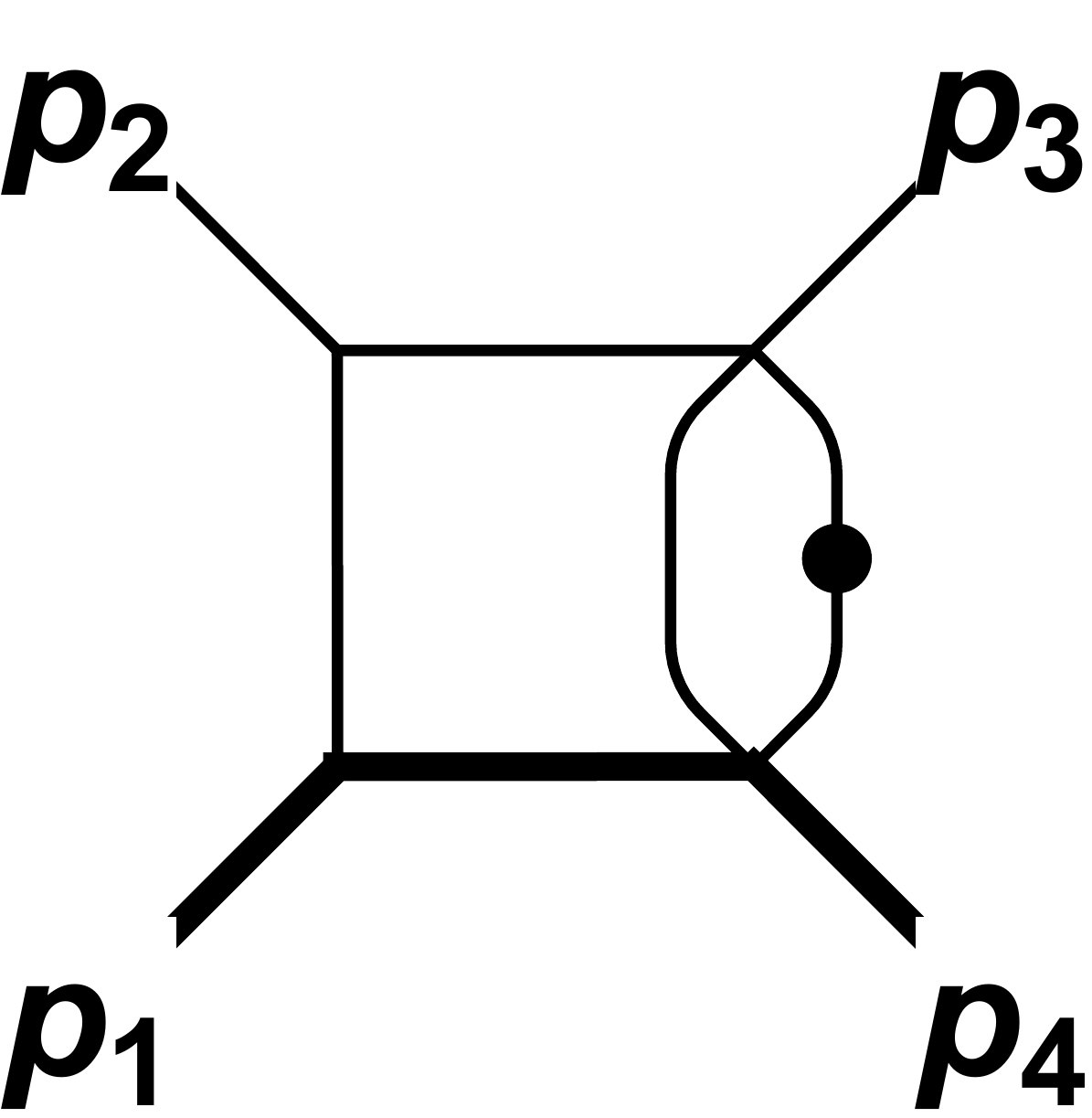}
  }
  \subfloat[$\mathcal{T}_{23}$]{%
    \includegraphics[width=0.11\textwidth]{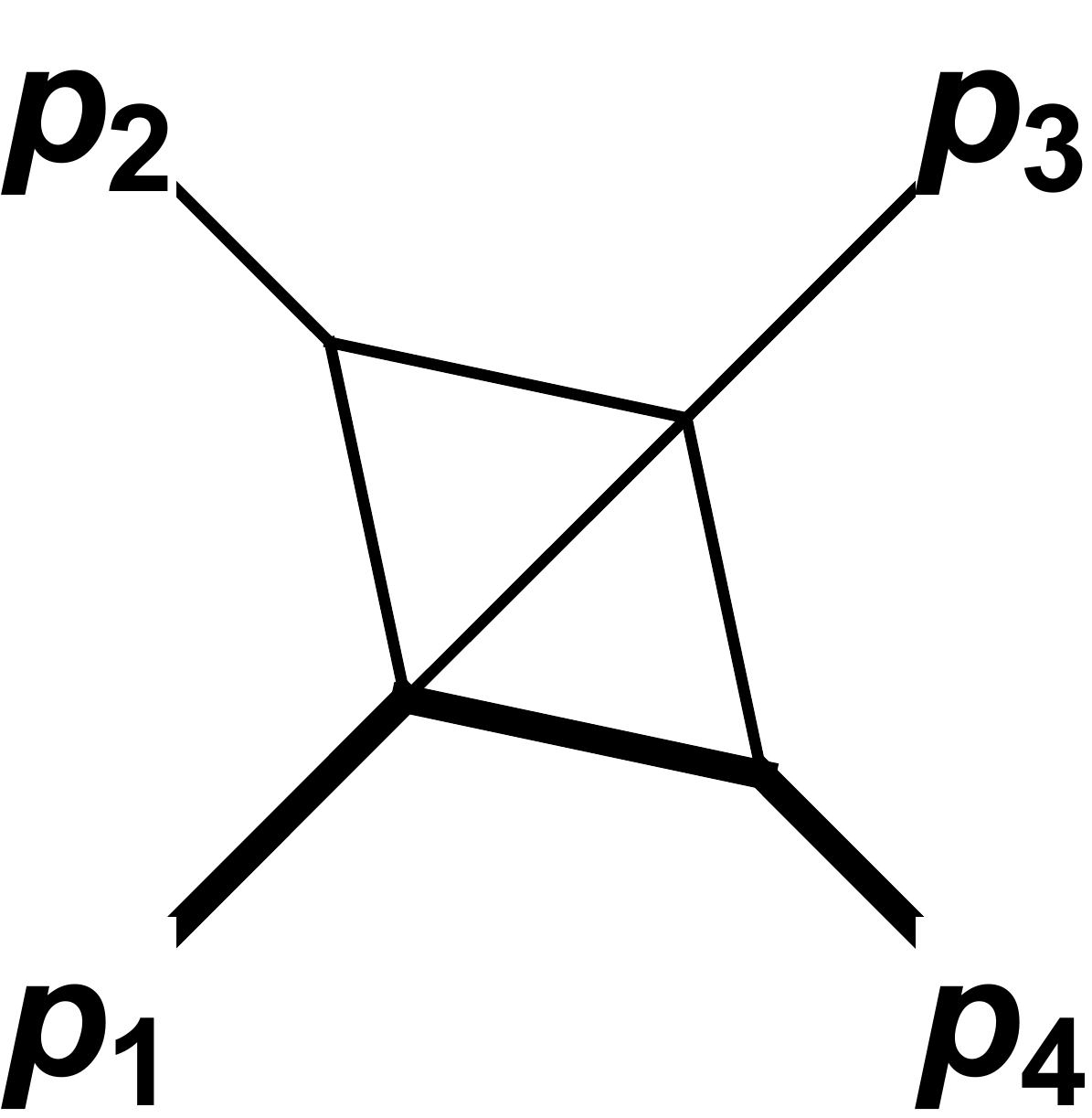}
  }
  \subfloat[$\mathcal{T}_{24}$]{%
    \includegraphics[width=0.11\textwidth]{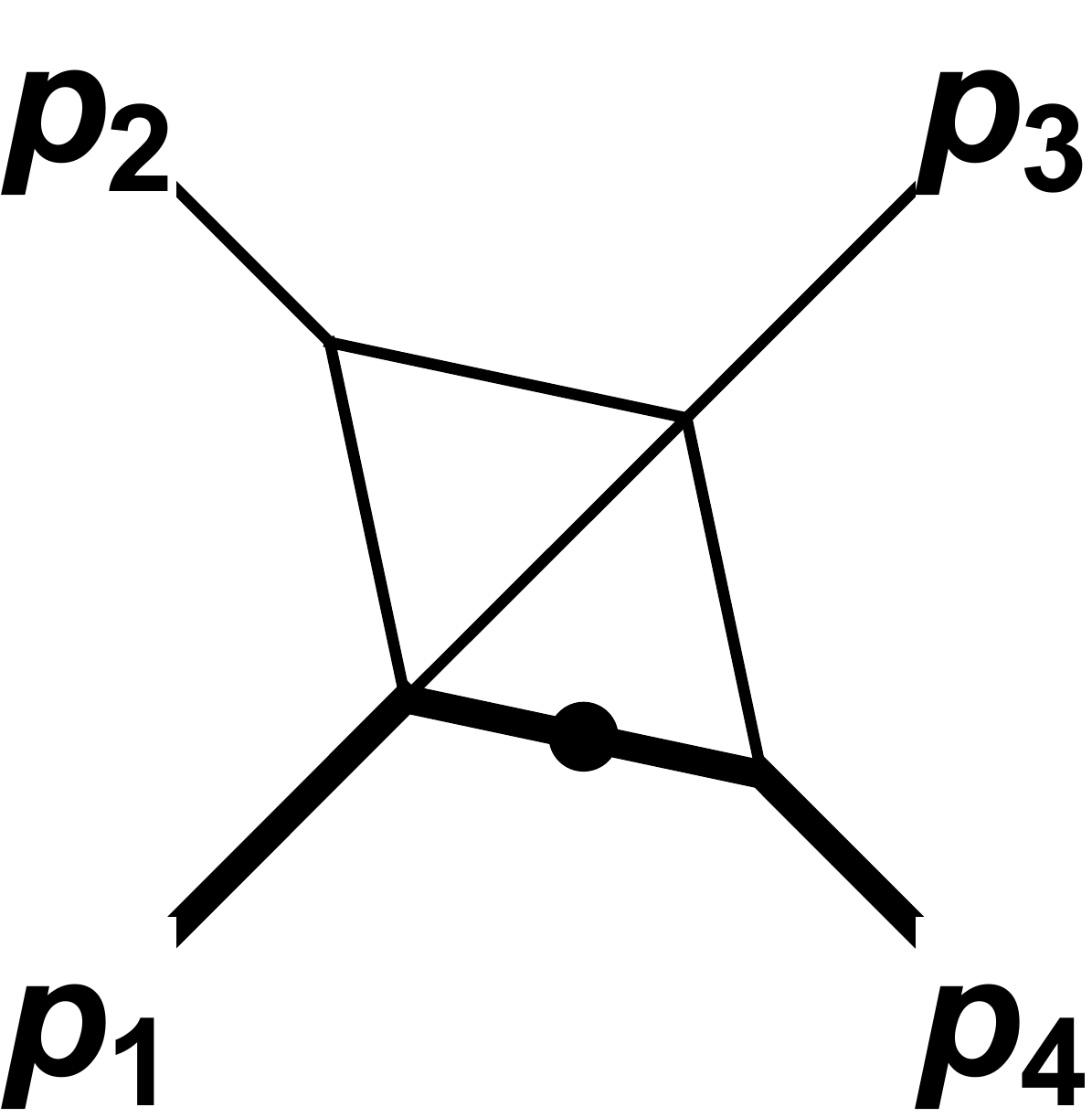}
  } \\
    \subfloat[$\mathcal{T}_{25}$]{%
    \includegraphics[width=0.11\textwidth]{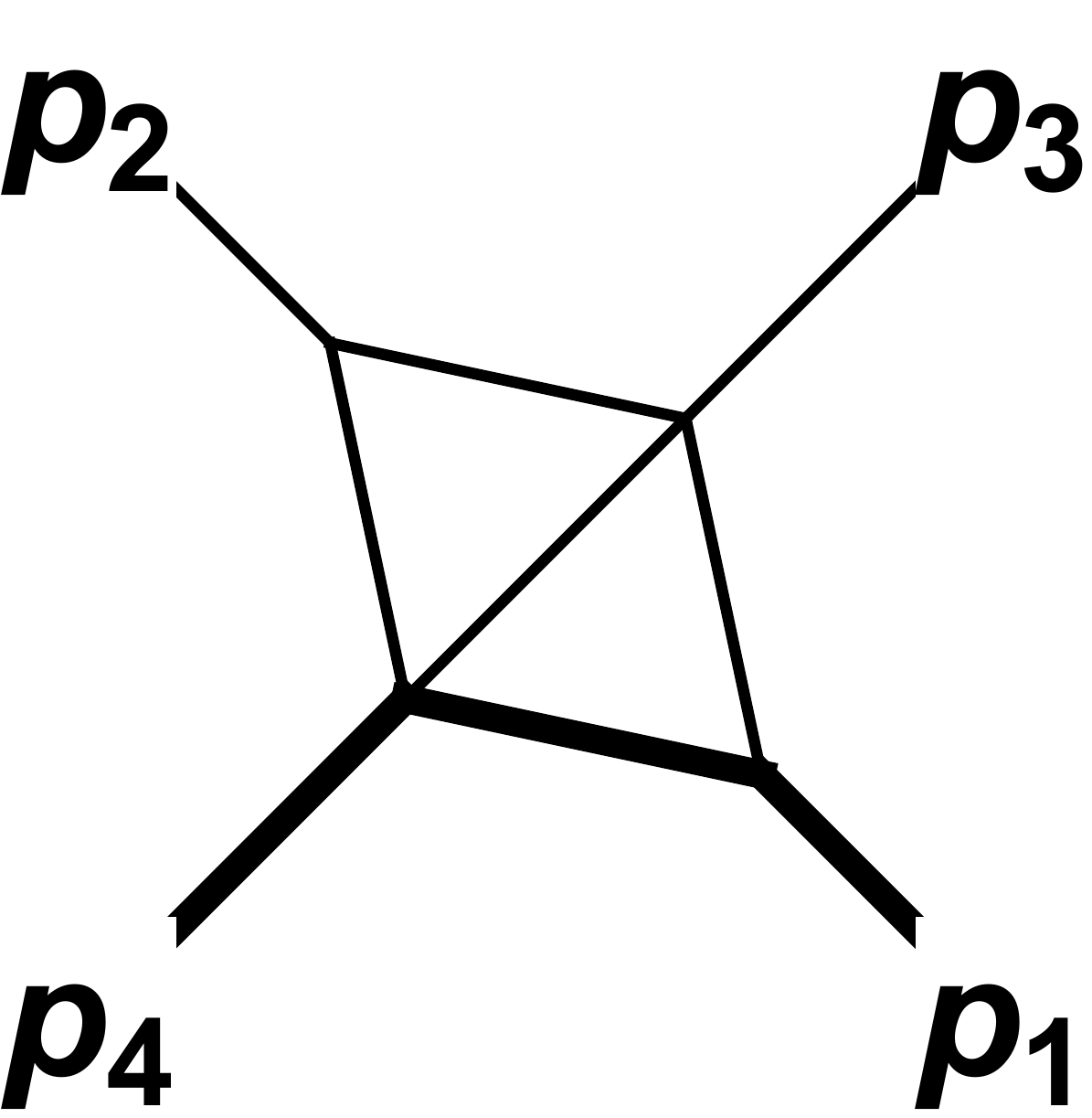}
  }
  \subfloat[$\mathcal{T}_{26}$]{%
    \includegraphics[width=0.11\textwidth]{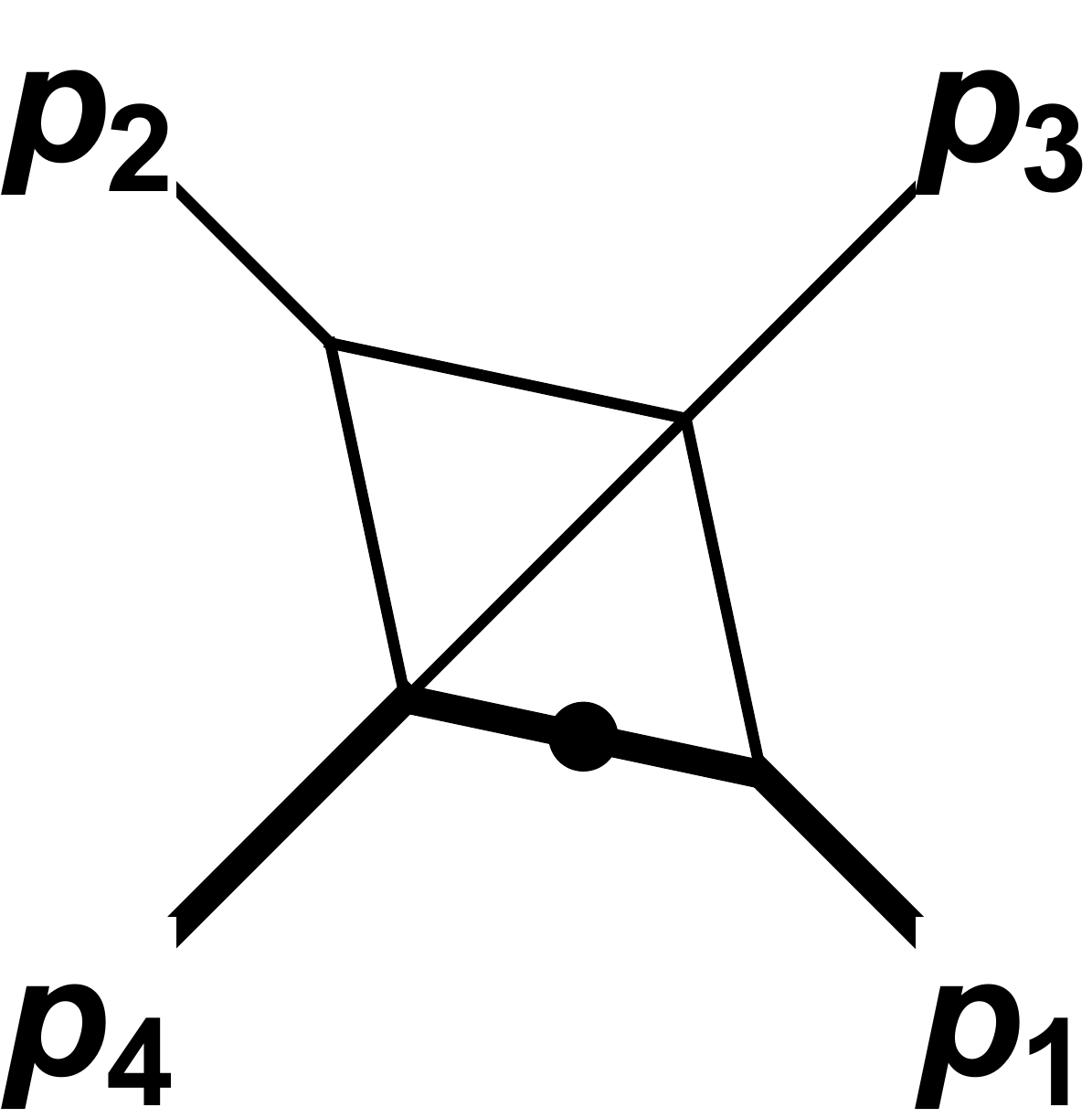}
  }
  \subfloat[$\mathcal{T}_{27}$]{%
    \includegraphics[width=0.11\textwidth]{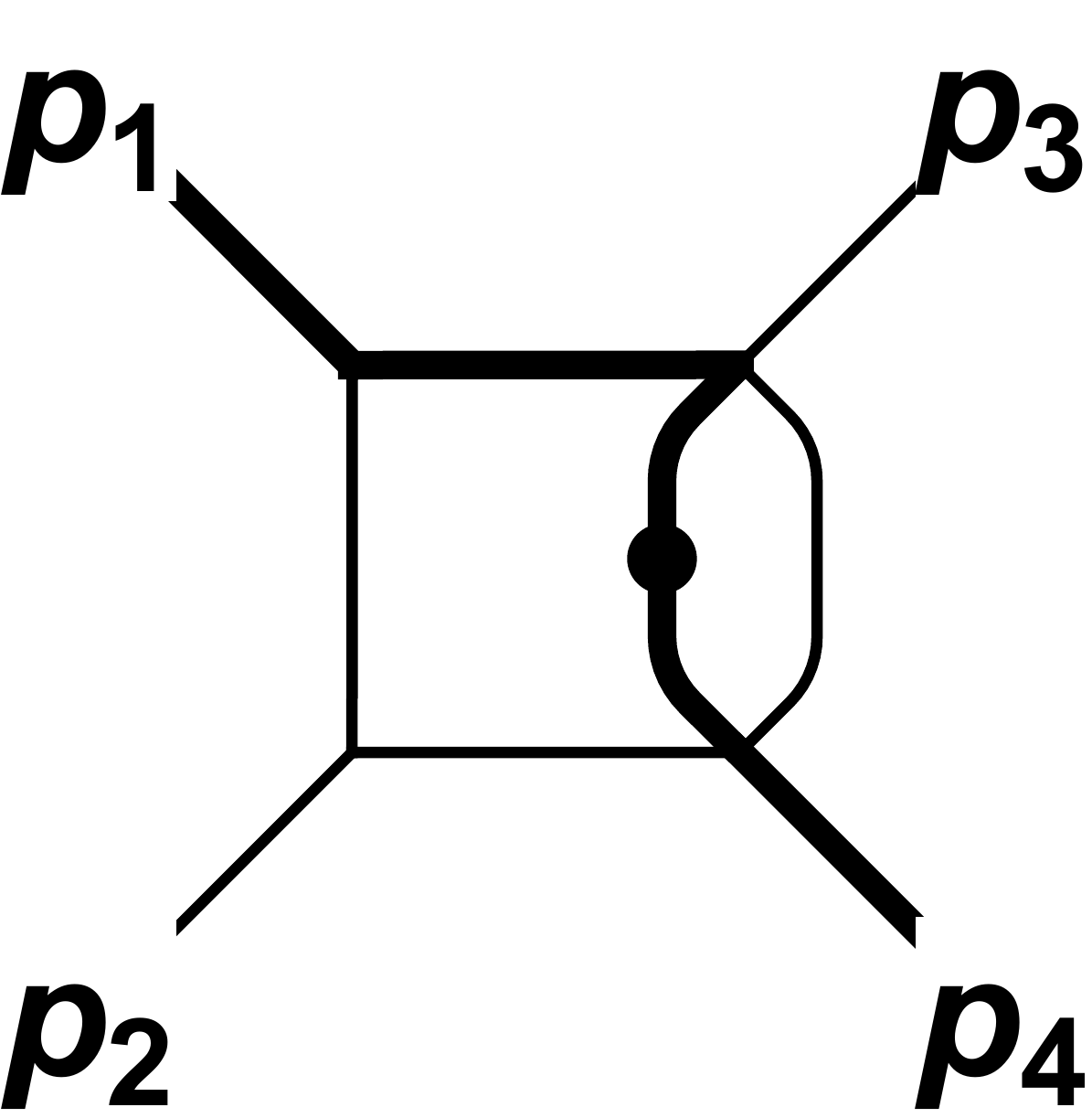}
  }
  \subfloat[$\mathcal{T}_{28}$]{%
    \includegraphics[width=0.11\textwidth]{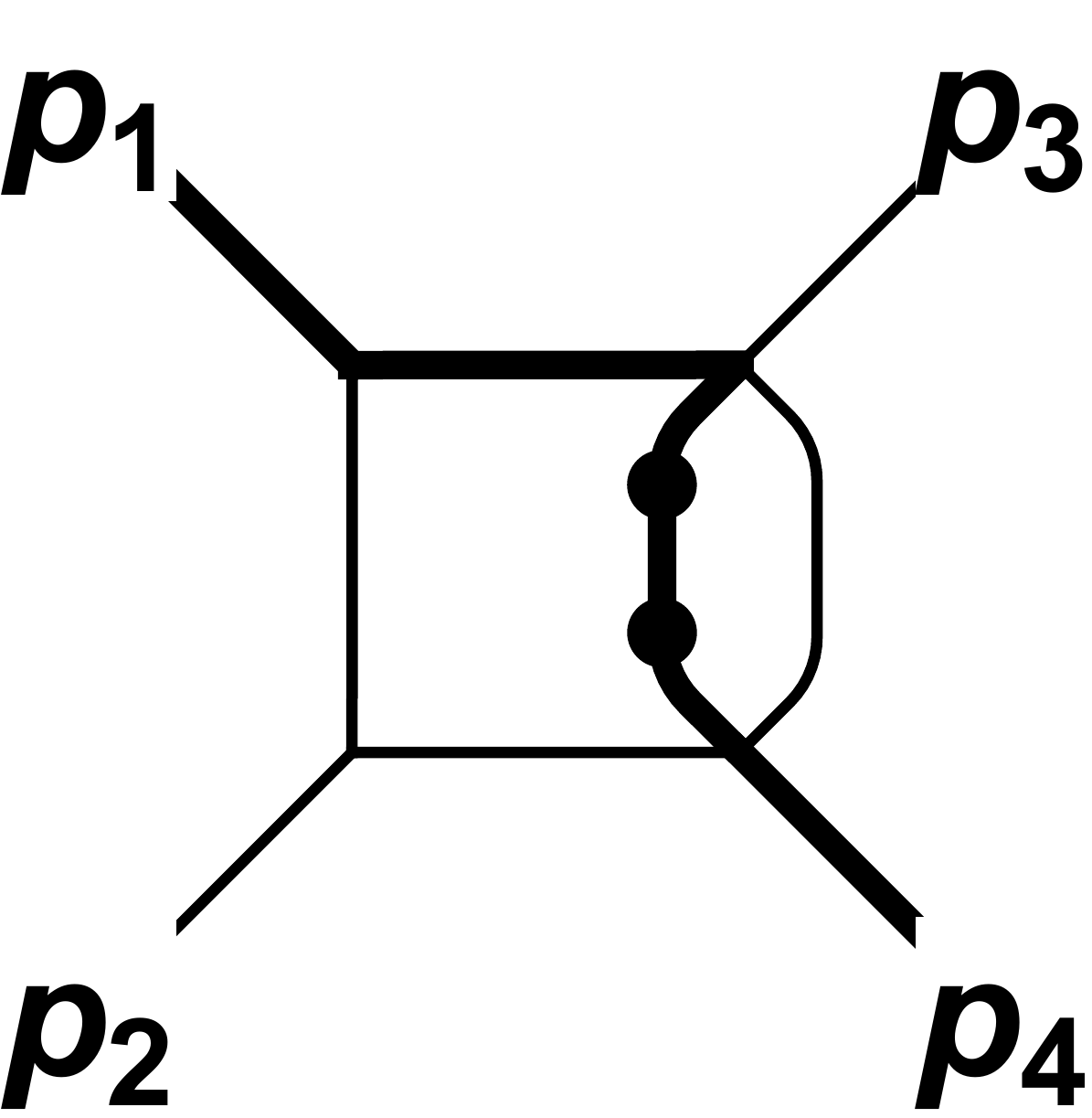}
  }
  \subfloat[$\mathcal{T}_{29}$]{%
    \includegraphics[width=0.11\textwidth]{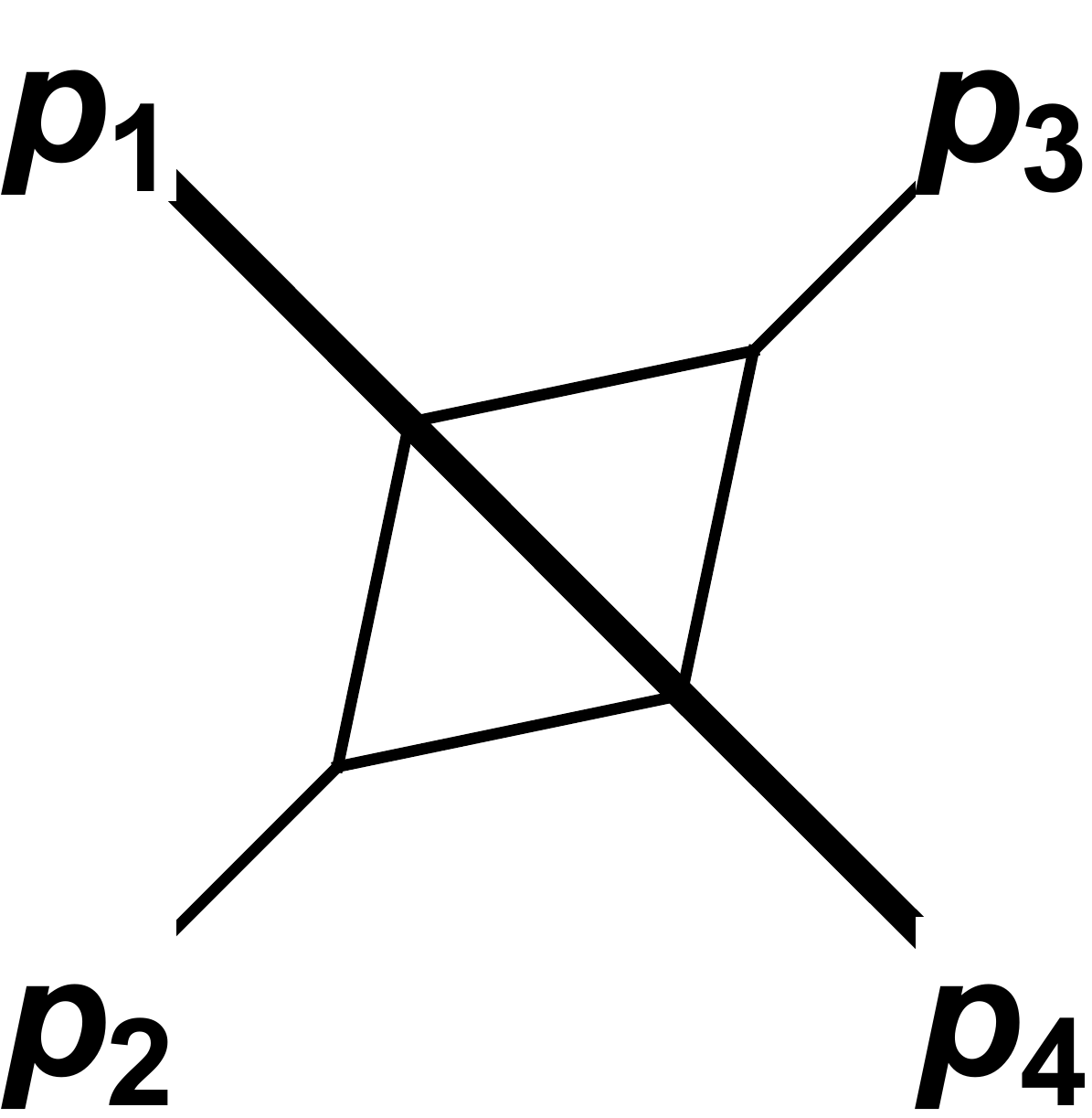}
  }
  \subfloat[$\mathcal{T}_{30}$]{%
    \includegraphics[width=0.11\textwidth]{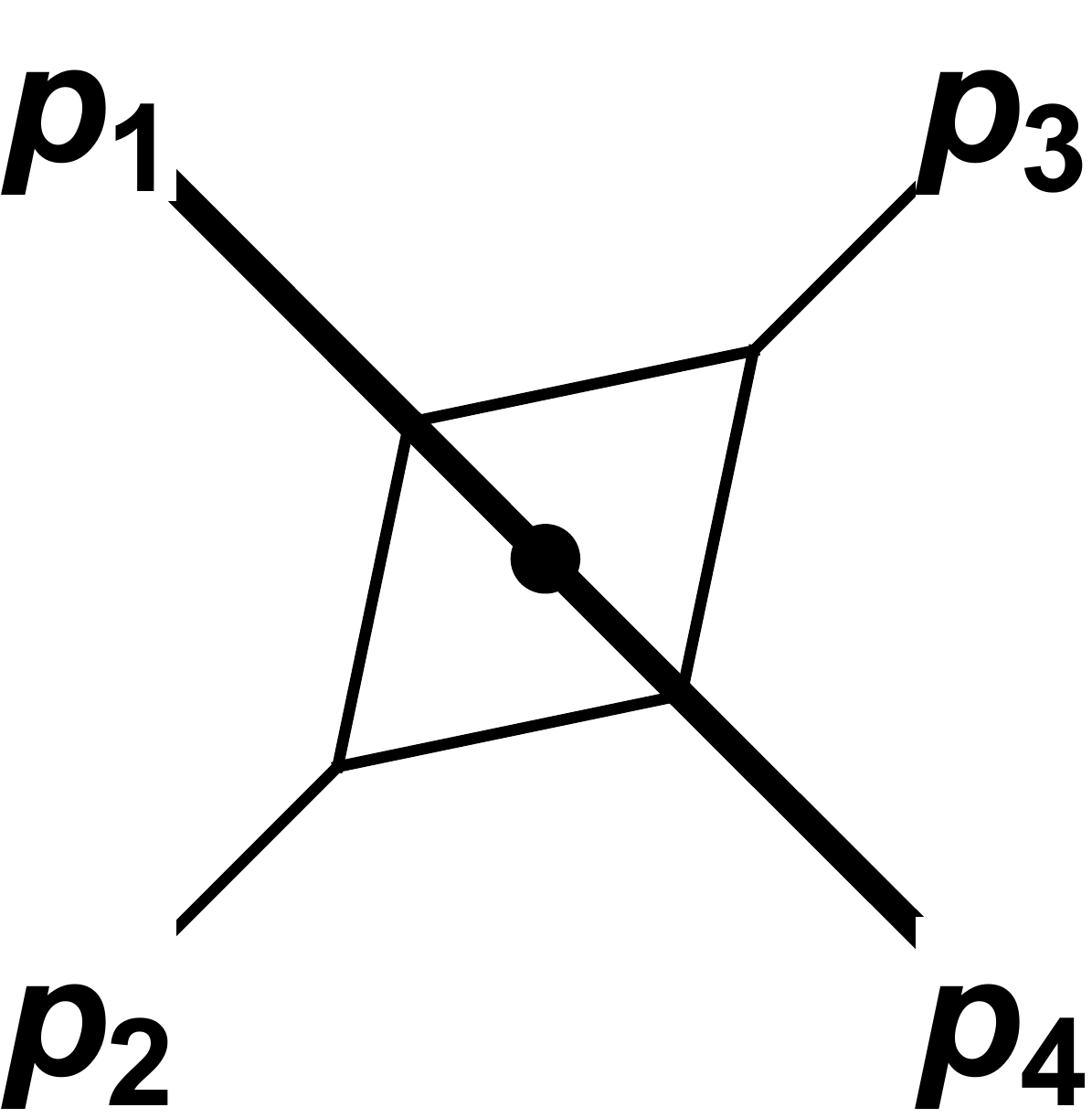}
  } \\
      \subfloat[$\mathcal{T}_{31}$]{%
    \includegraphics[width=0.11\textwidth]{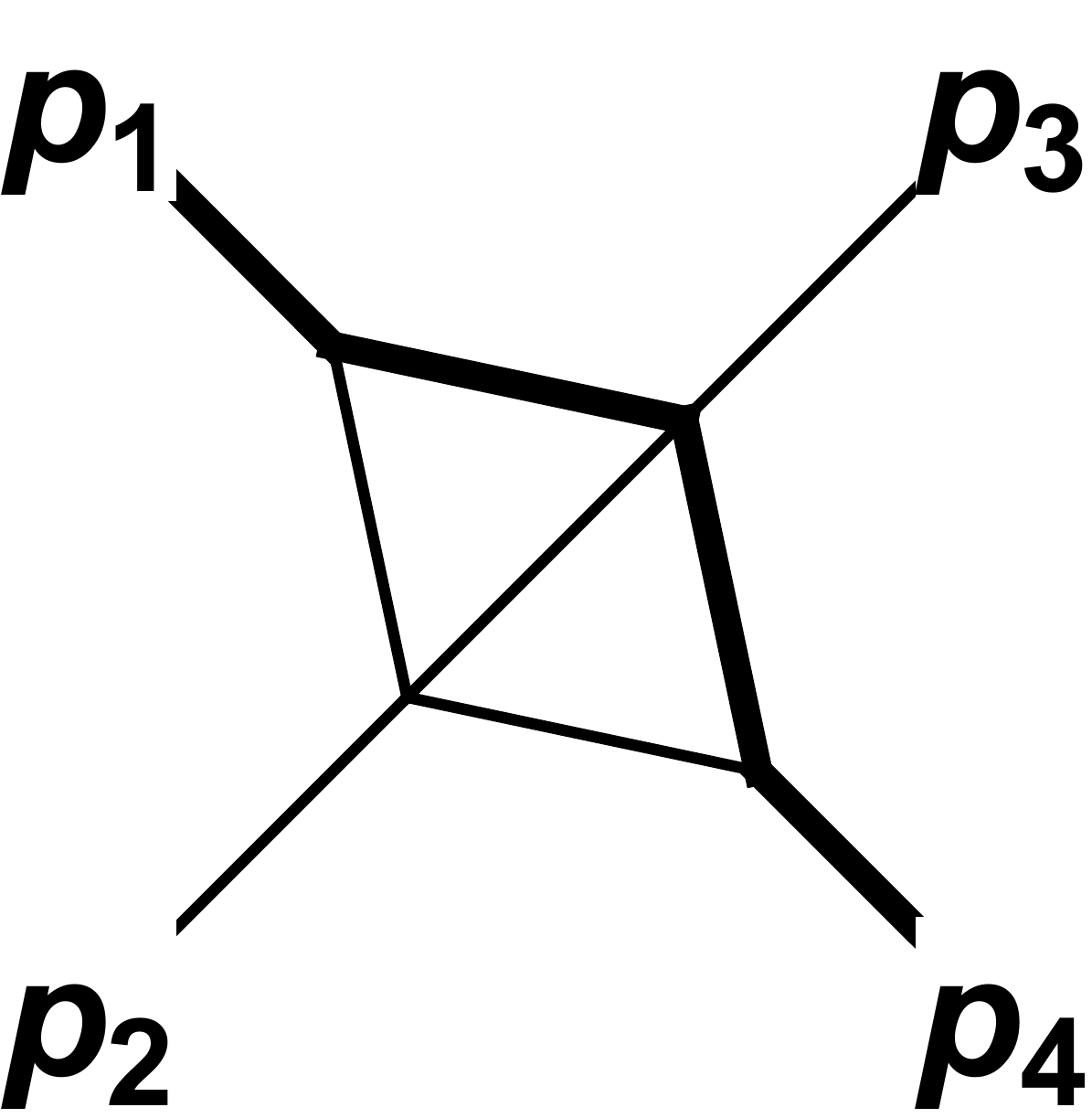}
  }
  \subfloat[$\mathcal{T}_{32}$]{%
    \includegraphics[width=0.11\textwidth]{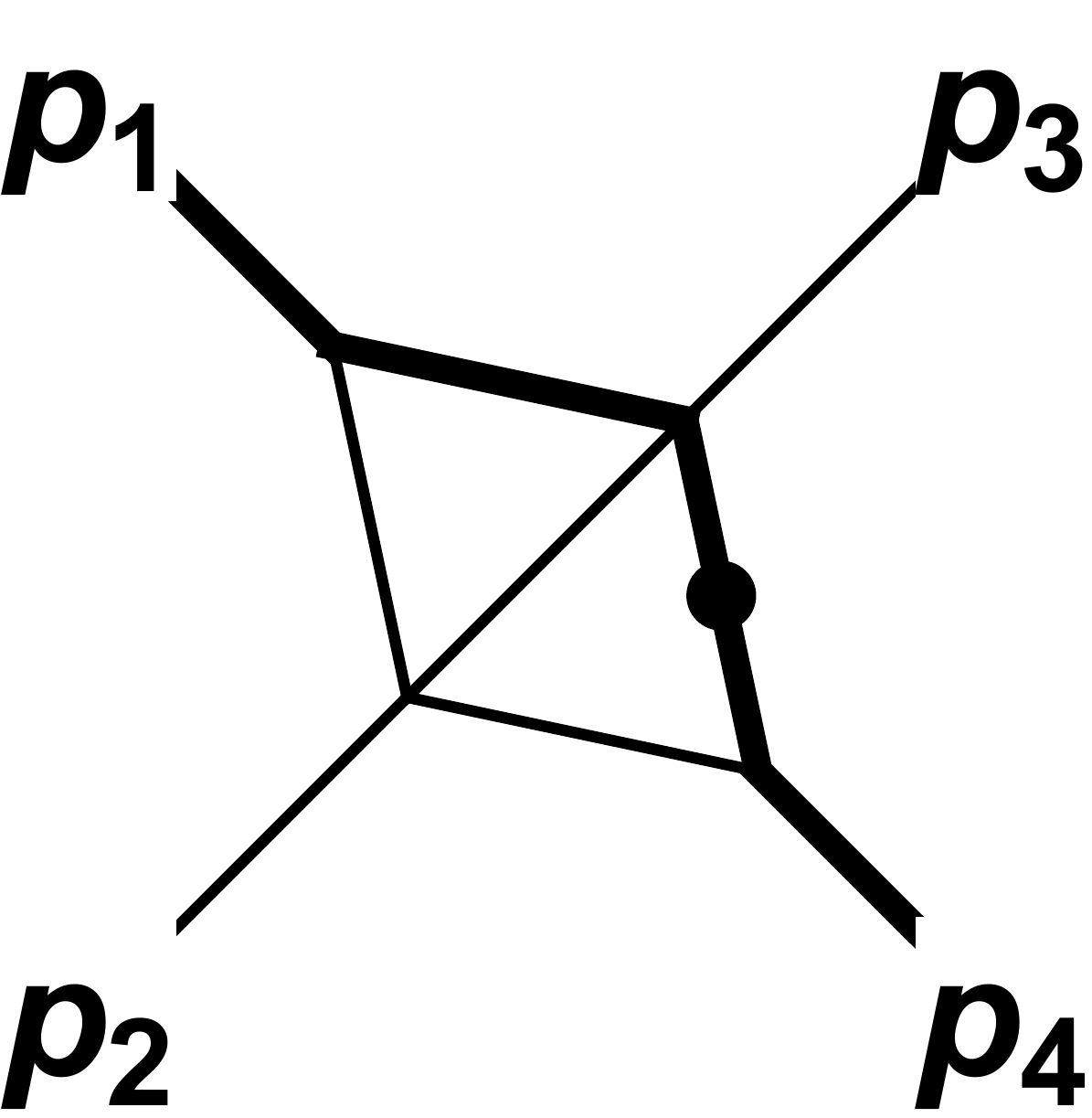}
  }
  \subfloat[$\mathcal{T}_{33}$]{%
    \includegraphics[width=0.11\textwidth]{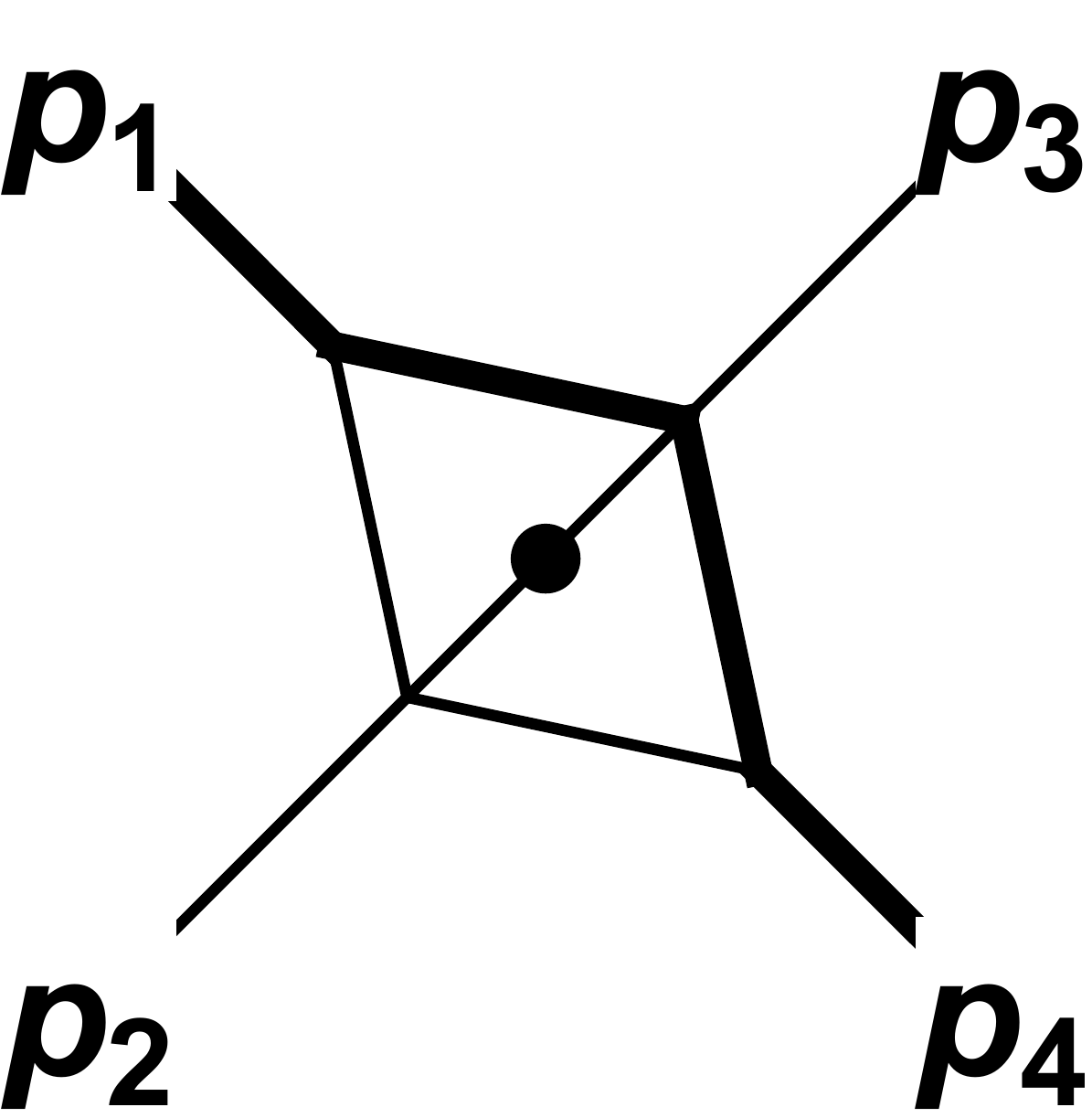}
  }
  \subfloat[$\mathcal{T}_{34}$]{%
    \includegraphics[width=0.11\textwidth]{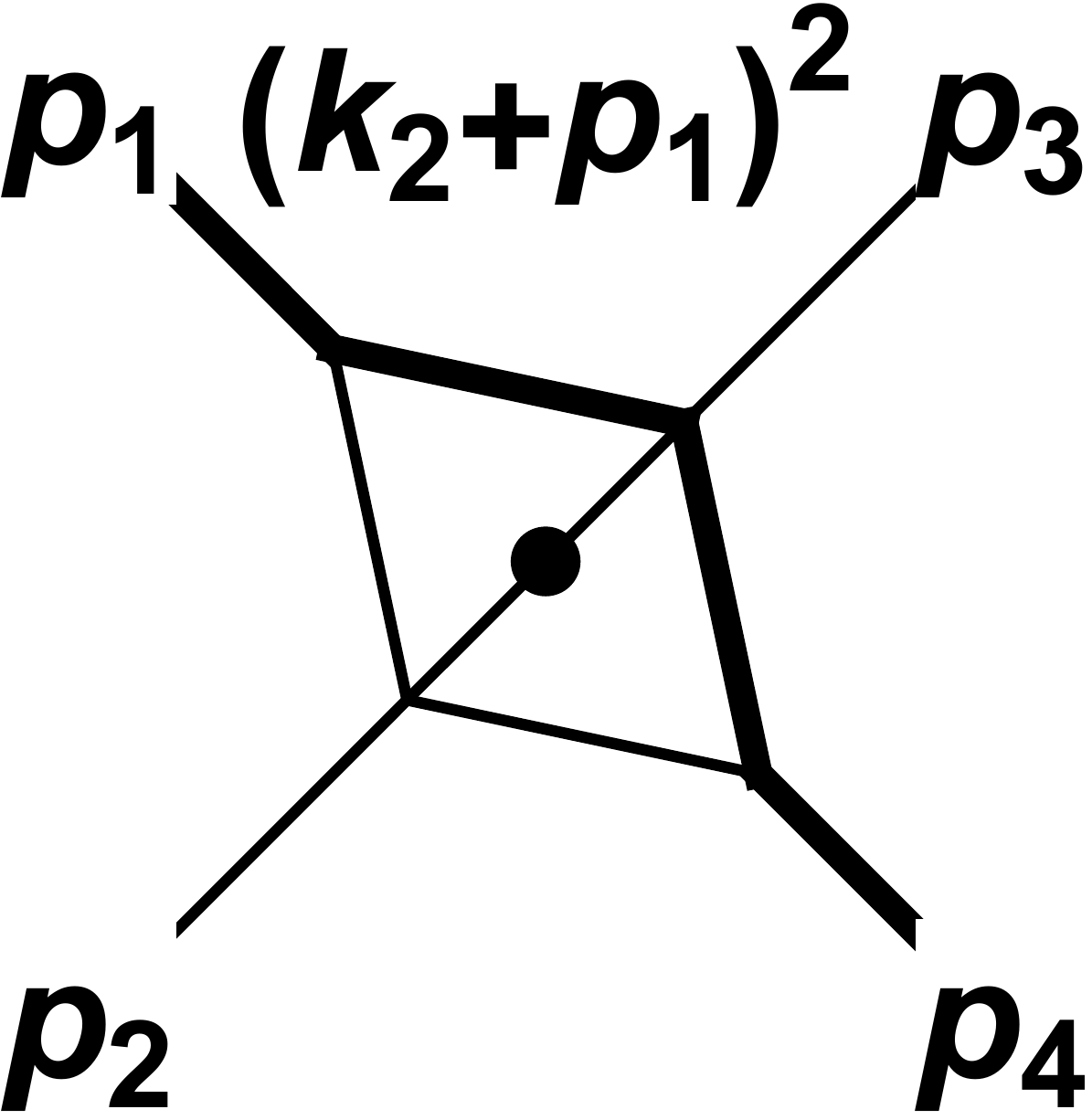}
  }
  \subfloat[$\mathcal{T}_{35}$]{%
    \includegraphics[width=0.11\textwidth]{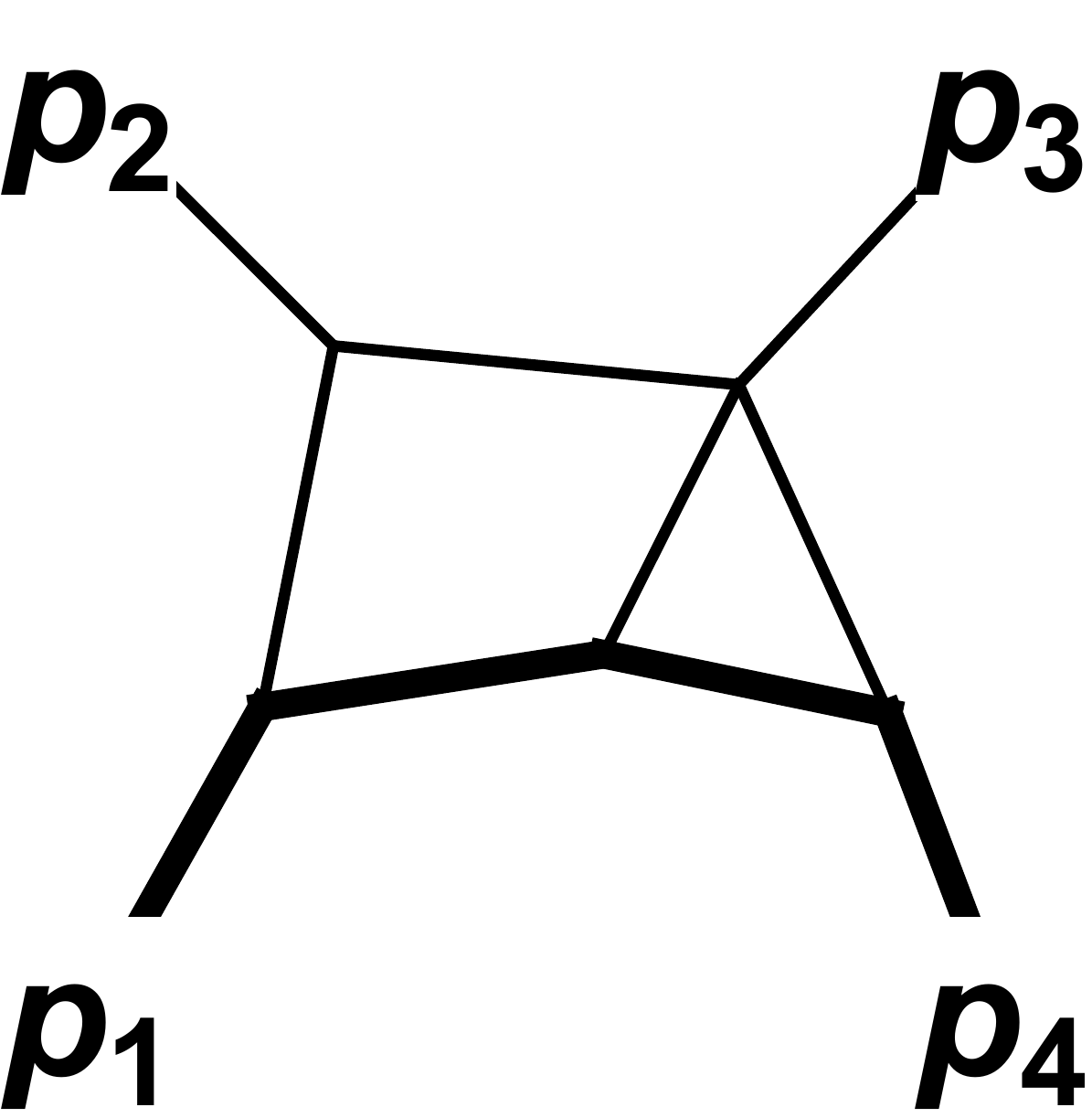}
  }
  \subfloat[$\mathcal{T}_{36}$]{%
    \includegraphics[width=0.11\textwidth]{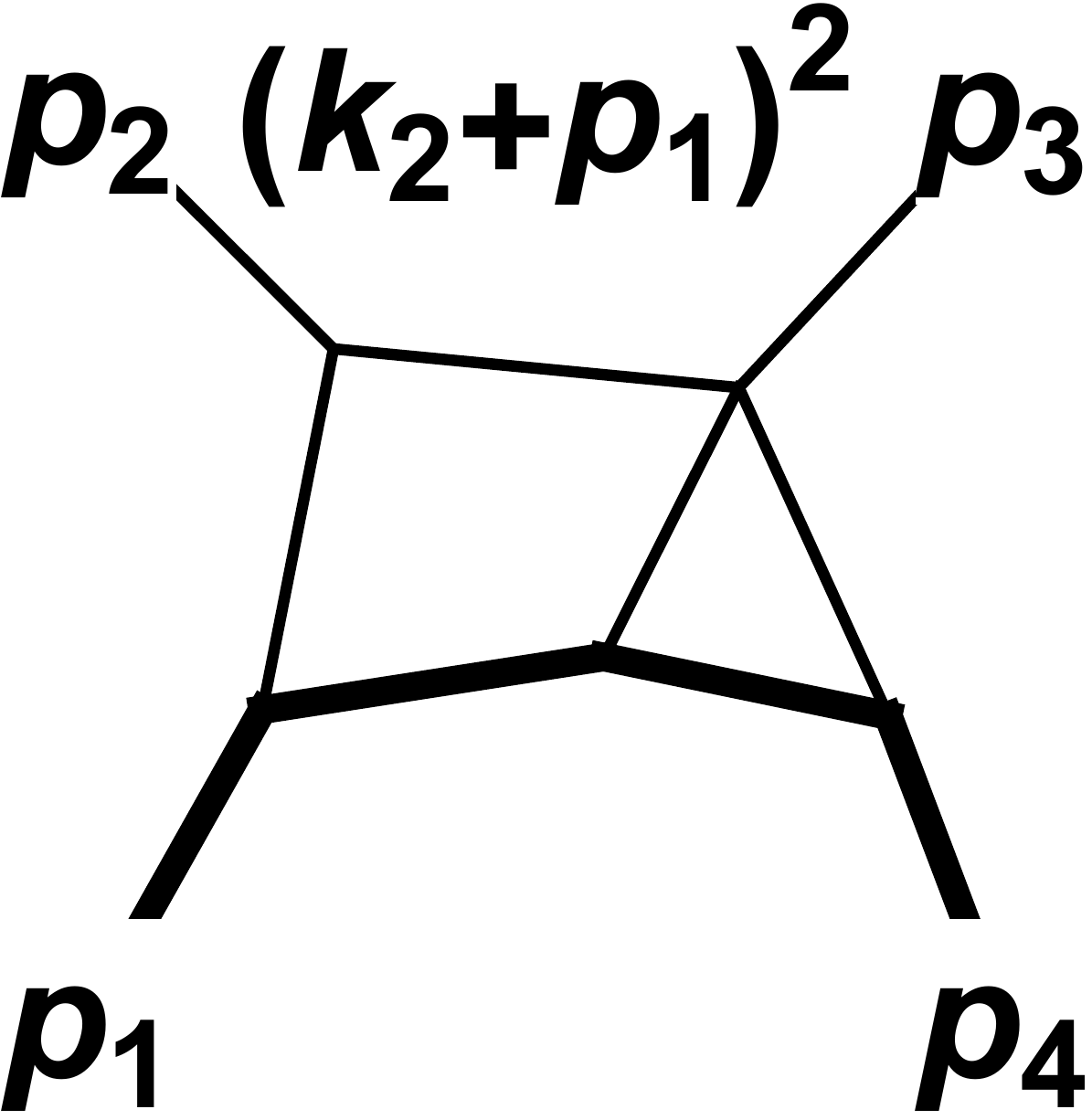}
  } \\
      \subfloat[$\mathcal{T}_{37}$]{%
    \includegraphics[width=0.11\textwidth]{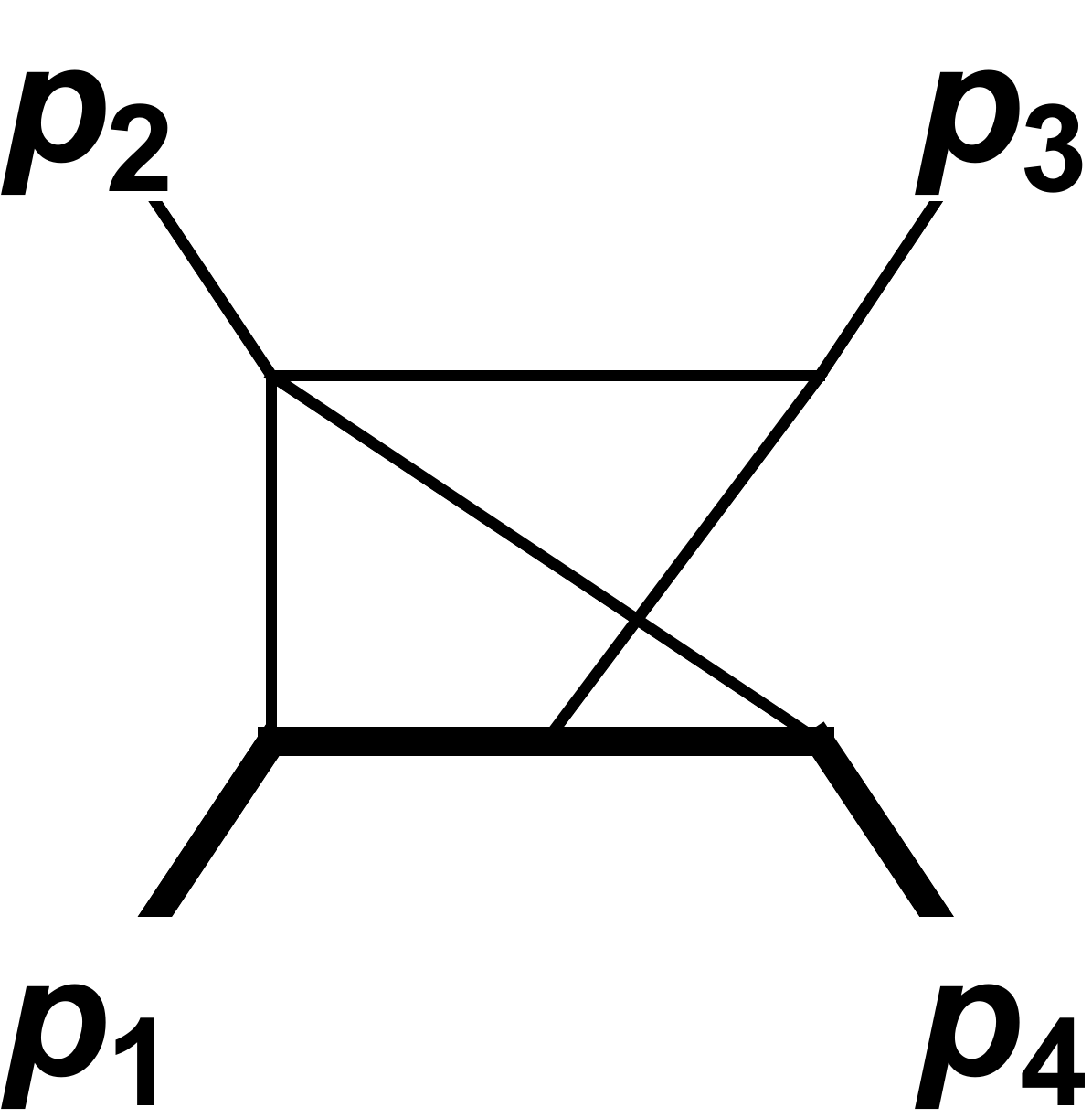}
  }
  \subfloat[$\mathcal{T}_{38}$]{%
    \includegraphics[width=0.11\textwidth]{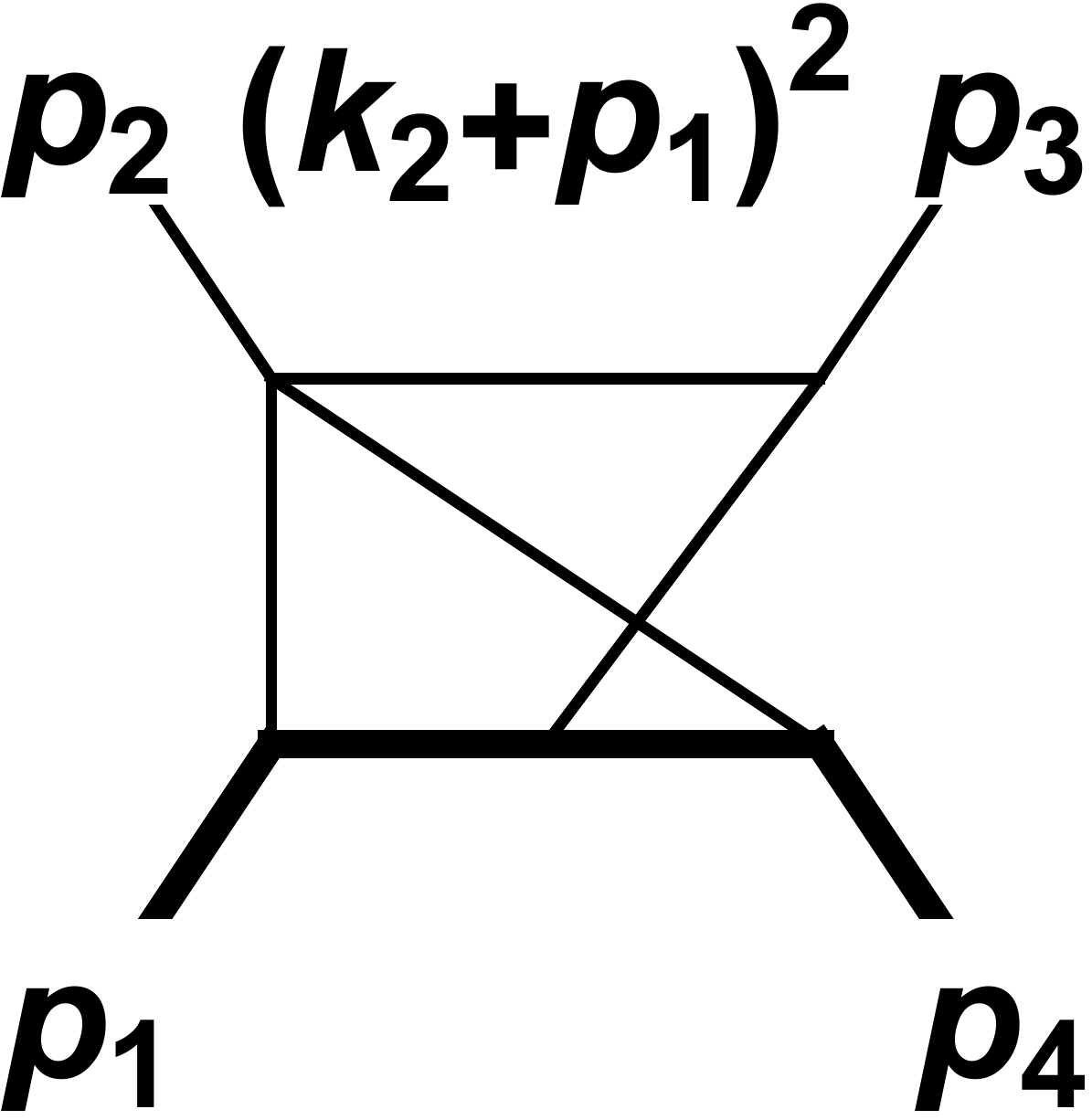}
  }
  \subfloat[$\mathcal{T}_{39}$]{%
    \includegraphics[width=0.11\textwidth]{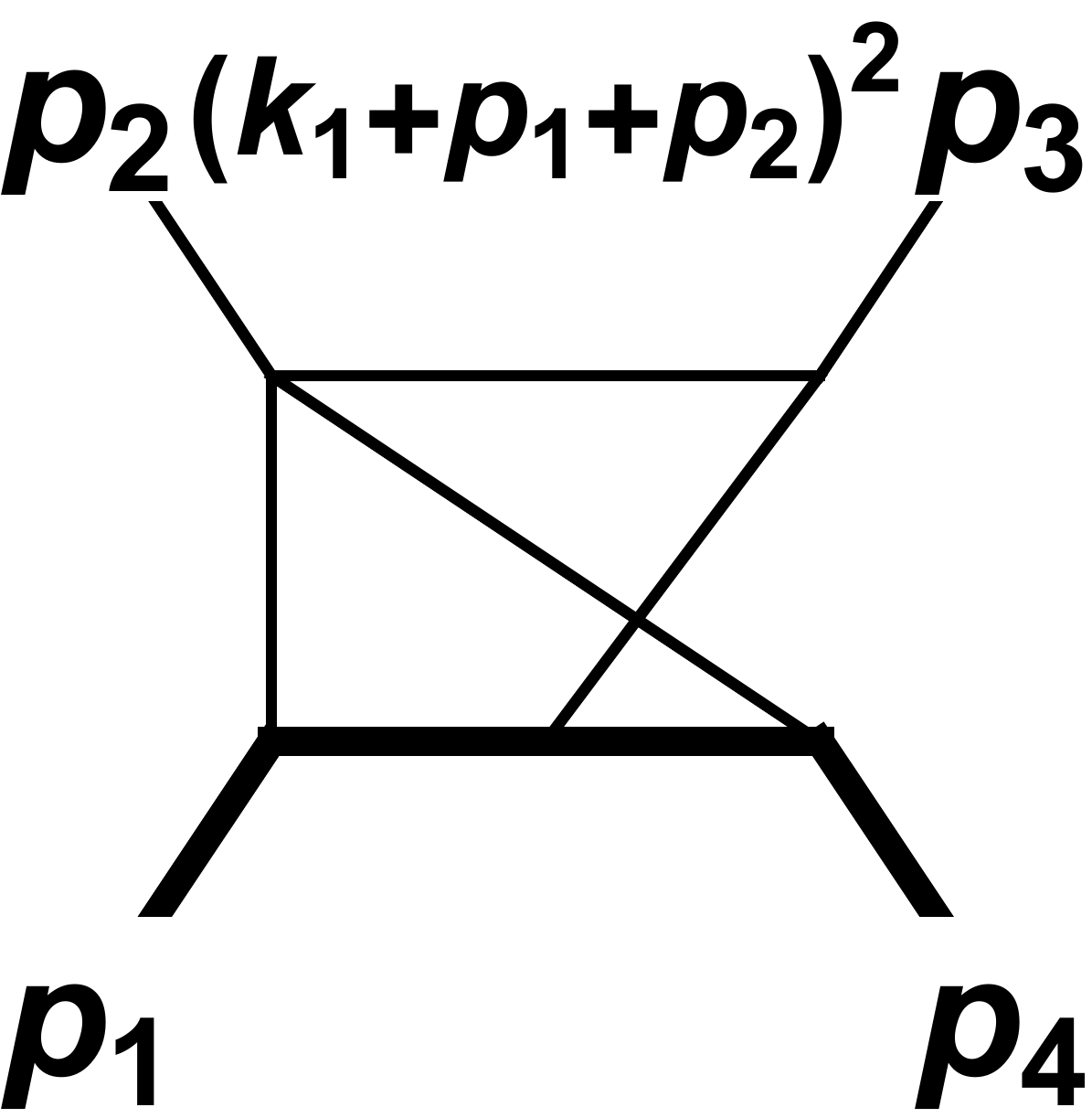}
  }
  \subfloat[$\mathcal{T}_{40}$]{%
    \includegraphics[width=0.11\textwidth]{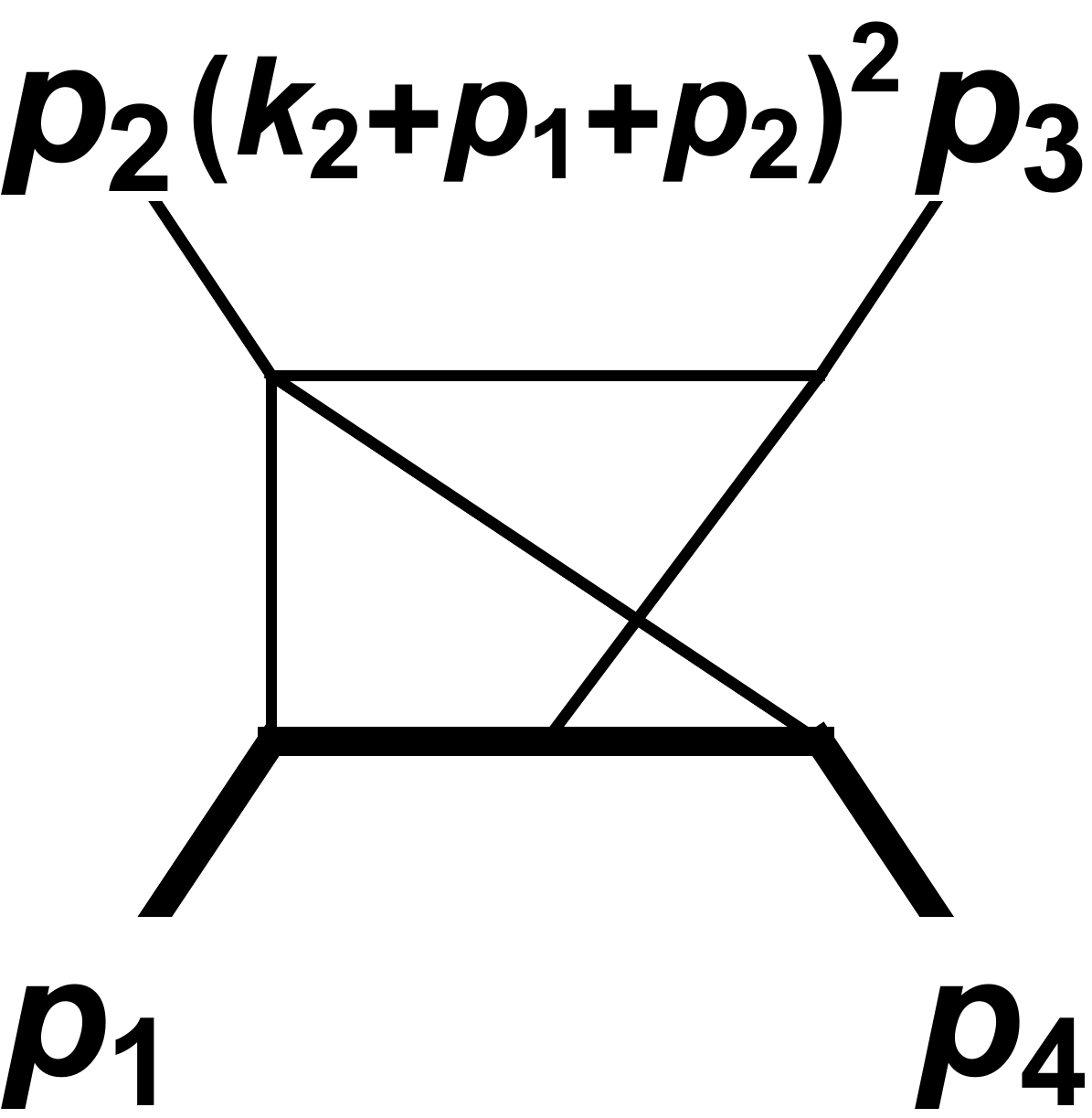}
  }
  \subfloat[$\mathcal{T}_{41}$]{%
    \includegraphics[width=0.11\textwidth]{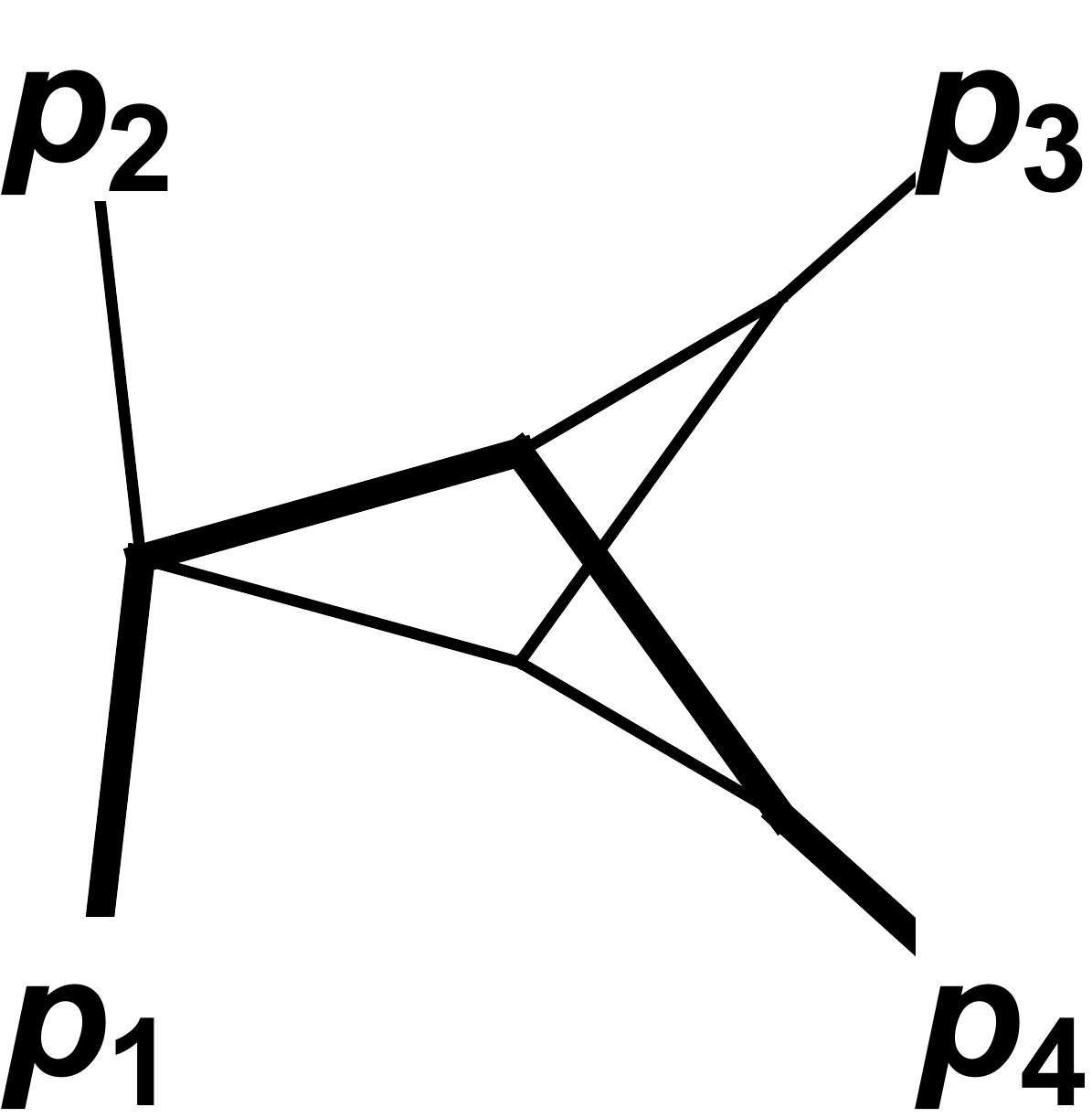}
  }
  \subfloat[$\mathcal{T}_{42}$]{%
    \includegraphics[width=0.11\textwidth]{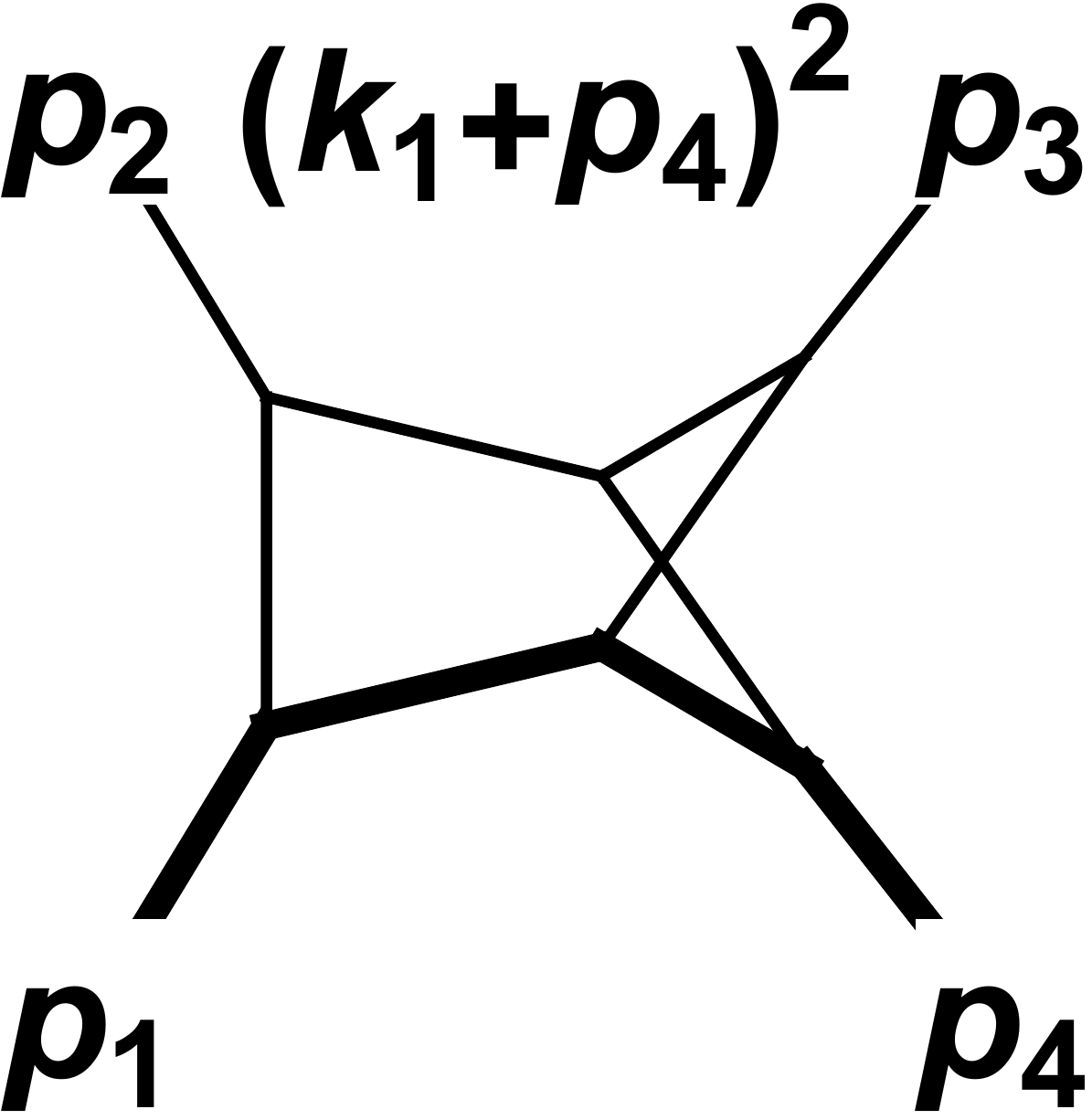}
  } \\
  \subfloat[$\mathcal{T}_{43}$]{%
    \includegraphics[width=0.11\textwidth]{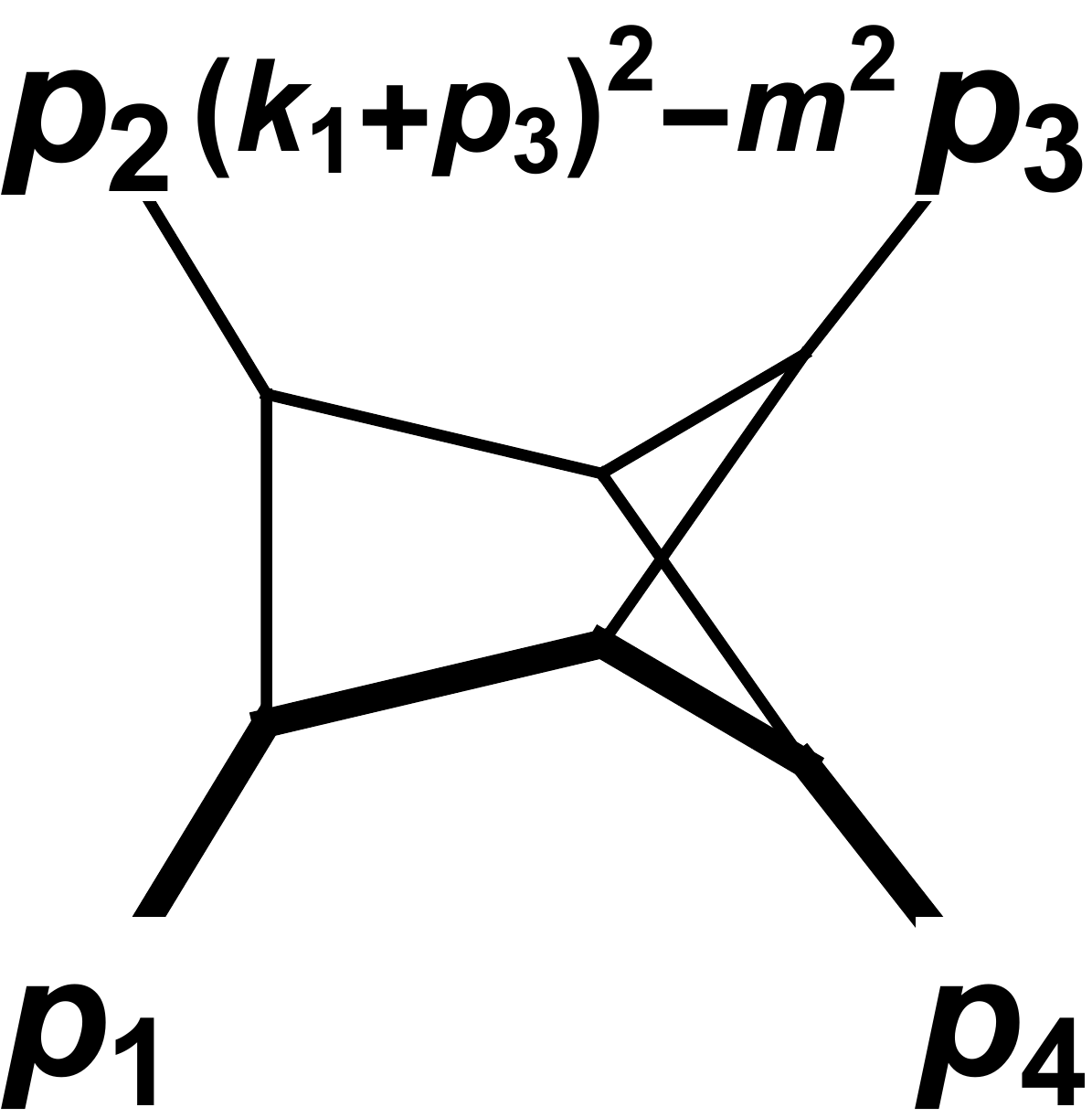}
  }
  \subfloat[$\mathcal{T}_{44}$]{%
    \includegraphics[width=0.11\textwidth]{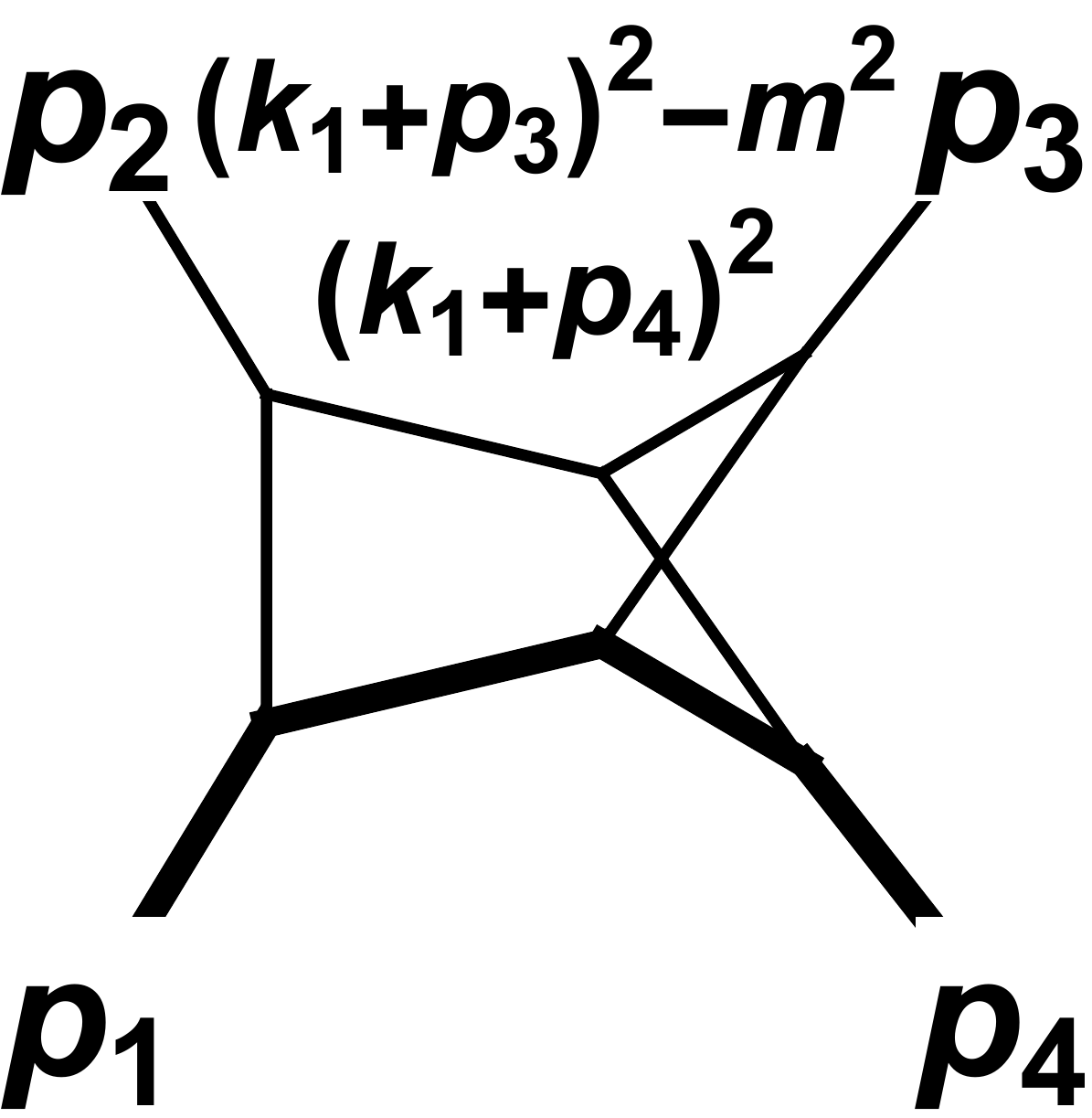}
  }
 \caption{
The 44 MIs $\mathcal{T}_{1,\ldots,44}$ for the two-loop non-planar topology $T_6$.
Thin lines represent massless propagators and thick lines stand for massive ones. Each dot indicates an additional power of the corresponding propagator. Numerator insertions are indicated explicitly on top of each diagram.}
 \label{fig:MIsT6}
\end{figure}

A canonical basis of MIs in $d=4-2\eps$ can be identified by making
use of the algorithm described in
\cite{Argeri:2014qva,DiVita:2014pza}. Namely, we start by choosing an
initial set of MIs $\FF_i$ that fulfill DEQs with linear dependence on
the dimensional regularisation parameter $\eps$,
\begin{align*}
\FF_{1}&=\eps^2 \, \top{1}\,,  &
\FF_{2}&=\eps^2 \, \top{2}\,,  &
\FF_{3}&=\eps^2 \, \top{3}\,,  \\
\FF_{4}&=\eps^2 \, \top{4}\,,  &
\FF_{5}&=\eps^2 \, \top{5}\,,  &
\FF_{6}&=\eps^2 \, \top{6}\,,  \\
\FF_{7}&=\eps^2 \, \top{7}\,,  &
\FF_{8}&=\eps^2 \, \top{8}\,,  &
\FF_{9}&=\eps^3 \, \top{9}\,,  \\
\FF_{10}&=\eps^3 \, \top{10}\,,  &
\FF_{11}&=\eps^3 \, \top{11}\,,  &
\FF_{12}&=\eps^2 \, \top{12}\,,  \\
\FF_{13}&=\eps^3 \, \top{13}\,,  &
\FF_{14}&=\eps^2 \, \top{14}\,,  &
\FF_{15}&=\eps^3 \, \top{15}\,,  \\
\FF_{16}&=\eps^2 \, \top{16}\,,  &
\FF_{17}&=\eps^2 \, \top{17}\,,  &
\FF_{18}&=\eps^4\, \top{18}\,,  \\
\FF_{19}&=\eps^3\, \top{19}\,,  &
\FF_{20}&=\eps^4\, \top{20}\,,  &
\FF_{21}&=\eps^2(1+2\eps) \, \top{21}\,, \\
\FF_{22}&=\eps^3 \, \top{22}\,, &
\FF_{23}&=\eps^4 \, \top{23}\,, &
\FF_{24}&=\eps^3 \, \top{24}\,, \\
\FF_{25}&=\eps^4 \, \top{25}\,, &
\FF_{26}&=\eps^3 \, \top{26}\,, &
\FF_{27}&=\eps^3 \, \top{27}\,, \\
\FF_{28}&=\eps^2 \, \top{28}\,, &
\FF_{29}&=\eps^4 \, \top{29}\,, &
\FF_{30}&=\eps^3 \, \top{30}\,, \\
\FF_{31}&=\eps^4 \, \top{31}\,, &
\FF_{32}&=\eps^3 \, \top{32}\,, &
\FF_{33}&=\eps^3 \, \top{33}\,, \\
\FF_{34}&=\eps^3 \, \top{34}\,, &
\FF_{35}&=\eps^4 \, \top{35}\,, &
\FF_{36}&=\eps^3(1-2\eps) \, \top{36}\,, \\
\FF_{37}&=\eps^4 \, \top{37}\,, &
\FF_{38}&=\eps^4 \, \top{38}\,, &
\FF_{39}&=\eps^4 \, \top{39}\,, \\
\FF_{40}&=\eps^4 \, \top{40}\,, &
\FF_{41}&=\eps^4 \, \top{41}\,, &
\FF_{42}&=\eps^4 \, \top{42}\,, \\
\FF_{43}&=\eps^4\, \top{43}\,, &
\FF_{44}&=\eps^4 \, \top{44}\,, &
\stepcounter{equation}\tag{\theequation}
\label{def:LBasisT6}
\end{align*}
where the $\mathcal{T}_i$ are the integrals depicted in figure~\ref{fig:MIsT6}.

Subsequently, we use the Magnus exponential in order rotate the integrals of eq.~\eqref{def:LBasisT6} into a new basis of MIs $\GG_i$ that satisfy canonical DEQs in both variables $w$ and $z$ (or, equivalently, in $s$ and $t$),

\begin{gather*}
\begin{alignedat}{2}
  \GG_{1}&= \FF_1\,, &
  \GG_{2}&= -s \,  \FF_2\,, \nn
  \GG_{3}&= m^2\, \FF_3\,, &
  \GG_{4}&= - s \FF_4 \,, \nn
  \GG_{5}&= (m^2-s)(2\,\FF_4+ \FF_5)\,, &
  \GG_{6}&= u\,  \FF_6\,,  \nn
  \GG_{7}&= -2m^2\,\FF_{6}+ \left( u-m^2 \right) \,\FF_{7}\,, &
  \GG_{8}&= -t\,\FF_{8}\,, \nn
  \GG_{9}&= (m^2-s)\, \FF_{9}\,,  &
  \GG_{10}&= \lambda_t \, \FF_{10}\,, \nn 
  \GG_{11}&= (u-m^2) \, \FF_{11}\,, &
  \GG_{12}&=~ m^2 \left(u-m^2 \right)\, \FF_{12}\,, \nn 
  \GG_{13}&= (m^2-s)\, \FF_{13}\,,  &
  \GG_{14}&= m^2 \left(m^2-s\right)\, \FF_{14}\,, \nn
  \GG_{15}&=(m^2-s) \, \FF_{15}\,,  &
  \GG_{16}&= m^2(m^2-s)\,\FF_{16}\,, \nn 
  \GG_{17}&=3m^2\,\FF_{15}+2m^4\,\FF_{16}+m^2(2m^2-s)\,\FF_{17}\,, \qquad&
  \GG_{18}&=(m^2-s)\FF_{18}\,,  \nn  
  \GG_{19}&= m^2(m^2-s)\, \FF_{19}\,, &
  \GG_{20}&=\lambda_t \, \FF_{20}\,,\nn
\end{alignedat}\\
  \GG_{21}= \left(\lambda_t -t \right)  \left(\frac{1}{2} \FF_{10}-2\,\FF_{20}\right)-m^2 \, t \, \FF_{21}\,,\nn
\begin{alignedat}{2}
  \GG_{22}&=  -(m^2-s)\,t\, \FF_{22}\,, &
  \GG_{23}&= (u-m^2) \,  \FF_{23}\,, \nn
  \GG_{24}&= -m^2\,t\, \FF_{24}\,, \qquad\qquad\qquad\qquad\qquad\qquad\quad\quad  &
  \GG_{25}&= (m^2-s)\FF_{25} \,, \nn
  \GG_{26}&= -m^2t\, \FF_{26}\,,  &
  \GG_{27}&= \left(m^2-s\right) \, u \, \FF_{27} \,, \nn
  \GG_{28}&= -m^2 \, (m^2-s)\left(\, \FF_{27}-(u-m^2)\,\FF_{28}\right)\,,&
  \GG_{29}&= -t \, \FF_{29} \,, \qquad\nn
  \GG_{30}&= (m^2-s) \, (u-m^2)\,\FF_{30}\,, &
  \GG_{31}&= \lambda_t \, \FF_{31} \,, \nn
  \GG_{32}&= m^2(m^2-s)\,\FF_{32}\,,  &
  \GG_{33}&= (m^2-s)(u-m^2)\FF_{33} \,,\nn
\end{alignedat}\\
%   \GG_{28}= -m^2 \, (m^2-s)\left(\, \FF_{27}-(u-m^2)\,\FF_{28}\right)\,,\nn
% \begin{alignedat}{2}
%   \GG_{29}&= -t \, \FF_{29} \,, \qquad\qquad\qquad\qquad\qquad\quad\qquad&
%   \GG_{30}&= (m^2-s) \, (u-m^2)\,\FF_{30}\,,  \nn
%   \GG_{31}&= \lambda_t \, \FF_{31} \,, &
%   \GG_{32}&= m^2(m^2-s)\,\FF_{32}\,,  \nn
% \end{alignedat}\\
%   \GG_{33}= (m^2-s)(u-m^2)\FF_{33} \,,\nn
  \GG_{34}= \frac{1}{6} \left(u-m^2\right)\left(2\,\FF_{4}+\FF_{5}-12\,\FF_{13}-12m^2\,\FF_{14}+6\,\FF_{34}\right)+ \left( u-m^2 \right)^2 \,\FF_{33} \,,\nn
\begin{alignedat}{2}
  \GG_{35}&=(m^2-s) \, \lambda_t \FF_{35}\,, &
  \GG_{36}&=(m^2-s) \left(2 F_{35} \left(t-\lambda _t\right)+2 F_{23}+F_{36}\right)\,,\nn
  \GG_{37}&=\sqrt{m^2} \, \sqrt{m^2-s} \, \sqrt{u-m^2} \sqrt{-t} \,\FF_{37}\,, \quad\;\; &
  \GG_{38}&=-(m^2-s)\left(\FF_{23}-\FF_{38}\right)\,,\nn
\end{alignedat}\\
\begin{aligned}
  \GG_{39} ={}& \left\{\frac{1}{4} \left[\FF_7-4 \left(m^2\, \FF_{12}+\FF_{11}+\FF_{20}-\FF_{31}+\FF_{40}\right)\right] \left(u-m^2+\lambda_t\right) -\FF_{25}+\FF_{39}\right\} \nn
  {}&  +m^2\, \FF_3 \left(1+\frac{\lambda_t}{u-m^2}\right) -(m^2-s)
  \lambda_t \, \FF_{37}+\frac{1}{2}\FF_6 \left[ u-(u+m^2) \left(1+\frac{\lambda_t}{u-m^2}\right) \right]\,, \nn  
  \GG_{40} ={}&  -m^2\, \FF_3 \left(1+\frac{\lambda_t}{u-m^2}\right)-\frac{1}{4} \left[\FF_7-4 \left(m^2\, \FF_{12}+\FF_{11}\right)\right] \left(u-m^2+\lambda_t\right) \nn
  {}& +(m^2-s) \lambda_t \, \FF_{37}+ \lambda_t\, \FF_{40}+\frac{\FF_6}{2}( u+m^2) \left(1 + \frac{\lambda_t}{u-m^2}\right) \,,\nn
\end{aligned}\\
\begin{alignedat}{2}
  \GG_{41}&=  (m^2-s)^2\, \FF_{41}\,, \qquad\qquad\qquad\qquad &
  \GG_{42}&= (m^2-s)\,(u-m^2) \,  \FF_{42}\,,\qquad\qquad\nn
  \GG_{43}&= -(m^2-s)\,t\, \FF_{43}\,,  &
  \GG_{44}&= (m^2-s)\FF_{44} \,,
\end{alignedat}
\label{def:CanonicalBasisT6}
\stepcounter{equation}\tag{\theequation}
\end{gather*}
where we introduced the abbreviation $\lambda_t=\sqrt{-t} \sqrt{4m^2-t}$.

By combining the two DEQs in $w$ and $z$ into a single total differential, we get
\begin{equation}
d \GGvec = \eps \dA \GGvec \, ,
\label{eq:canonicalDEQ}
\end{equation}
where $\GGvec$ is a vector that collects the 44 MIs and
\begin{equation}
\dA = \sum_{i=1}^{12} \MM_i  \dlog(\eta_i) \, ,
\end{equation}
with the $\MM_i$ being constant matrices with rational entries.
The arguments $\eta_i$ of this $\dlog$-form,
which define the so-called {\it alphabet} of the DEQs, are the following 12 \emph{letters}:
 \begin{align}
 \begin{alignedat}{2}
\eta_1 & =w\,,&\quad
\eta_2 & =1+w\,, \nn 
\eta_3 & =1-w\,, &\quad
\eta_4 & =z\,, \nn
\eta_5 & =1+z\,,&\quad
\eta_6 & =1-z\,,\nn
\eta_7 & =w+z  \,,&\quad
\eta_8 & =z-w  \,,\nn
\eta_9 & =z^2-w \,, &\quad
\eta_{10} & = 1-w+w^2-z^2\,,\nn
\eta_{11} & =1-3w+w^2+z^2 \,, &\quad
\eta_{12} & =z^2 - w^2 - w z^2 + w^2 \,z^2 \,.
\end{alignedat} \stepcounter{equation}\tag{\theequation}
\label{alphabet}
\end{align}
In the present work, we compute the MIs in the kinematic region where all the letters are real and positive, 
\begin{equation}
  \label{eq:positivityx}
  0<w<1\land\sqrt{w}<z<\sqrt{1-w+w^2}\,,
\end{equation}
which corresponds to the unphysical region 
\begin{equation}
  t<0\,\land\,s<0\,.
\end{equation}
The analytic continuation to the physical region for
$\mu e$ scattering
\begin{equation}
  s\geq m^2 \,\land\, -s\,\left(1-\frac{m^2}{s}\right)^2 \leq t\leq 0 \,\land\, 2m^2-s\leq u \leq \frac{m^4}{s} \,,
\end{equation}
can be obtained through the Feynman prescription, by adding a small
positive imaginary part $i\omega$ to the Mandelstam invariants
whenever they become positive.

Since all the integrals defined in eq.~\eqref{def:CanonicalBasisT6} are finite in the $\eps\to 0$ limit, the vector $\GGvec(\eps,w,z)$ admits a Taylor expansion in $\epsilon$ (of uniform transcendental weight),
\begin{align}
  \GGvec(\eps,w,z) =  \GGvec^{(0)}(w,z) + \epsilon\, \GGvec^{(1)}(w,z) + \epsilon^2 \GGvec^{(2)}(w,z) + \ldots\,,
\end{align}
with the $n$-th order coefficient given by
\begin{align}
\GGvec^{(n)}(w,z) = \sum_{i=0}^n \Delta^{(n-i)} (w,z; w_0,z_0)  \GGvec^{(i)}(w_0,z_0),
\end{align}
where $ \GGvec^{(i)}(w_0,z_0)$ is a constant vector and $\Delta^{(k)}$ the weight-$k$ operator
\begin{align}
\Delta^{(k)} (w,z; w_0,z_0) =\int_\gamma \dAk{k},\qquad \Delta^{(0)} (w,z; w_0,z_0) = 1\,,
\label{eq:delta}
\end{align}
that iterates $k$ ordered integrations of the 1-form $\dA$ along any
piecewise-smooth path
$\gamma:[0,1] \to M\subset \mathbb{C}\times\mathbb{C}$ such that
$\gamma(0)=(w_0,z_0)$ and $\gamma(1)=(w,z)$. If the singularities of
the integrand are removed from $M$, and suitable branch cuts are
imposed, the iterated integrals in eq.~\eqref{eq:delta} do not depend
on the chosen path (see e.g.\ ref.~\cite{DiVita:2017xlr} for a compact
review of the properties of the iterated path integrals).  Since the
rational alphabet given in eq.~\eqref{alphabet} has only algebraic
roots, we can directly express (by first integrating in $w$ and then
in $z$ or viceversa\footnote{
  The integration to GPLs of a rational
  $\dlog$ form, for instance first in $w$ and then in $z$, can be
  performed in two ways. First, one can use the iterated path-integral
  approach and choose $\gamma$ to be a sequence of straight lines,
  from $(w_0,z_0)$ to $(w,z_0)$ and then to $(w,z)$. Equivalently (see
  e.g.~ref.~\cite{DiVita:2014pza}) one can work directly on the
  associated canonical partial DEQs,
  $\partial_i \GGvec = \eps \partial_i\AA \GGvec $ order by order in
  $\eps$ ($i=w,z$). In particular, one first integrates the partial DEQ in $w$
  up to an unknown vector $\HH$ depending on $z$. An ordinary DEQ for
  $\HH$ is then obtained by explicitly taking the derivative of the
  integral of the partial DEQ in $w$ and matching to the partial DEQ in $z$. For
  definiteness, the latter is the strategy we followed to produce the
  results of this work.}) the iterated integrals of
eq.~\eqref{eq:delta} in terms of GPLs, which are defined as
\begin{align}
  G({\vec a}_{n} ; x) &\equiv G(a_1, {\vec a}_{n-1} ; x) \equiv
  \int_0^x dt \frac{1}{t-a_1}
  G({\vec a}_{n-1};t) , \\
  G(\vec{0}_n;x)& \equiv \frac{1}{n!}\text{log}^n(x) \,.
\end{align}
The length $n$ of the vector ${\vec a}_n$ corresponds the
transcendental {\it weight} of $G({\vec w}_{n} ; x)$ and it amounts to
the number of iterated integrations that define the GPL.  The GPLs in
our solution, which for definiteness we obtain by first integrating in
$w$ and then in $z$, are of two classes, namely GPLs in $w$, with
weights drawn from the set
\begin{align}
  \left\{
  0\,,\;
  \pm 1\,,\;
  \pm z\,,\;
  z^2\,,\;
  \frac{1}{2} \left(3\pm\sqrt{5-4 z^2}\right)\,,\;
  \frac{z \left(z\pm\sqrt{4-3 z^2}\right)}{2 \left(z^2-1\right)}\,,\;
  \frac{1}{2} \left(1\pm\sqrt{4 z^2-3}\right)
  \right\}\,,
\end{align}
and GPLs in $z$, with weights drawn from
\begin{align}
  \{
  0\,,\;
  \pm 1\,,\;
  \pm i
  \}\,.
\end{align}
The analytic structure of the canonical MIs is determined by the
letters at all orders in $\epsilon$, and the solution can in principle
be built up to any weight. We compute the MIs up to weight 4, which
will be enough for the two-loop virtual calculation.

In the region defined by eq.~\eqref{eq:positivityx}, the imaginary part of our solution $\GGvec(\eps,w,z)$ only originates from the integration constants $\GGvec^{(i)}(w_0,z_0)$.
\subsection{Boundary conditions}
The general solution of the system of DEQs in terms of GPLs, which is obtained from the integration of eq.~\eqref{eq:canonicalDEQ}, must be complemented by a suitable set of boundary conditions.
These boundary conditions can be determined either from the knowledge of the analytic expression of the MIs in special kinematic configurations or by imposing their regularity at pseudo-thresholds of the DEQs. For the problem under consideration, regularity conditions express the boundary constant as combinations of GPLs of argument $1$, with weights drawn from the set $a_i\in\{-1,-i,0,i,1\}$,
which arises from the kinematic limits imposed on the alphabet given in eq.~(\ref{alphabet}). 
We used \texttt{GiNaC} to numerically verify that for each MI, at each order in $\eps$, the corresponding combination of constant GPLs is proportional to a uniform combination of the transcendental constants $\pi$, $\zeta_k$ and $\log 2$. 
\\

In the following, we specify how the boundary constants of each integral have been obtained:
\begin{itemize}
\item The integrals $\GG_{1,...,5,8,9,10,13,...,18,20,...,24,29,30,41}$ are common to the two-loop topologies discussed in~ref~\cite{Mastrolia:2017pfy}, to which we refer the reader for the discussion of the boundary fixing. Furthermore, the integrals $\GG_{25,26}$ are related to $\GG_{23,24}$ by $s\leftrightarrow u$ crossing, so that their boundary constants can be inferred directly from the ones of $\GG_{23,24}$.

 \item The integrals $\GG_{6,7}$ are regular in the limit $u\to 0$, where they can be reduced, via IBPs, to a single two-loop vacuum diagram. From the analytic expression of the latter, we obtain the boundary values
  \begin{align}
  \GG_6|_{u=0}&=0\,,\nn
  \GG_7|_{u=0}&=1+\frac{\pi ^2}{3}\eps^2-2  \zeta_3\eps^3+\frac{\pi ^4}{10}\eps^4+\mathcal{O}\left(\eps^5\right)\,.
  \end{align}
  \item  The integrals $\GG_{11,12}$ are regular in the limit $u\to 0$.
  In particular, we observe  that their boundary values can be obtained as the limits
   \begin{align}
 \GG_{11}|_{u=0}=-\eps^3m^2\lim_{p_1^2\to m^2} \raisebox{-35pt}{\includegraphics[scale=0.18]{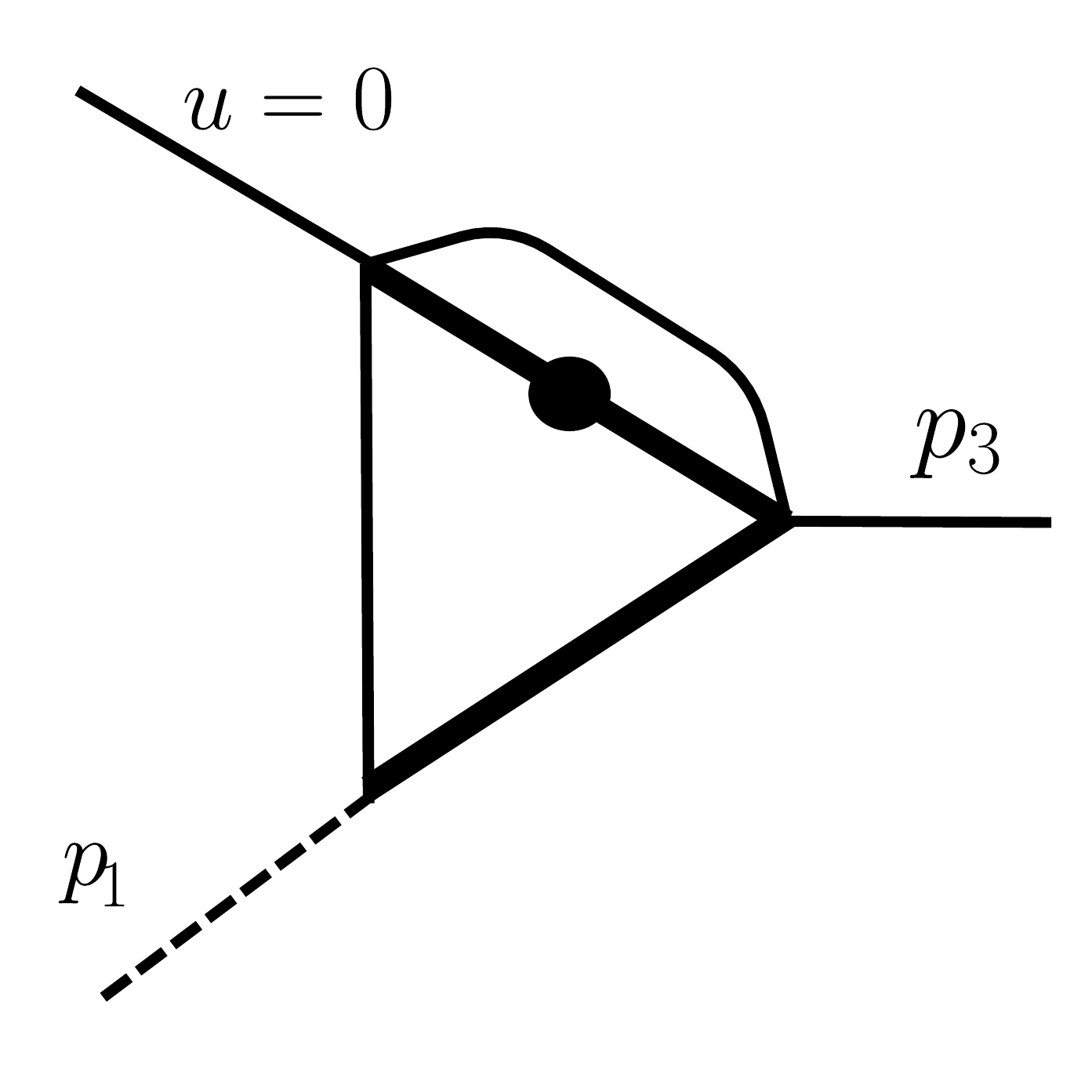}}\,,\quad
  \GG_{12}|_{u=0}=-\eps^2m^4\lim_{p_1^2\to m^2} \raisebox{-35pt}{\includegraphics[scale=0.18]{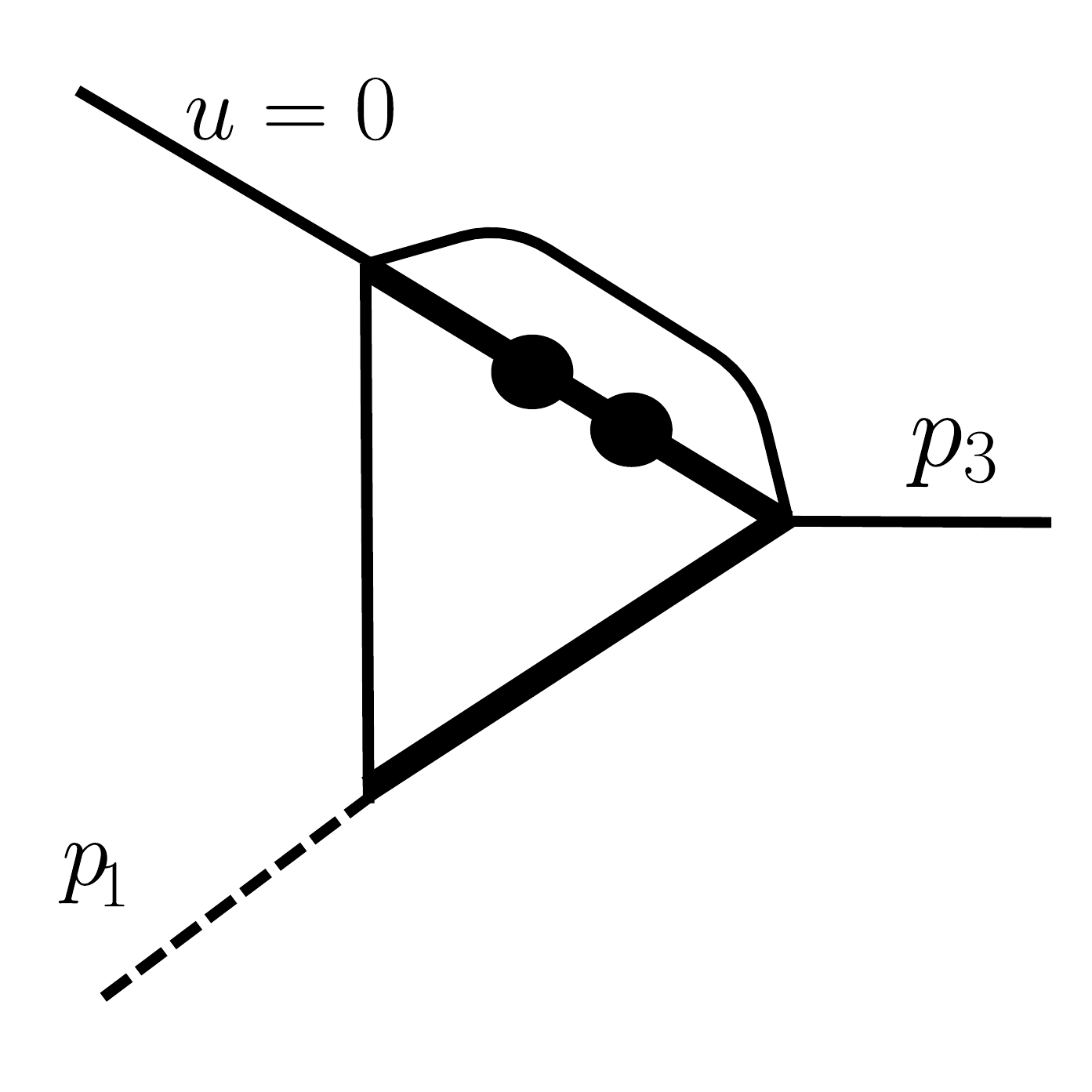}}\,.
 \label{eq:limit_i1112}
 \end{align}
Therefore, we can generate the DEQs for the analogous triangle integrals with $p_4^2=0$, and $u=0$ and an off-shell leg $p_1^2$, solve them by using as an integration base-point the regular point $p_1^2=0$, and finally extract the boundary values of $\GG_{11}$  and $\GG_{12}$ by means of eq.\eqref{eq:limit_i1112}.
The details of this computation are reported in appendix~\ref{app:A}.
In this way, we obtain
  \begin{align}
  \GG_{11}|_{u=0}=&2 \zeta_3 \eps^3+\frac{7 \pi ^4}{180} \eps^4+\mathcal{O}\left(\eps^5\right)\,,\nn
   \GG_{12}|_{u=0}=&-\frac{\pi ^2 }{12}\eps^2-\frac{5 \zeta_3}{2} \eps^3-\frac{1}{12} \pi ^4 \eps^4+\mathcal{O}\left(\eps^5\right)\,.
  \end{align}
  \item The boundary constants of the integrals $\GG_{27,28}$ can be fixed by imposing the regularity of their DEQs as $t\to 0$.
  \item The boundary constants of the integrals $\GG_{31,32,33}$ are obtained by demanding regularity at $t\to 0$, as well the reality of the integrals in the region $s\leq 0$, $u\leq 0$.
\item The boundary constants of the integrals $\GG_{19,34,35...40,42,43,44}$ are obtained by demanding their finiteness in the limit $s\rightarrow \frac{\sqrt{4m^2-t}-\sqrt{-t}}{\sqrt{4m^2-t}+\sqrt{-t}}$.
\end{itemize}
The analytic expressions of the MIs are given in electronic form in the ancillary files attached to the \texttt{arXiv} version of the manuscript.

%%%%%%%%%%%%%%%%%%
%%% Local Variables:
%%% TeX-master: "../main"
%%% End:

%% file: tex/Numerics.tex
\section{Numerical evaluation of the non-planar four point integrals}
\label{sec:Numerics}
The analytic expression of our MIs have been numerically evaluated in the region 
$s,t < 0$ by means of the {\tt GiNac} library, and successfully
checked against independent calculations.
In particular, the integrals $I_i$ with $i=1,\ldots,36,41$ were computed with the
package {\tt SecDec}.
For the most complex topologies, 
corresponding to the non-planar four-point integrals $I_i$ with
$i=37,\ldots,40,$ $42,43,44$, we adopted a different strategy. 
As the numerical evaluation of those integrals is challenging, we identified an alternative set of independent MIs that are {\it quasi finite}~\cite{vonManteuffel:2014qoa} in $d=6$.
The latter have been computed semi-numerically by means of an in-house algorithm: starting from
the Feynman parametrisation of the integrals, we carried out as many
analytic integration as possible, until we reached a form where the left
over multivariate integral could be numerically evaluated by means of Gauss quadrature.
Dimension-shifting identities~\cite{Tarasov:1996br,Lee:2009dh} and IBPs, implemented in \texttt{LiteRed}~\cite{Lee:2012cn,Lee:2013mka}, establish analytical relations between this set of integrals and the MIs we computed around $d=4$.

The definition of the 7 non-planar MIs that are quasi finite in $d=6$
dimensions, together with our results at the phase-space point
$s=-1/7$, $t=-1/3$, $m^2=1$, are collected in table~\ref{tab:numerics}.
We identified them through educated guesses or with the help of \texttt{Reduze}.
In the next subsection, we use the first of those integrals as an example to describe our evaluation strategy.

\begin{table}
  \begin{center}
    \begin{tabular}{ |c | c | r |}
      \hline
      {\bf graph} & $I^{[d]}-${\bf integral} & $ I^{[d=6-2\eps]}(s=-{1 \over 7}, t=-{1 \over 3}, m^2=1)$ \\
      \hline
      \hline
      \hline
      \raisebox{-19pt}{\includegraphics[scale=0.13]{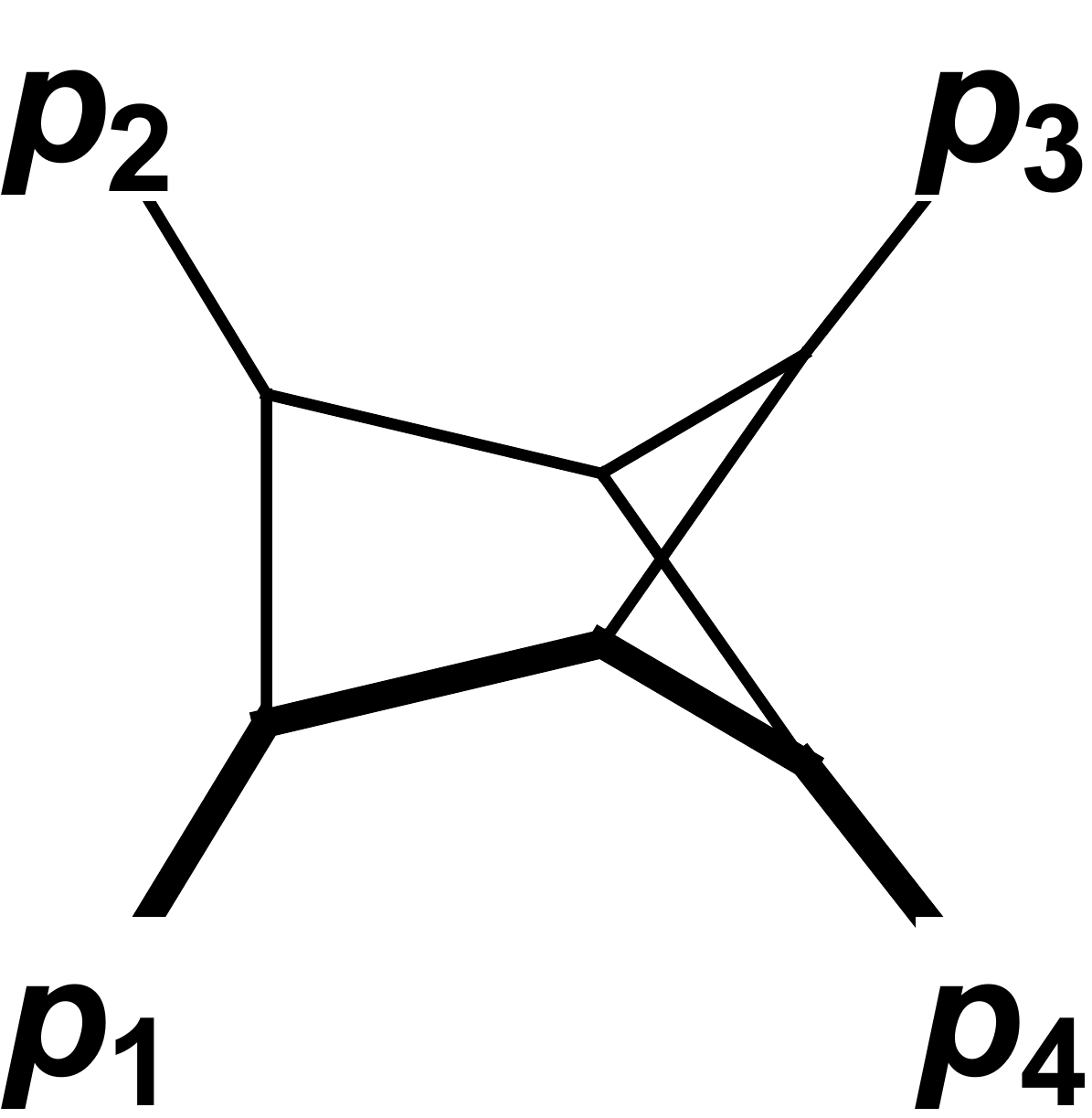}} 
                  & $I^{[d]}(1,1,1,0,1,0,1,1,1)$ & $-1.219372 - i\, 0.294408$ \\
      \hline
      \raisebox{-19pt}{\includegraphics[scale=0.13]{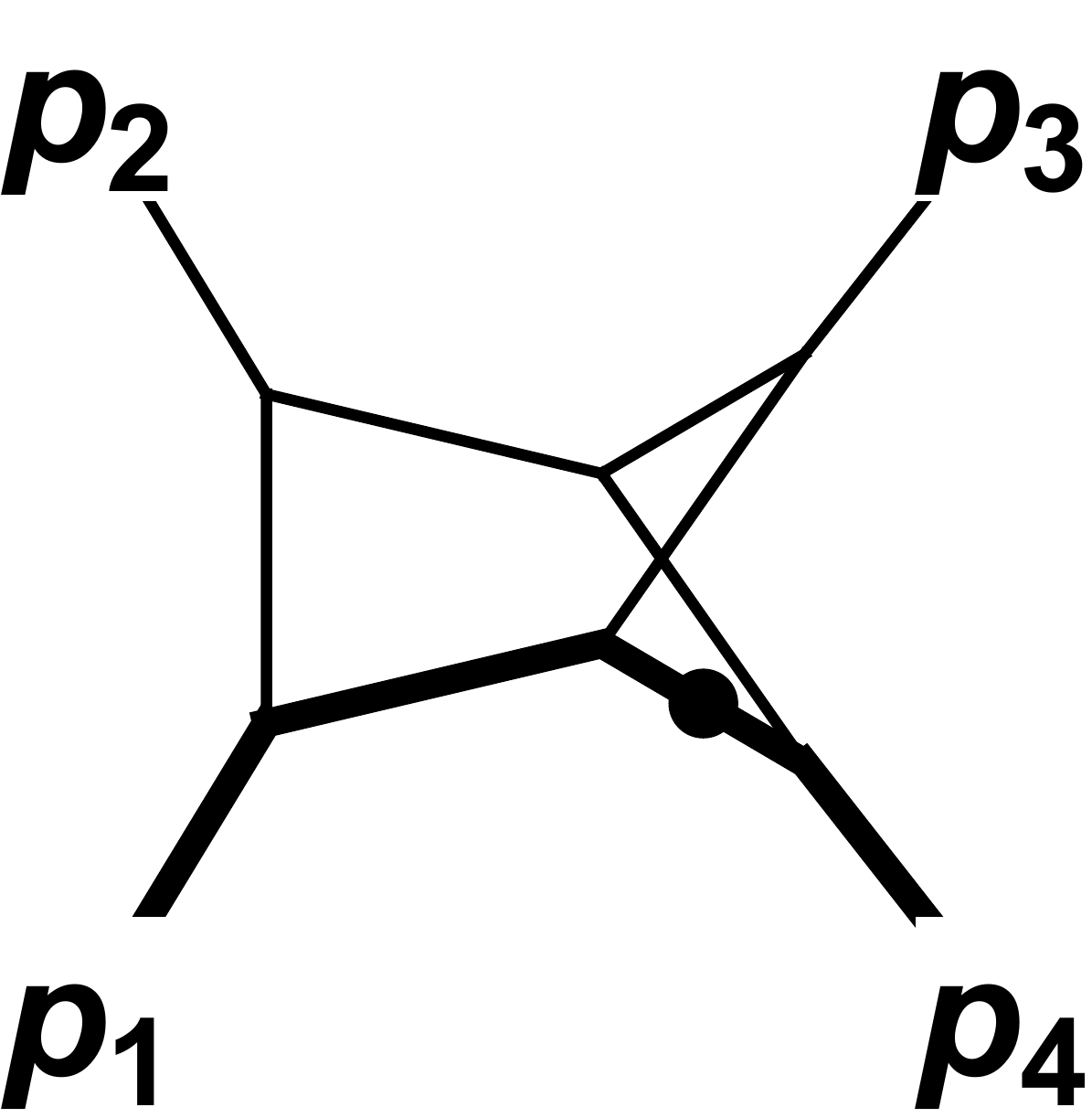}} 
                  & $I^{[d]}(1,2,1,0,1,0,1,1,1)$ & $0.98317 + i\, 1.00335$ \\
      \hline
      \raisebox{-19pt}{\includegraphics[scale=0.13]{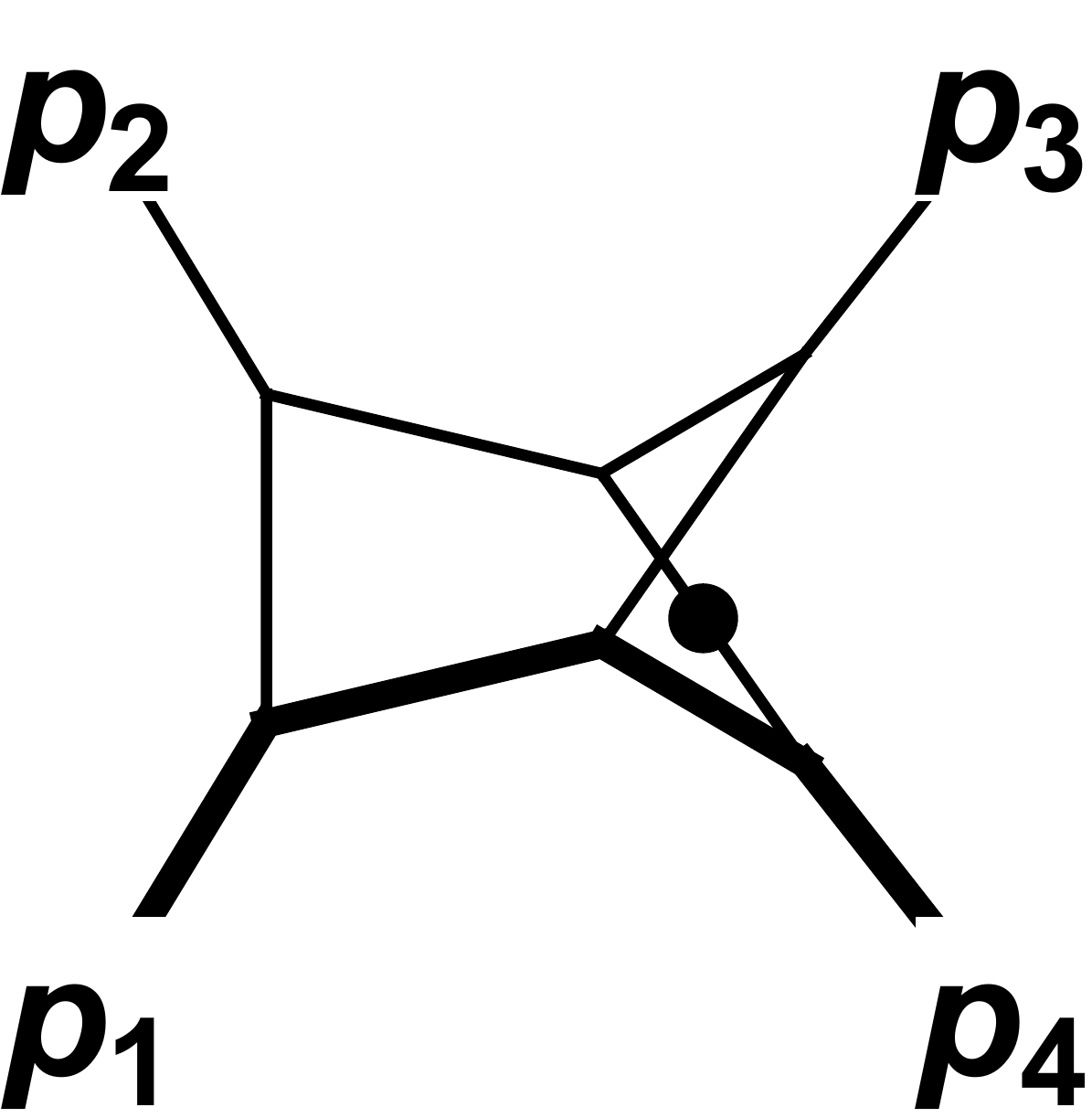}} 
                  & $I^{[d]}(1,1,1,0,1,0,1,2,1)$ & $12.039969 + i\, 6.660946$ \\
      \hline
      \raisebox{-19pt}{\includegraphics[scale=0.13]{figures/Topo6/T37}} 
                  & $I^{[d]}(1,1,1,0,0,0,1,1,1)$ & $\frac{1}{4\eps}+0.6798187 + i\, 0.0300909$ \\
      \hline
      \raisebox{-19pt}{\includegraphics[scale=0.13]{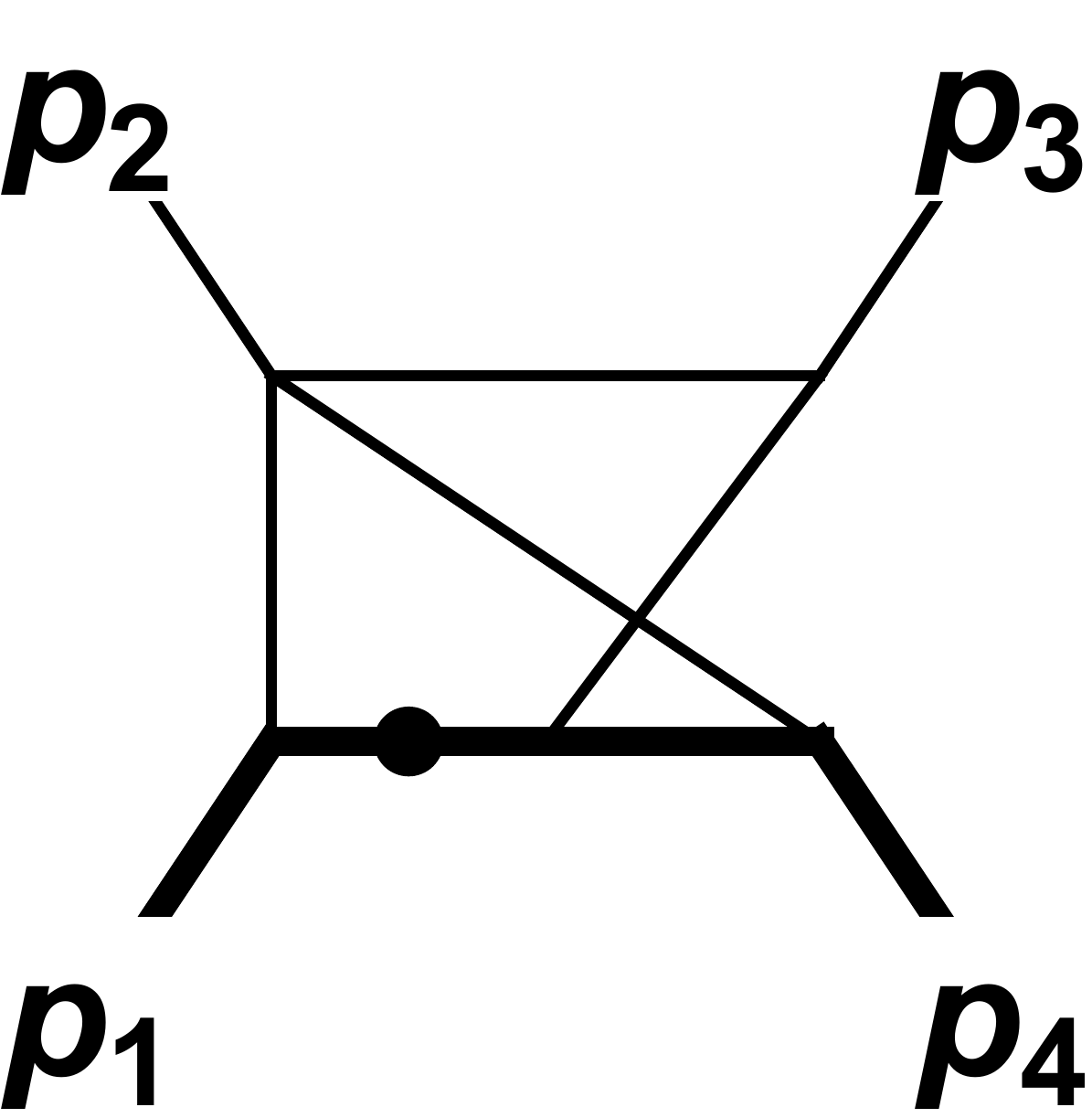}} 
                  & $I^{[d]}(2,1,1,0,0,0,1,1,1)$ & $-0.554605 - i\, 0.06984485$ \\
      \hline
      \raisebox{-19pt}{\includegraphics[scale=0.13]{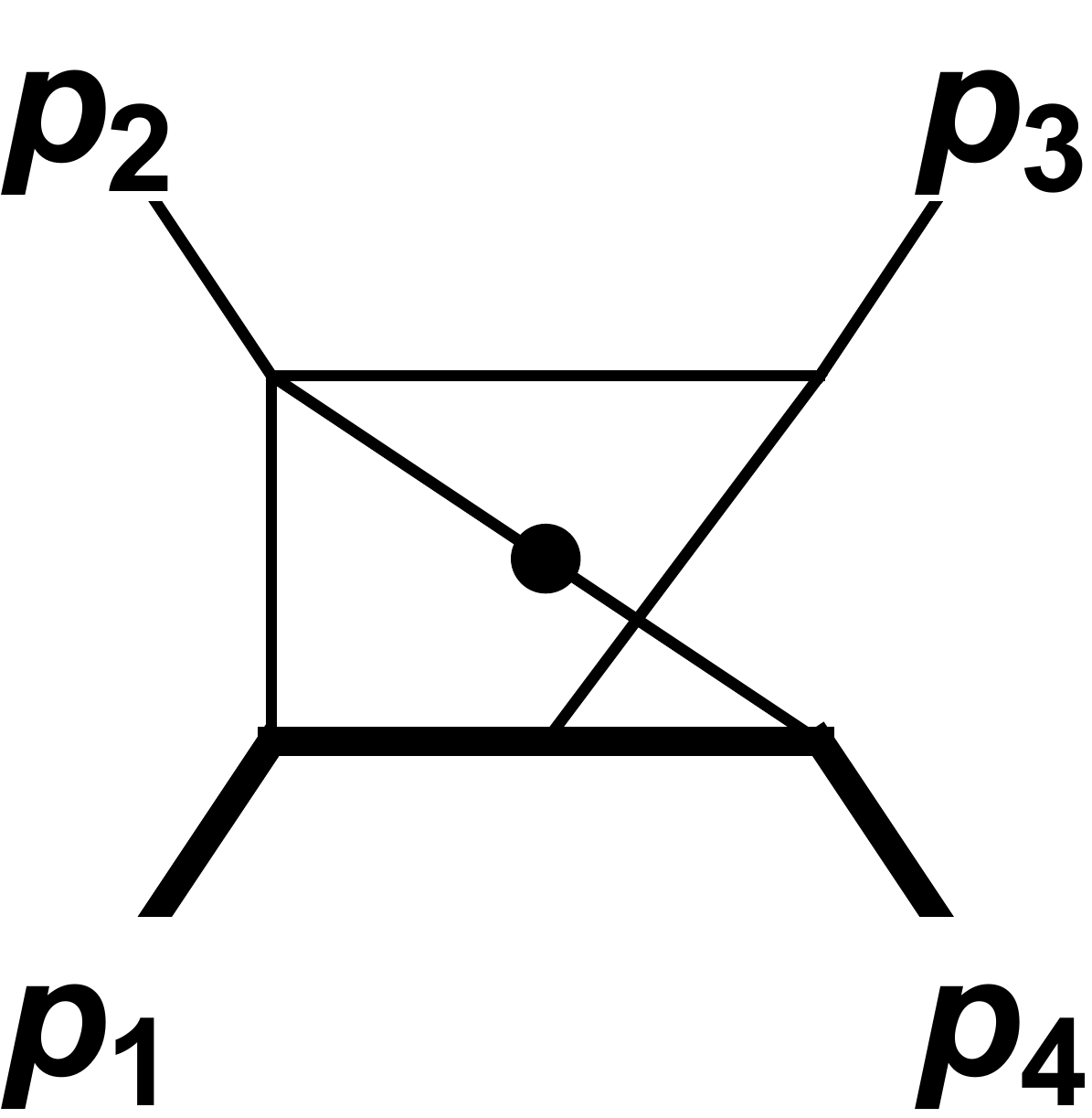}} 
                  & $I^{[d]}(1,1,1,0,0,0,1,2,1)$ & $-1.91103 + i\, 0.241649$ \\
      \hline
      \raisebox{-19pt}{\includegraphics[scale=0.13]{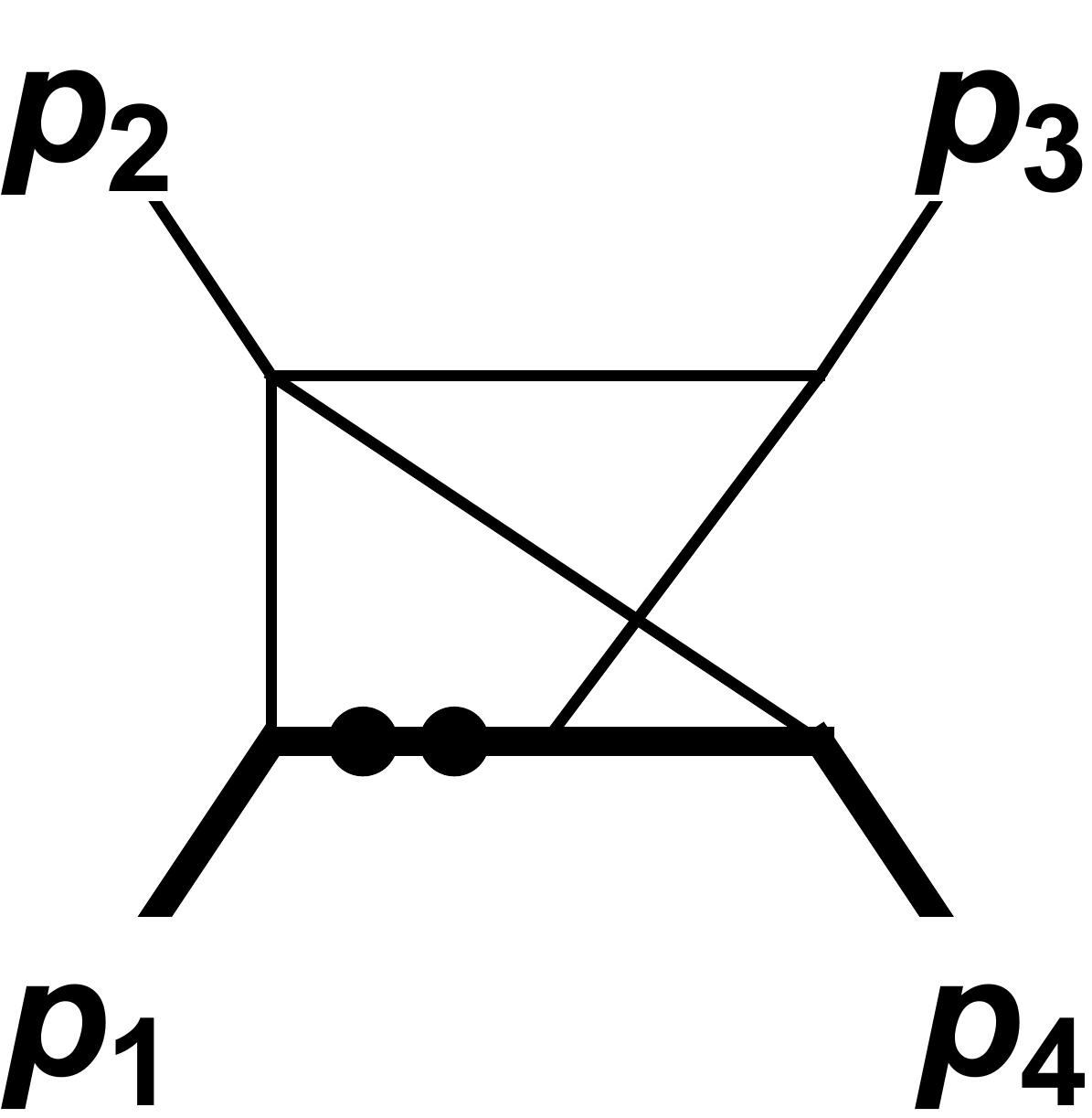}} 
                  & $I^{[d]}(3,1,1,0,0,0,1,1,1)$ & $0.525679 + i\, 0.248668$ \\
      \hline
    \end{tabular}
  \end{center}
  \caption{Numerical results for our set of quasi-finite non-planar MIs belonging to the 6- and 7-denominators topologies ($m^2=1$).}
  \label{tab:numerics}
\end{table}

\subsection{The non-planar box in $d=6$ dimensions}

As an example, we describe the numerical evaluation of the non-planar scalar integral
\begin{equation}
\raisebox{-19pt}{\includegraphics[scale=0.13]{figures/Topo6/npGraph}} 
=I^{[d]}(1,1,1,0,1,0,1,1,1)\ ,
\end{equation}
carried out in two steps.

\subsubsection{Analytic integrations}
By using Feynman parametrisation,
the integral can be written as %(keeping the Feynman prescription $+i\omega$)
\begin{equation}
  \raisebox{-19pt}{\includegraphics[scale=0.13]{figures/Topo6/npGraph}} = \Gamma(7) \int \widetilde{\dd^dk_1} \widetilde{\dd^dk_2} \int_0^1 dx_1 {\ldots} \int_0^1 dx_7 \,
\frac{\delta(1-x_{1234567})}{D_{tot}^7}\,,
\end{equation}
where
\begin{equation}
D_{tot}=x_1 D_1 + x_2 D_2 +x_3 D_3 +x_4 D_5 +x_5 D_7 +x_6 D_8 +x_7 D_9
+i\omega
\, .
\end{equation}
After integrating over  $k_1$ and $k_2$, one finds
%%%%%%%%%%%%%%%%%%%%%%%%%%%%%%%%%%%%%%%%%%%%%%%%%%%%%%%%%%%
\begin{equation}
\label{inte71}
\Gamma_\eps^2\raisebox{-19pt}{\includegraphics[scale=0.13]{figures/Topo6/npGraph}} 
=-\Gamma(7-d)\int_0^1 dx_1 {\ldots} \int_0^1 dx_7 
\frac{\delta(1-x_{1234567})}{A_0^{\frac{3d}{2}-7}\Delta^{7-d}} \,,
\end{equation}
%%%%%%%%%%%%%%%%%%%%%%%%%%%%%%%%%%%%%%%%%%%%%%%%%%%%%%%%%%%%%%%%%%%%%%%%%%%%%%
\begin{align}
A_0&= x_{34}x_{56} + x_5 x_6 + x_{346} x_7 +  x_2 x_{3457} + x_1
x_{2567} \,, \cr
\Delta&=  x_2^2 x_{3457} + x_1^2 x_{2567} + x_1 x_2 (x_2 + 2 x_{57})
+x_3 (-x_5 x_6 + x_2 x_7) t
\cr
&+(x_2 (-x_4 x_5 + x_3 x_7) - x_1 (x_4 x_{256} + x_{46} x_7)) (s-1)
-i \omega A_0 \,,
\end{align}
where we used the notation 
$x_{i_1 i_2 {\ldots} i_n} =x_{i_1}+x_{i_2}+{\ldots}+x_{i_n}$.
We perform as many analytic integrations as possible.
In particular, we integrate over $x_3$ eliminating the $\delta$-function,
and we make the changes of variables
$x_6 \to x_{26}-x_2$, $x_7 \to x_{57}-x_5$.
In this way, the polynomial $\Delta$ becomes linear in $x_4$ and $x_5$, so that
eq.~\eqref{inte71} becomes
\begin{align}
\label{inte6}
\Gamma_\eps^2\raisebox{-19pt}{\includegraphics[scale=0.13]{figures/Topo6/npGraph}} 
&=-\Gamma(7-d)
\int_0^1 dx_{26}
\int_0^{1-x_{26}} \frac{dx_{57}}{A^{\frac{3D}{2}-7}}
\int_0^{1-x_{2657}} \hspace{-1em} dx_1
\int_0^{x_{26}} \hspace{-0.5em} dx_2  
\int_0^{x_{57}} dx_5 
\times \cr &\phantom{=}
\int_0^{1-x_{12657}}  
\frac{dx_4}{(C_{41} x_4+C_{40})^{7-d}} \,,
\end{align}
where
\begin{align}
A&=x_{2657} (1 - x_{2657}) + x_{26} x_{57}\,, \cr
C_{41}&=t x_{26}x_5 - (s + t - 1) x_2 x_{57} - (s-1) x_1
x_{2657}\,,
\cr
C_{40}&=
x_1^2 x_{2657} + x_2 \left(x_2(1-x_{26}) + 2 x_1 x_{57}\right)
+t(-1 + x_{12657}) (x_{26} x_5 - x_2 x_{57})
-i\omega A
\cr
&\phantom{=}
+(s-1)(x_5 - x_{57}) \left(x_1 x_{26} + x_2 (-1 + x_{2657})\right) \,.
\end{align}
The integral over $x_4$ in eq.~\eqref{inte6} is finite for $d\to 6$ and, in this limit, we get
%After integrating over $x_4$ and taking the limit $d\to 6$, where the
%integral is finite, eq.~\eqref{inte6} becomes  
\begin{align}
\label{inte5}
\Gamma_\eps^2\raisebox{-19pt}{\includegraphics[scale=0.13]{figures/Topo6/npGraph}} 
\stackrel{d \to 6}{=}&
-
\int_0^1 dx_{26}
\int_0^{1-x_{26}} \frac{dx_{57}}{A^{2}}
\int_0^{1-x_{2657}} \hspace{-1em} dx_1
\int_0^{x_{26}} \hspace{-0.5em} dx_2  \times \nn
&\int_0^{x_{57}}  
\frac{dx_5}{x_5 t x_{26}-f_3}
\ln \left(\frac{f_4 x_5+P_4}{f_2 x_5 +P_3}\right) \,,
\end{align}
where 
\begin{align}
P_1&=x_2^2 (1 - x_{26}) + 2 x_1 x_2 x_{57} + x_1^2 x_{2657} + t x_{57}
(x_{26}-x_2)(-1 + x_{12657})-i\omega A \,,\cr
P_2&=x_2^2 (1 - x_{26})  +  2 x_1 x_2 x_{57} + 
    x_1^2 x_{2657} +  (s-1) (-1 + x_{12657})(x_1
    x_{2657}+ x_2 x_{57})-i\omega A
    \,, \cr
P_3&=P_1-f_2 x_{57}  \,, \cr
P_4&=P_2-f_4 x_{57}  \,, 
\end{align}
and
\begin{align}
f_1&=f_3- t x_{26} x_{57} \,, \cr
f_2&=f_4- t x_{26} (1- x_{12657}) \,,\cr
f_3&= (s+t-1) x_2 x_{57} + (s-1) x_1 x_{2657} \,, \cr
f_4&= (s-1) \left( x_1 x_{26} + x_2 (-1 + x_{2657}) \right) \,. 
\end{align}
Finally, we integrate over $x_5$, and reduce eq.~\eqref{inte5} to 
\begin{align}
\label{inte4}
\Gamma_\eps^2\raisebox{-19pt}{\includegraphics[scale=0.13]{figures/Topo6/npGraph}} 
\stackrel{d \to 6}{=}
&-
\int_0^1 \frac{dx_{26}}{t x_{26}}
\int_0^{1-x_{26}} \frac{dx_{57}}{A^{2}}
\int_0^{1-x_{2657}} dx_1
\int_0^{x_{26}} dx_2  \ \times \cr
&\left(
 {\mathrm{Li}}_2\left(\frac{Q_1}{R}\right)
-{\mathrm{Li}}_2\left(\frac{Q_2}{R}\right)
-{\mathrm{Li}}_2\left(\frac{Q_3}{R}\right)
+{\mathrm{Li}}_2\left(\frac{Q_4}{R}\right)
\right) \,,
\end{align}
where 
\begin{align}
Q_1&=f_1 f_2 \,, \quad Q_2=f_1 f_4 \,, \quad Q_3=f_3 f_2 \,, \quad Q_4=f_3 f_4 \,, \cr
R&= Q_i+P_i t x_{26}, \quad \forall i, \ i=1,{\ldots},4
\,.
\end{align}

\subsubsection{Numerical integrations} 
The four remaining integration variables in eq.~\eqref{inte4} 
are rescaled, and mapped onto a four-dimensional hypercube of unit side,
\begin{equation}
x_{26}=t_1\;,\ 
x_{57}=(1-x_{26})t_2  \; ,\ 
x_{1}=(1-x_{2657})t_3  \; ,\ 
x_{2}=(x_{26})t_4  \; ;
\end{equation}
so that the new variables $t_i$ have to be integrated over $[0,1]$.
At this point, we have to consider the branch points of the dilogarithms that
appear in eq.~\eqref{inte4}, which correspond the hypersurfaces defined by the equations
\begin{equation}
\label{disc1}
R(t_1,t_2,t_3,t_4)=0\,, \quad P_i(t_1,t_2,t_3,t_4)=0, \quad i=1,{\ldots} 4 \,.
\end{equation}
It is necessary to sample
carefully the integrand near these branch points.
Therefore, for the integration over $t_4$,  we split the
integration interval 
at the $N_4(t_1,t_2,t_3)$ real solutions $z_{4j}(t_1,t_2,t_3)$ of
eq.~\eqref{disc1} which are on the interval $[0,1]$,
\begin{equation}
\int_0^1 dt_4 = \sum_{j=0}^{N_4-1}
\int_{z_{4j}(t_1,t_2,t_3)}^{z_{4,j+1}(t_1,t_2,t_3)} dt_4\,, 
z_{40}=0 \;, \ z_{4{N_4}}=1\,.
\end{equation}
Analogously, for the integration over $t_3$, we split the integration
interval at the $N_3(t_1,t_2)$ real zeros $z_{3j}(t_1,t_2)$ of the discriminants
(polynomials in $(t_1,t_2,t_3))$  that appear in the zeros $z_{4j}$.
These are the points where the hypersurfaces of eq.~\eqref{disc1} are tangent to the
hyperplane $t_4=\text{constant}$,
\begin{equation}
\int_0^1 dt_3 = \sum_{j=0}^{{N_3}-1} \int_{z_{3j}}^{z_{3,j+1}} dt_3\,, 
\quad z_{30}=0 \;,\  z_{3{N_3}}=1\,.
\end{equation}
Analogously, for the integration over $t_2$, we split the integration
interval at the $N_2(t_1)$ zeros $z_{2j}(t_1)$ of the discriminants
(polynomials in $(t_1,t_2))$  that appear in the zeros $z_{3j}$,
\begin{equation}
\int_0^1 dt_2 = \sum_{j=0}^{{N_2}-1} \int_{z_{2j}}^{z_{2,j+1}} dt_2\,, 
\quad z_{20}=0 \;, \ z_{2{N_2}}=1\,.
\end{equation}
We proceed in a similar way for the last integration,
\begin{equation}
\int_0^1 dt_1 = \sum_{j=0}^{{N_1}-1} \int_{z_{1j}}^{z_{1,j+1}} dt_1\,, 
\quad z_{10}=0 \;, \ z_{1{N_1}}=1\,.
\end{equation}

To carry out the integration over a generic interval $[t_a,t_b]$, we perform the change
of variables $t_i \to u_i$, with
\begin{equation}
t_i=t_{ai}+\frac{e^{u_i^3}}{e^{u_i^3}+1} (t_{bi}-t_{ai}) \,,
\quad i=1,{\ldots},4 \,,
\end{equation}
in order to deal with possible singularities at the endpoints. The variable
$u_i$ should be integrated in $(-\infty,\infty)$ but we
actually  truncate the integration domain to $(-M,+M)$, with $M$ suitably large
(typically $M\sim 4$), 
and we use Gauss-Legendre integration over 16 points.
Note that all the singularities in the integrands are logarithmic,
and therefore integrable, so we can safely set a very small
value of $\omega$, like $10^{-30}$.  

By using $16$ subdivisions in each interval and in every variable we
find that our integral, in the phase space point $s=-1/7$, $t=-1/3$, $m^2=1$,
amounts to

\begin{equation}
\raisebox{-19pt}{\includegraphics[scale=0.13]{figures/Topo6/npGraph}} 
\stackrel{d \to 6}{=}
 -1.219372 - i\, 0.294408 \,. \\ 
\end{equation}

A similar procedure is adopted for the other integrals in
Tab.~\ref{tab:numerics}.  Case-by-case, after the analytic
integrations, the corresponding integrands, in the $d \to 6$ limit,
are found to be combinations of logarithms, so that the decomposition
of the integration domain, and the numerical integration can be
carried out along the same lines as for the non-planar scalar box
integral.

\clearpage
%%%%%%%%%%%%%%%%%%
%%% Local Variables:
%%% TeX-master: "../main"
%%% End:

%% file: tex/Conclusions.tex
\section{Conclusions}

In this work, we presented the analytic evaluation of the two-loop master integrals needed to compute the non-planar Feynman diagrams contributing to $\mu e$
elastic scattering in QED at NNLO.
We adopted the same computational strategy previously applied to planar diagrams, and presented in the companion article \cite{Mastrolia:2017pfy}. Namely,
we employed the method of differential equations and of the Magnus exponential to identify a canonical set of master integrals and we derived boundary conditions either from the regularity requirements at pseudothresholds or from the knowledge of the integrals at special kinematic points, possibly evaluated by means of auxiliary, simpler systems of differential equations.
The considered master integrals were expressed as a Taylor series around four space-time dimensions, whose coefficients are written as a combination of generalised polylogarithms. We worked in the massless electron approximation, while keeping full dependence on the muon mass. 

The scattering of high-energy muons on atomic electrons has been recently proposed as an ideal framework to determine, in a novel way, the leading hadronic contribution to the anomalous magnetic moment of the muon. The ambitious experimental goal of the MUonE project, namely measuring the differential cross section of the $\mu e \to \mu e$ process with an accuracy of 10ppm, requires, on the theoretical side, the knowledge of the QED corrections at NNLO.
The results of the planar and non-planar master integrals we obtained represent an important step towards the evaluation of the virtual corrections at the required order. 

By crossing symmetry, our results are also relevant for muon-pair production at $e^+ e^-$-colliders operating well below the $Z$-pole, such as Belle II and VEPP-2000, as well as for the QCD corrections to heavy-quark pair production at hadron colliders. The former application is particularly interesting, as a precise knowledge of the differential cross section in QED could be exploited to constrain non-standard $ee\mu\mu$ interactions via the measurement of a forward-backward asymmetry.

%%%%%%%%%%%%%%%%%%
%%% Local Variables:
%%% TeX-master: "../main"
%%% End:

%% file: tex/AuxVert.tex
\section{Evaluation of the auxiliary vertex integrals for eq.~\eqref{eq:limit_i1112}}
\label{app:A}
In this appendix, we discuss the solution of the DEQs for the vertex integrals that we used in eq.~\eqref{eq:limit_i1112} as an input for the determination of the boundary constants of the MIs $\GG_{11,\,12}$.
\begin{figure}[t]
  \centering
  \includegraphics[width=0.15\textwidth]{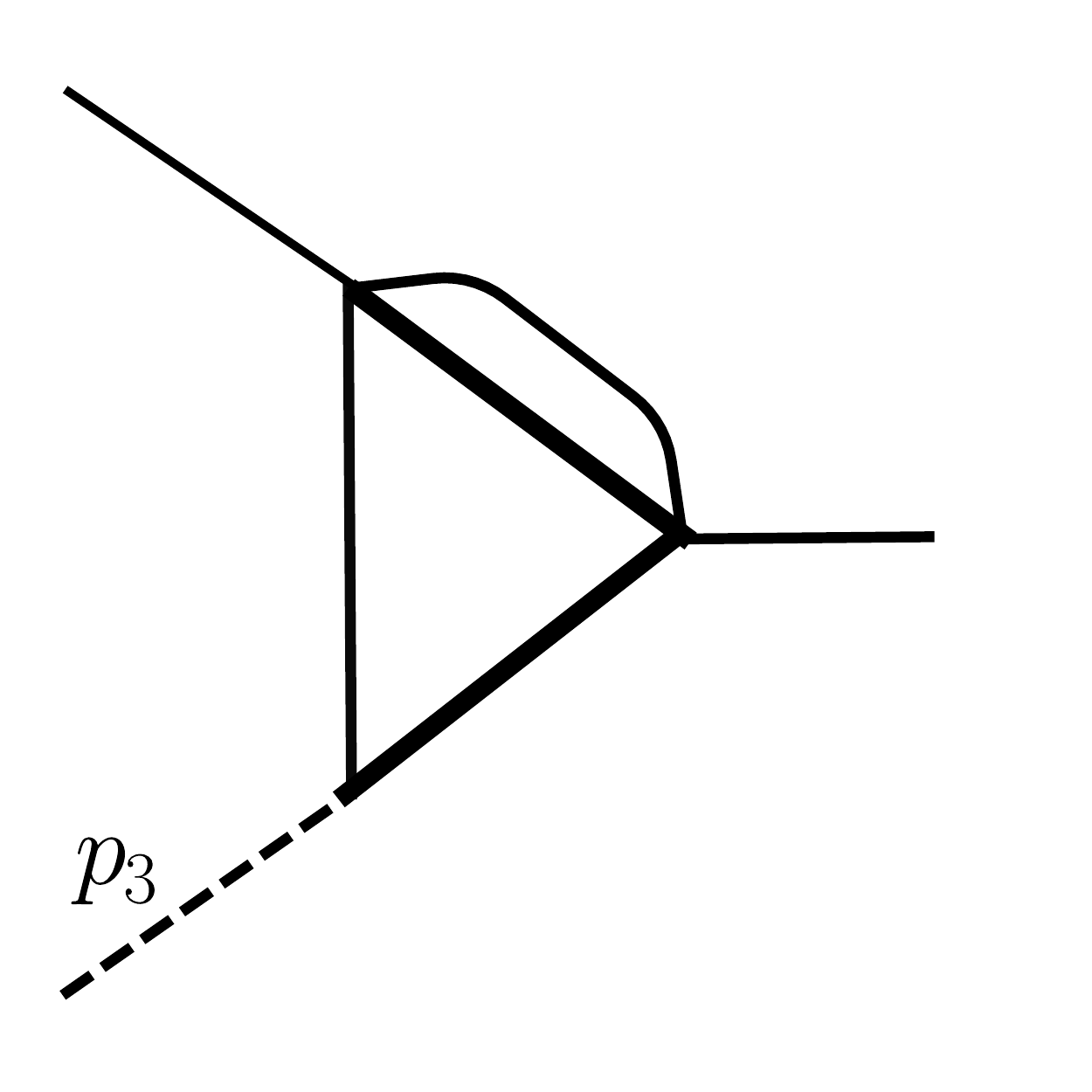}
  \caption{Auxiliary three-point integral family for eq.~\eqref{eq:limit_i1112}. Thick and thin lines represent, respectively, massive and massless propagators. The dashed line corresponds to an external leg with arbitrary squared momentum.}
  \label{fig:Aux_Topo12}
\end{figure}

The required input integrals belong to the integral family
\begin{gather}
  \int \widetilde{\dd^d k_1}\widetilde{\dd^d k_2}\,
  \frac{\Den_{5}^{n_5}\Den_{6}^{n_6}\Den_{7}^{n_7}}{\Den_{1}^{n_1}\Den_{2}^{n_3}\Den_{3}^{n_3}\Den_{4}^{n_4}}\,,\quad n_i\geq0\,,
  \label{eq:Aux1fam}
\end{gather}
which is identified by the set of denominators
\begin{gather}
\Den_1 = k_1^2,\quad
\Den_2 = k_2^2-m^2,\quad
\Den_3 = (k_1+k_2+p_1)^2,\quad
\Den_4 = (k_1+p_1+p_2)^2-m^2, \nonumber\\
\Den_5 = (k_1+p_2)^2,\quad
\Den_6 = (k_2+p_1)^2,\quad
\Den_7 = (k_2+p_2)^2,
\end{gather} 
and by the external momenta
\begin{align}
p_1^2=p_2^2=0\,,\quad p_3^2=(p_1+p_2)^2\,.
\label{eq:kinaux}
\end{align}
A representative 4-propagator integral  of this family is depicted in figure~\ref{fig:Aux_Topo12}.
IBPs reduce the integral family of eq.~\eqref{eq:Aux1fam} to a set of 5 MIs, whose dependence on $p_3^2$ is parametrised in terms of the dimensionless variable
\begin{align}
x=-\frac{p_3^2}{m^2}\,.
\end{align}
The integral basis
  \begin{align}
 \GG_{1}=&\eps^2 \raisebox{-19pt}{\includegraphics[scale=0.23]{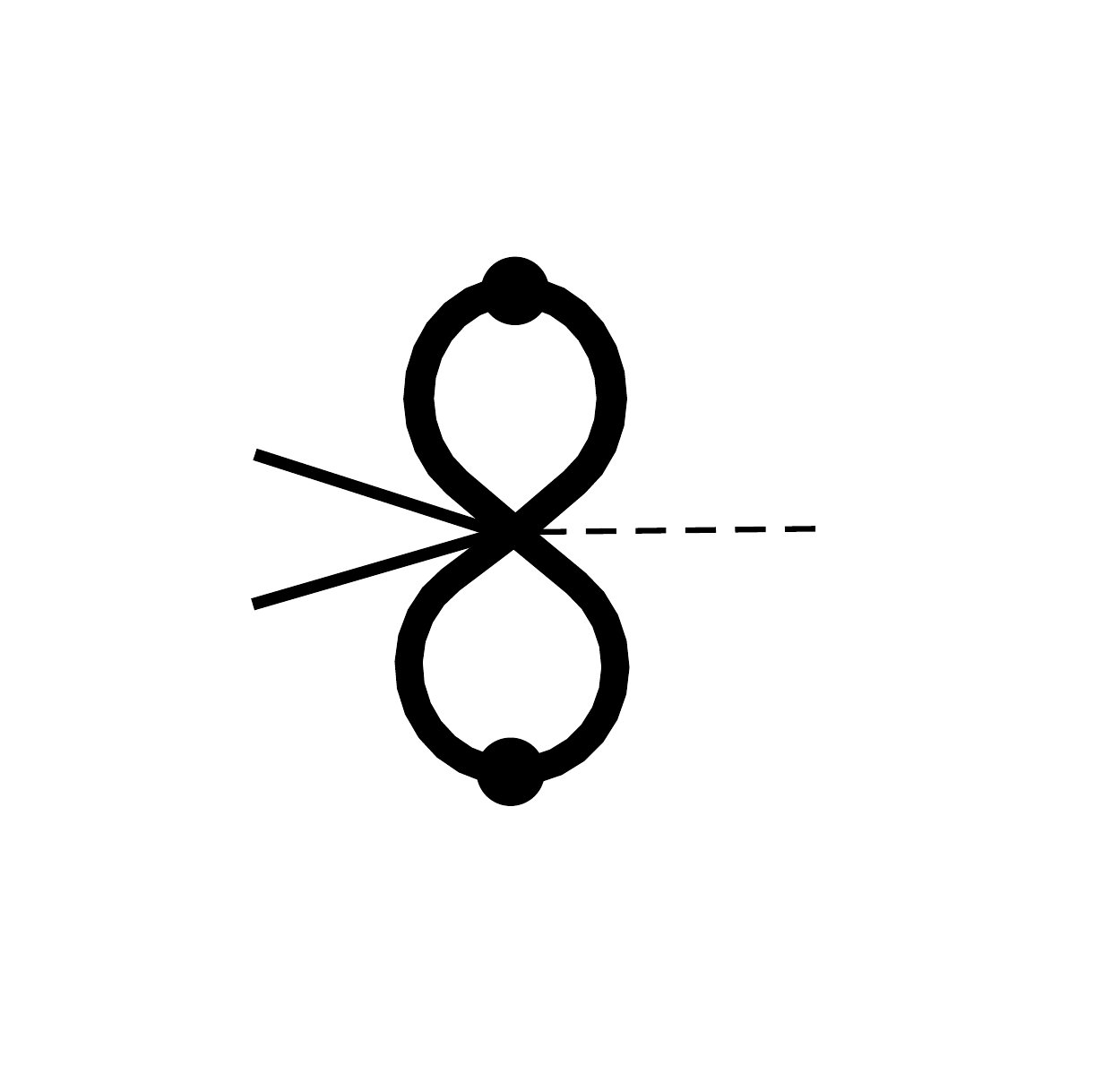}}\,,\quad \GG_{2}=-\eps(1-\eps)m^2 \raisebox{-12pt}{\includegraphics[scale=0.23]{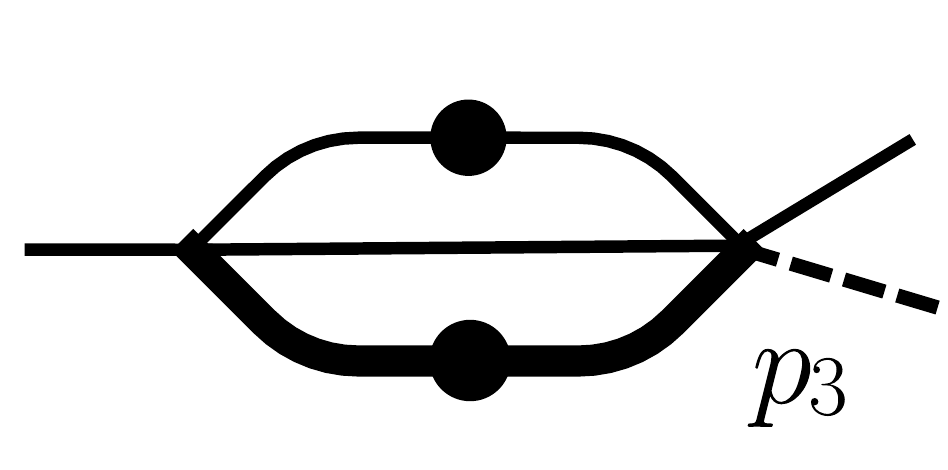}}\,,\quad 
  \GG_{3}=-\eps^2p_3^2 \raisebox{-14pt}{\includegraphics[scale=0.23]{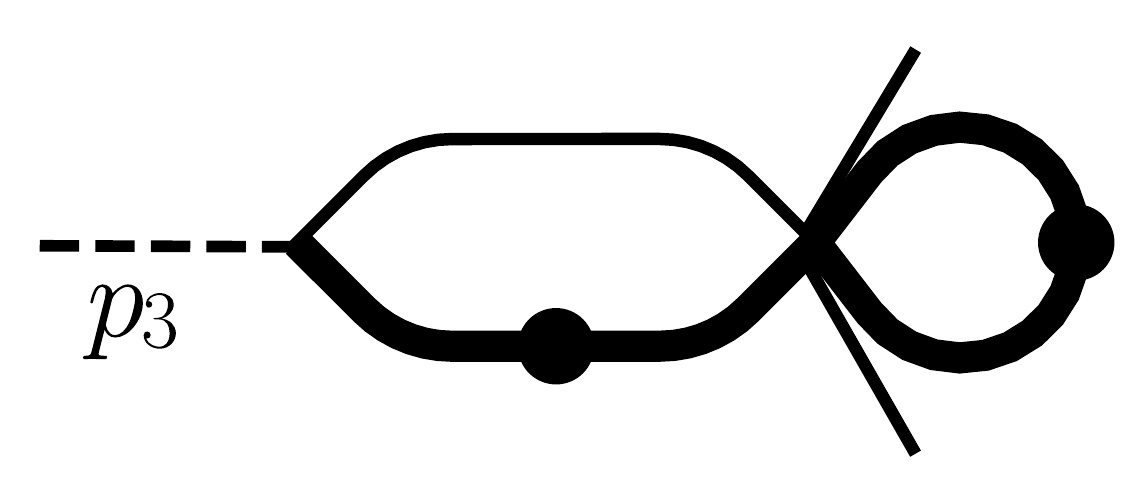}}\nn
&\qquad \GG_{4}=-\eps^3p_3^2 \raisebox{-30pt}{\includegraphics[scale=0.20]{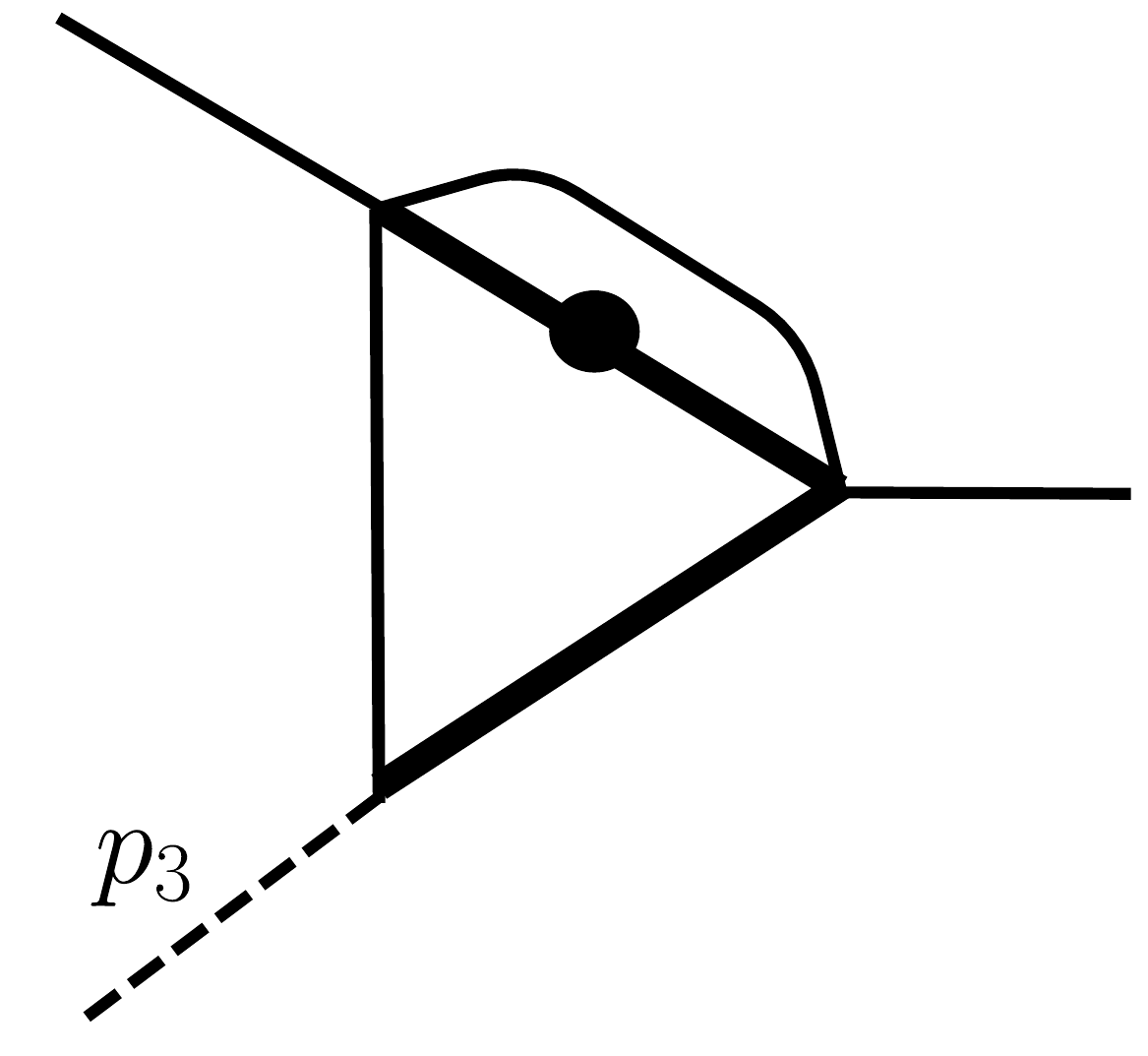}}\,,\quad
\GG_{5}=-\eps^2m^2p_3^2 \raisebox{-30pt}{\includegraphics[scale=0.18]{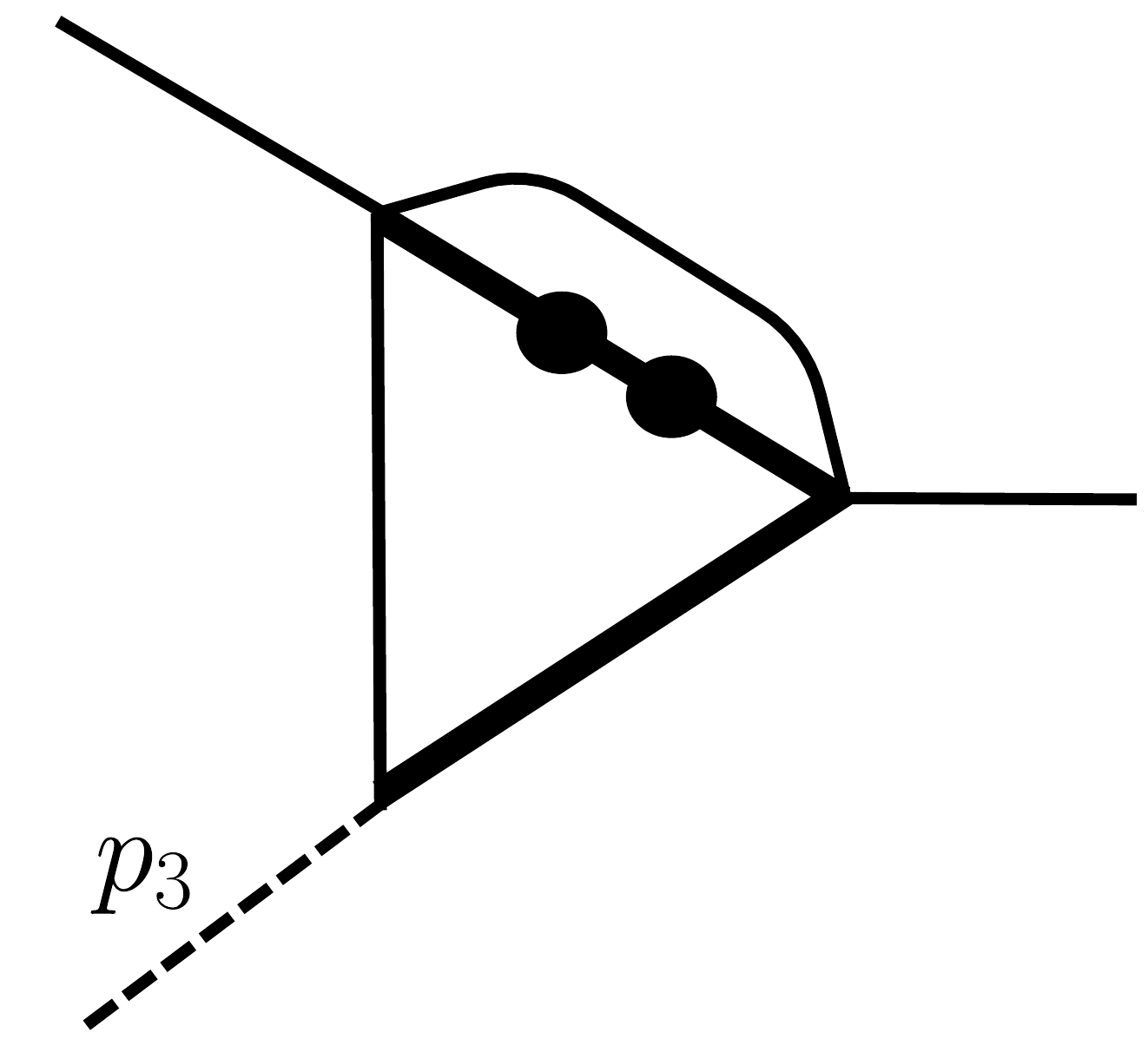}}\,,
  \end{align}
   fulfils canonical DEQs,
  \begin{align}
  d \GGvec = \eps \, \dA \, \GGvec \ ,
\end{align}
where 
\begin{align}
\dA =  \MM_0 \, \dlog x+  \MM_{-1} \, \dlog (x+1)\,,
\label{eq:canaux}
\end{align}
with
\begin{align}
 \MM_0=\left(
\begin{array}{ccccc}
 0 &\pminus 0 &\pminus 0 &\pminus 0 &\pminus 0 \\
 0 &\pminus 0 &\pminus 0&\pminus 0 &\pminus 0 \\
 0 &\pminus 0 &\pminus 1&\pminus 0 &\pminus 0 \\
 0 &\pminus0&\pminus 0 &\pminus 1 &\pminus 0 \\
 0 &\pminus 0 & -\frac{1}{2} &-3 &-2 \\
\end{array}
\right)\,,\quad
\MM_1=\left(
\begin{array}{ccccc}
 0 &\pminus 0 &\pminus 0 &\pminus 0 &\pminus 0 \\
 0 &\pminus 0&\pminus0&\pminus 0 &\pminus 0 \\
 -1 &\pminus0&\pminus -2&\pminus 0 &\pminus 0 \\
\frac{1}{2}&-\frac{1}{2}&0& -3& -2\\
-\frac{1}{2}&\pminus \frac{1}{2}&\pminus0&\pminus3 &\pminus2 \\
\end{array}
\right)\,.
\end{align}
The general solution of the DEQs can be expressed in terms of harmonic polylogarithms (HPLs), i.e. GPLs with weights $a_i\in\{-1,0,1\}$. The integrals $\GG_{1,2}$, which are independent of $x$, are determined by direct integration,
\begin{align}
\GG_{1}=1\,,\quad \GG_{2}(\eps)=1+\frac{\pi^2}{3}\eps^2-2\zeta_3\eps^3+\frac{\pi^4}{10}\eps^4+\mathcal{O}(\eps^5)\,,
\end{align}
whereas the boundary constants of $\GG_{3,4,5}$ are obtained by demanding their vanishing in the regular limit $x\to 0$. In particular, for the two triangle integrals, we obtain
\begin{align}
\GG_{4}(\eps,x)=& \left(\frac{ \pi ^2}{6} G(-1;x)-G(-1;x)G(0,-1;x)+2 G(0,-1,-1;x)\right)\eps^3\nn
&+\bigg(\zeta_3 G(-1;x)+\frac{ \pi ^2}{12}G^2(-1;x)-\frac{ \pi ^2}{6}G(0,-1;x)+\frac{1}{2}G^2(-1;x)G(0,-1;x)
\nn&+G(-1;x)G(0,0,-1;x)-3 G(0,-1,-1,-1;x)+2 G(0,-1,0,-1;x)\nn
&-2G(0,0,-1,-1;x)\bigg)\eps^4 +\mathcal{O}\left(\eps^5\right)\,,\nn
\GG_{5}(\eps,x)=&\frac{1}{2}
   G(0,-1;x)\eps^2+\bigg(\frac{ \pi ^2}{6}  G(-1;x)+G(-1;x)G(0,-1;x)-3
   G(0,-1,-1;x)\nn
   &-\frac{1}{2} G(0,0,-1;x)\bigg)\eps^3+\bigg(-\zeta_3 G(-1;x)-\frac{\pi^2}{12}G^2(-1;x)+\frac{ \pi ^2}{6} G(0,-1;x)\nn
   &-\frac{1}{2}G^2(-1;x)G(0,-1;x)-G(-1,x)G(0,0,-1;x)+5 G(0,-1,-1,-1;x)\nn
   &+
   G(0,-1,0,-1;x)+3 G(0,0,-1,-1;x)+\frac{3}{2} G(0,0,0,-1;x)\bigg)\eps^4+\mathcal{O}\left(\eps^5\right)\,.
 \end{align}
 These solutions are real valued in the interval $0<x<1$. By analytic continuation to the region $x<0$, we can extract the values of the two integral at $p_3^2=m^2$,
 \begin{align}
\GG_{4}(\eps,-1)=& 2 \zeta_3 \eps^3+\frac{7 \pi ^4}{180} \eps^4+\mathcal{O}\left(\eps^5\right) \,,\nn
\GG_{5}(\eps,-1)=&-\frac{\pi ^2 }{12}\eps^2-\frac{5 \zeta_3}{2} \eps^3-\frac{1}{12} \pi ^4 \eps^4+\mathcal{O}\left(\eps^5\right)\,,
 \end{align}
 which can then be used in eq.~\eqref{eq:limit_i1112} .

%%% Local Variables:
%%% TeX-master: "../main"
%%% End:

%% file: tex/dlogforms.tex
\section[dlog forms]{$\dlog$ forms}
\label{app:C}
In this appendix we collect the coefficient matrices of the
$\dlog$-form
\begin{align}
  \dA = {} & \MM_1 \, \dlog(w) + \MM_2 \, \dlog(1+w) + \MM_3 \, \dlog(1-w) \nn
  &+ \MM_4 \, \dlog(z) + \MM_5 \, \dlog(1+z)  + \MM_6\, \dlog(1-z) \nn
  &+ \MM_7\, \dlog(w+z) + \MM_8 \, \dlog\left(z-w\right) \nn
  &+ \MM_9\, \dlog\left(z^2-w\right)+ \MM_{10}\, \dlog\left(1-w+w^2-z^2\right) \nn
   &+ \MM_{11}\, \dlog\left(1-3w+w^2+z^2\right)+ \MM_{12}\, \dlog\left(z^2-w^2-w\, z^2+w^2z^2 \right) \nn
\end{align}
for the master integrals in the non-planar integral family, defined in eqs.~\eqref{eq:def:ourintegrals} and~(\ref{eq:2Lfamily}):
% \newpage
\vspace*{\fill}
\begin{align}
\MM_1 = \scalemath{0.5}{
  \input{matrices/EMu_np_1}
  }\,,
\end{align}
\vspace*{\fill}
\newpage
\vspace*{\fill}
\begin{align}
\MM_2 = \scalemath{0.53}{
  \input{matrices/EMu_np_2}
  }\,,
\end{align}
\vspace*{\fill}
\newpage
\vspace*{\fill}
\begin{align}
\MM_3 = \scalemath{0.53}{
  \input{matrices/EMu_np_3}
  }\,,
\end{align}
\vspace*{\fill}
\newpage
\vspace*{\fill}
\begin{align}
\MM_4 = \scalemath{0.53}{
  \input{matrices/EMu_np_4}
  }\,,
\end{align}
\vspace*{\fill}
\newpage
\vspace*{\fill}
\begin{align}
\MM_5 = \scalemath{0.53}{
  \input{matrices/EMu_np_5}
  }\,,
\end{align}
\vspace*{\fill}
\newpage
\vspace*{\fill}
\begin{align}
\MM_6 = \scalemath{0.53}{
  \input{matrices/EMu_np_6}
  }\,,
\end{align}
\vspace*{\fill}
\newpage
\vspace*{\fill}
\begin{align}
\MM_7 = \scalemath{0.53}{
	\input{matrices/EMu_np_7}
}\,,
\end{align}
\vspace*{\fill}
\newpage
\vspace*{\fill}
\begin{align}
\MM_8 = \scalemath{0.53}{
	\input{matrices/EMu_np_8}
}\,,
\end{align}
\vspace*{\fill}
\newpage
\vspace*{\fill}
\begin{align}
\MM_9 = \scalemath{0.53}{
	\input{matrices/EMu_np_9}
}\,,
\end{align}
\vspace*{\fill}
\newpage
\vspace*{\fill}
\begin{align}
\MM_{10} = \scalemath{0.53}{
	\input{matrices/EMu_np_10}
}\,,
\end{align}
\vspace*{\fill}
\newpage
\vspace*{\fill}
\begin{align}
\MM_{11} = \scalemath{0.53}{
	\input{matrices/EMu_np_11}
}\,,
\end{align}
\vspace*{\fill}
\newpage
\vspace*{\fill}
\begin{align}
\MM_{12} = \scalemath{0.53}{
	\input{matrices/EMu_np_12}
}\,.
\end{align}
\vspace*{\fill}

%%% Local Variables:
%%% TeX-master: "../main"
%%% End:

%% file: matrices/EMu_np_1.tex
\left(
\begin{array}{cccccccccccccccccccccccccccccccccccccccccccc}
 \pminus 0 & \pminus 0 & \pminus 0 & \pminus 0 & \pminus 0 & \pminus 0 & \pminus 0 & \pminus 0 & \pminus 0 & \pminus 0 & \pminus 0 & \pminus 0 & \pminus 0 & \pminus 0 & \pminus 0 & \pminus 0 & \pminus 0 & \pminus 0 & \pminus 0 & \pminus 0 & \pminus 0 & \pminus 0 & \pminus 0 & \pminus 0 & \pminus 0 & \pminus 0 & \pminus 0 & \pminus 0 & \pminus 0 & \pminus 0 & \pminus 0 & \pminus 0 & \pminus 0 & \pminus 0 & \pminus 0 & \pminus 0 & \pminus 0 & \pminus 0 & \pminus 0 & \pminus 0 & \pminus 0 & \pminus 0 & \pminus 0 & \pminus 0 \\
 \pminus 0 & \pminus 0 & \pminus 0 & \pminus 0 & \pminus 0 & \pminus 0 & \pminus 0 & \pminus 0 & \pminus 0 & \pminus 0 & \pminus 0 & \pminus 0 & \pminus 0 & \pminus 0 & \pminus 0 & \pminus 0 & \pminus 0 & \pminus 0 & \pminus 0 & \pminus 0 & \pminus 0 & \pminus 0 & \pminus 0 & \pminus 0 & \pminus 0 & \pminus 0 & \pminus 0 & \pminus 0 & \pminus 0 & \pminus 0 & \pminus 0 & \pminus 0 & \pminus 0 & \pminus 0 & \pminus 0 & \pminus 0 & \pminus 0 & \pminus 0 & \pminus 0 & \pminus 0 & \pminus 0 & \pminus 0 & \pminus 0 & \pminus 0 \\
 \pminus 0 & \pminus 0 & \pminus 0 & \pminus 0 & \pminus 0 & \pminus 0 & \pminus 0 & \pminus 0 & \pminus 0 & \pminus 0 & \pminus 0 & \pminus 0 & \pminus 0 & \pminus 0 & \pminus 0 & \pminus 0 & \pminus 0 & \pminus 0 & \pminus 0 & \pminus 0 & \pminus 0 & \pminus 0 & \pminus 0 & \pminus 0 & \pminus 0 & \pminus 0 & \pminus 0 & \pminus 0 & \pminus 0 & \pminus 0 & \pminus 0 & \pminus 0 & \pminus 0 & \pminus 0 & \pminus 0 & \pminus 0 & \pminus 0 & \pminus 0 & \pminus 0 & \pminus 0 & \pminus 0 & \pminus 0 & \pminus 0 & \pminus 0 \\
\end{array}
\right)

%% file: matrices/EMu_np_2.tex
\left(
\begin{array}{cccccccccccccccccccccccccccccccccccccccccccc}
 \pminus 0 & \pminus 0 & \pminus 0 & \pminus 0 & \pminus 0 & \pminus 0 & \pminus 0 & \pminus 0 & \pminus 0 & \pminus 0 & \pminus 0 & \pminus 0 & \pminus 0 & \pminus 0 & \pminus 0 & \pminus 0 & \pminus 0 & \pminus 0 & \pminus 0 & \pminus 0 & \pminus 0 & \pminus 0 & \pminus 0 & \pminus 0 & \pminus 0 & \pminus 0 & \pminus 0 & \pminus 0 & \pminus 0 & \pminus 0 & \pminus 0 & \pminus 0 & \pminus 0 & \pminus 0 & \pminus 0 & \pminus 0 & \pminus 0 & \pminus 0 & \pminus 0 & \pminus 0 & \pminus 0 & \pminus 0 & \pminus 0 & \pminus 0 \\
 \pminus 0 & \pminus 0 & \pminus 0 & \pminus 0 & \pminus 0 & \pminus 0 & \pminus 0 & \pminus 0 & \pminus 0 & \pminus 0 & \pminus 0 & \pminus 0 & \pminus 0 & \pminus 0 & \pminus 0 & \pminus 0 & \pminus 0 & \pminus 0 & \pminus 0 & \pminus 0 & \pminus 0 & \pminus 0 & \pminus 0 & \pminus 0 & \pminus 0 & \pminus 0 & \pminus 0 & \pminus 0 & \pminus 0 & \pminus 0 & \pminus 0 & \pminus 0 & \pminus 0 & \pminus 0 & \pminus 0 & \pminus 0 & \pminus 0 & \pminus 0 & \pminus 0 & \pminus 0 & \pminus 0 & \pminus 0 & \pminus 0 & \pminus 0 \\
 \pminus 0 & \pminus 0 & \pminus 0 & \pminus 0 & \pminus 0 & \pminus 0 & \pminus 0 & \pminus 0 & \pminus 0 & \pminus 0 & \pminus 0 & \pminus 0 & \pminus 0 & \pminus 0 & \pminus 0 & \pminus 0 & \pminus 0 & \pminus 0 & \pminus 0 & \pminus 0 & \pminus 0 & \pminus 0 & \pminus 0 & \pminus 0 & \pminus 0 & \pminus 0 & \pminus 0 & \pminus 0 & \pminus 0 & \pminus 0 & \pminus 0 & \pminus 0 & \pminus 0 & \pminus 0 & \pminus 0 & \pminus 0 & \pminus 0 & \pminus 0 & \pminus 0 & \pminus 0 & \pminus 0 & \pminus 0 & \pminus 0 & \pminus 0 \\
\end{array}
\right)

%% file: matrices/EMu_np_3.tex
\left(
\begin{array}{cccccccccccccccccccccccccccccccccccccccccccc}
 \pminus 0 & \pminus 0 & \pminus 0 & \pminus 0 & \pminus 0 & \pminus 0 & \pminus 0 & \pminus 0 & \pminus 0 & \pminus 0 & \pminus 0 & \pminus 0 & \pminus 0 & \pminus 0 & \pminus 0 & \pminus 0 & \pminus 0 & \pminus 0 & \pminus 0 & \pminus 0 & \pminus 0 & \pminus 0 & \pminus 0 & \pminus 0 & \pminus 0 & \pminus 0 & \pminus 0 & \pminus 0 & \pminus 0 & \pminus 0 & \pminus 0 & \pminus 0 & \pminus 0 & \pminus 0 & \pminus 0 & \pminus 0 & \pminus 0 & \pminus 0 & \pminus 0 & \pminus 0 & \pminus 0 & \pminus 0 & \pminus 0 & \pminus 0 \\
 \minus  2 & \minus  4 & \pminus 0 & \pminus 0 & \pminus 0 & \pminus 0 & \pminus 0 & \pminus 0 & \pminus 0 & \pminus 0 & \pminus 0 & \pminus 0 & \pminus 0 & \pminus 0 & \pminus 0 & \pminus 0 & \pminus 0 & \pminus 0 & \pminus 0 & \pminus 0 & \pminus 0 & \pminus 0 & \pminus 0 & \pminus 0 & \pminus 0 & \pminus 0 & \pminus 0 & \pminus 0 & \pminus 0 & \pminus 0 & \pminus 0 & \pminus 0 & \pminus 0 & \pminus 0 & \pminus 0 & \pminus 0 & \pminus 0 & \pminus 0 & \pminus 0 & \pminus 0 & \pminus 0 & \pminus 0 & \pminus 0 & \pminus 0 \\
 \pminus 0 & \pminus 0 & \pminus 0 & \pminus 0 & \pminus 0 & \pminus 0 & \pminus 0 & \pminus 0 & \pminus 0 & \pminus 0 & \pminus 0 & \pminus 0 & \pminus 0 & \pminus 0 & \pminus 0 & \pminus 0 & \pminus 0 & \pminus 0 & \pminus 0 & \pminus 0 & \pminus 0 & \pminus 0 & \pminus 0 & \pminus 0 & \pminus 0 & \pminus 0 & \pminus 0 & \pminus 0 & \pminus 0 & \pminus 0 & \pminus 0 & \pminus 0 & \pminus 0 & \pminus 0 & \pminus 0 & \pminus 0 & \pminus 0 & \pminus 0 & \pminus 0 & \pminus 0 & \pminus 0 & \pminus 0 & \pminus 0 & \pminus 0 \\
\end{array}
\right)

%% file: matrices/EMu_np_4.tex
\left(
\begin{array}{cccccccccccccccccccccccccccccccccccccccccccc}
 \pminus 0 & \pminus 0 & \pminus 0 & \pminus 0 & \pminus 0 & \pminus 0 & \pminus 0 & \pminus 0 & \pminus 0 & \pminus 0 & \pminus 0 & \pminus 0 & \pminus 0 & \pminus 0 & \pminus 0 & \pminus 0 & \pminus 0 & \pminus 0 & \pminus 0 & \pminus 0 & \pminus 0 & \pminus 0 & \pminus 0 & \pminus 0 & \pminus 0 & \pminus 0 & \pminus 0 & \pminus 0 & \pminus 0 & \pminus 0 & \pminus 0 & \pminus 0 & \pminus 0 & \pminus 0 & \pminus 0 & \pminus 0 & \pminus 0 & \pminus 0 & \pminus 0 & \pminus 0 & \pminus 0 & \pminus 0 & \pminus 0 & \pminus 0 \\
 \pminus 0 & \pminus 0 & \pminus 0 & \pminus 0 & \pminus 0 & \pminus 0 & \pminus 0 & \pminus 0 & \pminus 0 & \pminus 0 & \pminus 0 & \pminus 0 & \pminus 0 & \pminus 0 & \pminus 0 & \pminus 0 & \pminus 0 & \pminus 0 & \pminus 0 & \pminus 0 & \pminus 0 & \pminus 0 & \pminus 0 & \pminus 0 & \pminus 0 & \pminus 0 & \pminus 0 & \pminus 0 & \pminus 0 & \pminus 0 & \pminus 0 & \pminus 0 & \pminus 0 & \pminus 0 & \pminus 0 & \pminus 0 & \pminus 0 & \pminus 0 & \pminus 0 & \pminus 0 & \pminus 0 & \pminus 0 & \pminus 0 & \pminus 0 \\
 \pminus 0 & \pminus 0 & \pminus 0 & \pminus 0 & \pminus 0 & \pminus 0 & \pminus 0 & \pminus 0 & \pminus 0 & \pminus 0 & \pminus 0 & \pminus 0 & \pminus 0 & \pminus 0 & \pminus 0 & \pminus 0 & \pminus 0 & \pminus 0 & \pminus 0 & \pminus 0 & \pminus 0 & \pminus 0 & \pminus 0 & \pminus 0 & \pminus 0 & \pminus 0 & \pminus 0 & \pminus 0 & \pminus 0 & \pminus 0 & \pminus 0 & \pminus 0 & \pminus 0 & \pminus 0 & \pminus 0 & \pminus 0 & \pminus 0 & \pminus 0 & \pminus 0 & \pminus 0 & \pminus 0 & \pminus 0 & \pminus 0 & \pminus 0 \\
\end{array}
\right)

%% file: matrices/EMu_np_5.tex
\left(
\begin{array}{cccccccccccccccccccccccccccccccccccccccccccc}
 \pminus 0 & \pminus 0 & \pminus 0 & \pminus 0 & \pminus 0 & \pminus 0 & \pminus 0 & \pminus 0 & \pminus 0 & \pminus 0 & \pminus 0 & \pminus 0 & \pminus 0 & \pminus 0 & \pminus 0 & \pminus 0 & \pminus 0 & \pminus 0 & \pminus 0 & \pminus 0 & \pminus 0 & \pminus 0 & \pminus 0 & \pminus 0 & \pminus 0 & \pminus 0 & \pminus 0 & \pminus 0 & \pminus 0 & \pminus 0 & \pminus 0 & \pminus 0 & \pminus 0 & \pminus 0 & \pminus 0 & \pminus 0 & \pminus 0 & \pminus 0 & \pminus 0 & \pminus 0 & \pminus 0 & \pminus 0 & \pminus 0 & \pminus 0 \\
 \pminus 0 & \pminus 0 & \pminus 0 & \pminus 0 & \pminus 0 & \pminus 0 & \pminus 0 & \pminus 0 & \pminus 0 & \pminus 0 & \pminus 0 & \pminus 0 & \pminus 0 & \pminus 0 & \pminus 0 & \pminus 0 & \pminus 0 & \pminus 0 & \pminus 0 & \pminus 0 & \pminus 0 & \pminus 0 & \pminus 0 & \pminus 0 & \pminus 0 & \pminus 0 & \pminus 0 & \pminus 0 & \pminus 0 & \pminus 0 & \pminus 0 & \pminus 0 & \pminus 0 & \pminus 0 & \pminus 0 & \pminus 0 & \pminus 0 & \pminus 0 & \pminus 0 & \pminus 0 & \pminus 0 & \pminus 0 & \pminus 0 & \pminus 0 \\
 \pminus 0 & \pminus 0 & \pminus 0 & \pminus 0 & \pminus 0 & \pminus 0 & \pminus 0 & \pminus 0 & \pminus 0 & \pminus 0 & \pminus 0 & \pminus 0 & \pminus 0 & \pminus 0 & \pminus 0 & \pminus 0 & \pminus 0 & \pminus 0 & \pminus 0 & \pminus 0 & \pminus 0 & \pminus 0 & \pminus 0 & \pminus 0 & \pminus 0 & \pminus 0 & \pminus 0 & \pminus 0 & \pminus 0 & \pminus 0 & \pminus 0 & \pminus 0 & \pminus 0 & \pminus 0 & \pminus 0 & \pminus 0 & \pminus 0 & \pminus 0 & \pminus 0 & \pminus 0 & \pminus 0 & \pminus 0 & \pminus 0 & \pminus 0 \\
\end{array}
\right)

%% file: matrices/EMu_np_6.tex
\left(
\begin{array}{cccccccccccccccccccccccccccccccccccccccccccc}
 \pminus 0 & \pminus 0 & \pminus 0 & \pminus 0 & \pminus 0 & \pminus 0 & \pminus 0 & \pminus 0 & \pminus 0 & \pminus 0 & \pminus 0 & \pminus 0 & \pminus 0 & \pminus 0 & \pminus 0 & \pminus 0 & \pminus 0 & \pminus 0 & \pminus 0 & \pminus 0 & \pminus 0 & \pminus 0 & \pminus 0 & \pminus 0 & \pminus 0 & \pminus 0 & \pminus 0 & \pminus 0 & \pminus 0 & \pminus 0 & \pminus 0 & \pminus 0 & \pminus 0 & \pminus 0 & \pminus 0 & \pminus 0 & \pminus 0 & \pminus 0 & \pminus 0 & \pminus 0 & \pminus 0 & \pminus 0 & \pminus 0 & \pminus 0 \\
 \pminus 0 & \pminus 0 & \pminus 0 & \pminus 0 & \pminus 0 & \pminus 0 & \pminus 0 & \pminus 0 & \pminus 0 & \pminus 0 & \pminus 0 & \pminus 0 & \pminus 0 & \pminus 0 & \pminus 0 & \pminus 0 & \pminus 0 & \pminus 0 & \pminus 0 & \pminus 0 & \pminus 0 & \pminus 0 & \pminus 0 & \pminus 0 & \pminus 0 & \pminus 0 & \pminus 0 & \pminus 0 & \pminus 0 & \pminus 0 & \pminus 0 & \pminus 0 & \pminus 0 & \pminus 0 & \pminus 0 & \pminus 0 & \pminus 0 & \pminus 0 & \pminus 0 & \pminus 0 & \pminus 0 & \pminus 0 & \pminus 0 & \pminus 0 \\
 \pminus 0 & \pminus 0 & \pminus 0 & \pminus 0 & \pminus 0 & \pminus 0 & \pminus 0 & \pminus 0 & \pminus 0 & \pminus 0 & \pminus 0 & \pminus 0 & \pminus 0 & \pminus 0 & \pminus 0 & \pminus 0 & \pminus 0 & \pminus 0 & \pminus 0 & \pminus 0 & \pminus 0 & \pminus 0 & \pminus 0 & \pminus 0 & \pminus 0 & \pminus 0 & \pminus 0 & \pminus 0 & \pminus 0 & \pminus 0 & \pminus 0 & \pminus 0 & \pminus 0 & \pminus 0 & \pminus 0 & \pminus 0 & \pminus 0 & \pminus 0 & \pminus 0 & \pminus 0 & \pminus 0 & \pminus 0 & \pminus 0 & \pminus 0 \\
\end{array}
\right)

%% file: matrices/EMu_np_7.tex
\left(
\begin{array}{cccccccccccccccccccccccccccccccccccccccccccc}
 \pminus 0 & \pminus 0 & \pminus 0 & \pminus 0 & \pminus 0 & \pminus 0 & \pminus 0 & \pminus 0 & \pminus 0 & \pminus 0 & \pminus 0 & \pminus 0 & \pminus 0 & \pminus 0 & \pminus 0 & \pminus 0 & \pminus 0 & \pminus 0 & \pminus 0 & \pminus 0 & \pminus 0 & \pminus 0 & \pminus 0 & \pminus 0 & \pminus 0 & \pminus 0 & \pminus 0 & \pminus 0 & \pminus 0 & \pminus 0 & \pminus 0 & \pminus 0 & \pminus 0 & \pminus 0 & \pminus 0 & \pminus 0 & \pminus 0 & \pminus 0 & \pminus 0 & \pminus 0 & \pminus 0 & \pminus 0 & \pminus 0 & \pminus 0 \\
 \pminus 0 & \pminus 0 & \pminus 0 & \pminus 0 & \pminus 0 & \pminus 0 & \pminus 0 & \pminus 0 & \pminus 0 & \pminus 0 & \pminus 0 & \pminus 0 & \pminus 0 & \pminus 0 & \pminus 0 & \pminus 0 & \pminus 0 & \pminus 0 & \pminus 0 & \pminus 0 & \pminus 0 & \pminus 0 & \pminus 0 & \pminus 0 & \pminus 0 & \pminus 0 & \pminus 0 & \pminus 0 & \pminus 0 & \pminus 0 & \pminus 0 & \pminus 0 & \pminus 0 & \pminus 0 & \pminus 0 & \pminus 0 & \pminus 0 & \pminus 0 & \pminus 0 & \pminus 0 & \pminus 0 & \pminus 0 & \pminus 0 & \pminus 0 \\
 \pminus 0 & \pminus 0 & \pminus 0 & \pminus 0 & \pminus 0 & \pminus 0 & \pminus 0 & \pminus 0 & \pminus 0 & \pminus 0 & \pminus 0 & \pminus 0 & \pminus 0 & \pminus 0 & \pminus 0 & \pminus 0 & \pminus 0 & \pminus 0 & \pminus 0 & \pminus 0 & \pminus 0 & \pminus 0 & \pminus 0 & \pminus 0 & \pminus 0 & \pminus 0 & \pminus 0 & \pminus 0 & \pminus 0 & \pminus 0 & \pminus 0 & \pminus 0 & \pminus 0 & \pminus 0 & \pminus 0 & \pminus 0 & \pminus 0 & \pminus 0 & \pminus 0 & \pminus 0 & \pminus 0 & \pminus 0 & \pminus 0 & \pminus 0 \\
\end{array}
\right)

%% file: matrices/EMu_np_8.tex
\left(
\begin{array}{cccccccccccccccccccccccccccccccccccccccccccc}
 \pminus 0 & \pminus 0 & \pminus 0 & \pminus 0 & \pminus 0 & \pminus 0 & \pminus 0 & \pminus 0 & \pminus 0 & \pminus 0 & \pminus 0 & \pminus 0 & \pminus 0 & \pminus 0 & \pminus 0 & \pminus 0 & \pminus 0 & \pminus 0 & \pminus 0 & \pminus 0 & \pminus 0 & \pminus 0 & \pminus 0 & \pminus 0 & \pminus 0 & \pminus 0 & \pminus 0 & \pminus 0 & \pminus 0 & \pminus 0 & \pminus 0 & \pminus 0 & \pminus 0 & \pminus 0 & \pminus 0 & \pminus 0 & \pminus 0 & \pminus 0 & \pminus 0 & \pminus 0 & \pminus 0 & \pminus 0 & \pminus 0 & \pminus 0 \\
 \pminus 0 & \pminus 0 & \pminus 0 & \pminus 0 & \pminus 0 & \pminus 0 & \pminus 0 & \pminus 0 & \pminus 0 & \pminus 0 & \pminus 0 & \pminus 0 & \pminus 0 & \pminus 0 & \pminus 0 & \pminus 0 & \pminus 0 & \pminus 0 & \pminus 0 & \pminus 0 & \pminus 0 & \pminus 0 & \pminus 0 & \pminus 0 & \pminus 0 & \pminus 0 & \pminus 0 & \pminus 0 & \pminus 0 & \pminus 0 & \pminus 0 & \pminus 0 & \pminus 0 & \pminus 0 & \pminus 0 & \pminus 0 & \pminus 0 & \pminus 0 & \pminus 0 & \pminus 0 & \pminus 0 & \pminus 0 & \pminus 0 & \pminus 0 \\
 \pminus 0 & \pminus 0 & \pminus 0 & \pminus 0 & \pminus 0 & \pminus 0 & \pminus 0 & \pminus 0 & \pminus 0 & \pminus 0 & \pminus 0 & \pminus 0 & \pminus 0 & \pminus 0 & \pminus 0 & \pminus 0 & \pminus 0 & \pminus 0 & \pminus 0 & \pminus 0 & \pminus 0 & \pminus 0 & \pminus 0 & \pminus 0 & \pminus 0 & \pminus 0 & \pminus 0 & \pminus 0 & \pminus 0 & \pminus 0 & \pminus 0 & \pminus 0 & \pminus 0 & \pminus 0 & \pminus 0 & \pminus 0 & \pminus 0 & \pminus 0 & \pminus 0 & \pminus 0 & \pminus 0 & \pminus 0 & \pminus 0 & \pminus 0 \\
\end{array}
\right)

%% file: matrices/EMu_np_9.tex
\left(
\begin{array}{cccccccccccccccccccccccccccccccccccccccccccc}
 \pminus 0 & \pminus 0 & \pminus 0 & \pminus 0 & \pminus 0 & \pminus 0 & \pminus 0 & \pminus 0 & \pminus 0 & \pminus 0 & \pminus 0 & \pminus 0 & \pminus 0 & \pminus 0 & \pminus 0 & \pminus 0 & \pminus 0 & \pminus 0 & \pminus 0 & \pminus 0 & \pminus 0 & \pminus 0 & \pminus 0 & \pminus 0 & \pminus 0 & \pminus 0 & \pminus 0 & \pminus 0 & \pminus 0 & \pminus 0 & \pminus 0 & \pminus 0 & \pminus 0 & \pminus 0 & \pminus 0 & \pminus 0 & \pminus 0 & \pminus 0 & \pminus 0 & \pminus 0 & \pminus 0 & \pminus 0 & \pminus 0 & \pminus 0 \\
 \pminus 1 & \pminus 1 & \pminus 0 & \pminus 0 & \pminus 0 & \pminus 0 & \pminus 0 & \pminus 0 & \pminus 0 & \pminus 0 & \pminus 0 & \pminus 0 & \pminus 0 & \pminus 0 & \pminus 0 & \pminus 0 & \pminus 0 & \pminus 0 & \pminus 0 & \pminus 0 & \pminus 0 & \pminus 0 & \pminus 0 & \pminus 0 & \pminus 0 & \pminus 0 & \pminus 0 & \pminus 0 & \pminus 0 & \pminus 0 & \pminus 0 & \pminus 0 & \pminus 0 & \pminus 0 & \pminus 0 & \pminus 0 & \pminus 0 & \pminus 0 & \pminus 0 & \pminus 0 & \pminus 0 & \pminus 0 & \pminus 0 & \pminus 0 \\
 \pminus 0 & \pminus 0 & \pminus 0 & \pminus 0 & \pminus 0 & \pminus 0 & \pminus 0 & \pminus 0 & \pminus 0 & \pminus 0 & \pminus 0 & \pminus 0 & \pminus 0 & \pminus 0 & \pminus 0 & \pminus 0 & \pminus 0 & \pminus 0 & \pminus 0 & \pminus 0 & \pminus 0 & \pminus 0 & \pminus 0 & \pminus 0 & \pminus 0 & \pminus 0 & \pminus 0 & \pminus 0 & \pminus 0 & \pminus 0 & \pminus 0 & \pminus 0 & \pminus 0 & \pminus 0 & \pminus 0 & \pminus 0 & \pminus 0 & \pminus 0 & \pminus 0 & \pminus 0 & \pminus 0 & \pminus 0 & \pminus 0 & \pminus 0 \\
\end{array}
\right)

%% file: matrices/EMu_np_10.tex
\left(
\begin{array}{cccccccccccccccccccccccccccccccccccccccccccc}
 \pminus 0 & \pminus 0 & \pminus 0 & \pminus 0 & \pminus 0 & \pminus 0 & \pminus 0 & \pminus 0 & \pminus 0 & \pminus 0 & \pminus 0 & \pminus 0 & \pminus 0 & \pminus 0 & \pminus 0 & \pminus 0 & \pminus 0 & \pminus 0 & \pminus 0 & \pminus 0 & \pminus 0 & \pminus 0 & \pminus 0 & \pminus 0 & \pminus 0 & \pminus 0 & \pminus 0 & \pminus 0 & \pminus 0 & \pminus 0 & \pminus 0 & \pminus 0 & \pminus 0 & \pminus 0 & \pminus 0 & \pminus 0 & \pminus 0 & \pminus 0 & \pminus 0 & \pminus 0 & \pminus 0 & \pminus 0 & \pminus 0 & \pminus 0 \\
 \pminus 0 & \pminus 1 & \pminus 0 & \pminus 0 & \pminus 0 & \pminus 0 & \pminus 0 & \pminus 0 & \pminus 0 & \pminus 0 & \pminus 0 & \pminus 0 & \pminus 0 & \pminus 0 & \pminus 0 & \pminus 0 & \pminus 0 & \pminus 0 & \pminus 0 & \pminus 0 & \pminus 0 & \pminus 0 & \pminus 0 & \pminus 0 & \pminus 0 & \pminus 0 & \pminus 0 & \pminus 0 & \pminus 0 & \pminus 0 & \pminus 0 & \pminus 0 & \pminus 0 & \pminus 0 & \pminus 0 & \pminus 0 & \pminus 0 & \pminus 0 & \pminus 0 & \pminus 0 & \pminus 0 & \pminus 0 & \pminus 0 & \pminus 0 \\
 \pminus 0 & \pminus 0 & \pminus 0 & \pminus 0 & \pminus 0 & \pminus 0 & \pminus 0 & \pminus 0 & \pminus 0 & \pminus 0 & \pminus 0 & \pminus 0 & \pminus 0 & \pminus 0 & \pminus 0 & \pminus 0 & \pminus 0 & \pminus 0 & \pminus 0 & \pminus 0 & \pminus 0 & \pminus 0 & \pminus 0 & \pminus 0 & \pminus 0 & \pminus 0 & \pminus 0 & \pminus 0 & \pminus 0 & \pminus 0 & \pminus 0 & \pminus 0 & \pminus 0 & \pminus 0 & \pminus 0 & \pminus 0 & \pminus 0 & \pminus 0 & \pminus 0 & \pminus 0 & \pminus 0 & \pminus 0 & \pminus 0 & \pminus 0 \\
\end{array}
\right)

%% file: matrices/EMu_np_11.tex
\left(
\begin{array}{cccccccccccccccccccccccccccccccccccccccccccc}
 \pminus 0 & \pminus 0 & \pminus 0 & \pminus 0 & \pminus 0 & \pminus 0 & \pminus 0 & \pminus 0 & \pminus 0 & \pminus 0 & \pminus 0 & \pminus 0 & \pminus 0 & \pminus 0 & \pminus 0 & \pminus 0 & \pminus 0 & \pminus 0 & \pminus 0 & \pminus 0 & \pminus 0 & \pminus 0 & \pminus 0 & \pminus 0 & \pminus 0 & \pminus 0 & \pminus 0 & \pminus 0 & \pminus 0 & \pminus 0 & \pminus 0 & \pminus 0 & \pminus 0 & \pminus 0 & \pminus 0 & \pminus 0 & \pminus 0 & \pminus 0 & \pminus 0 & \pminus 0 & \pminus 0 & \pminus 0 & \pminus 0 & \pminus 0 \\
 \pminus 0 & \pminus 0 & \pminus 0 & \pminus 0 & \pminus 0 & \pminus 0 & \pminus 0 & \pminus 0 & \pminus 0 & \pminus 0 & \pminus 0 & \pminus 0 & \pminus 0 & \pminus 0 & \pminus 0 & \pminus 0 & \pminus 0 & \pminus 0 & \pminus 0 & \pminus 0 & \pminus 0 & \pminus 0 & \pminus 0 & \pminus 0 & \pminus 0 & \pminus 0 & \pminus 0 & \pminus 0 & \pminus 0 & \pminus 0 & \pminus 0 & \pminus 0 & \pminus 0 & \pminus 0 & \pminus 0 & \pminus 0 & \pminus 0 & \pminus 0 & \pminus 0 & \pminus 0 & \pminus 0 & \pminus 0 & \pminus 0 & \pminus 0 \\
 \pminus 0 & \pminus 0 & \pminus 0 & \pminus 0 & \pminus 0 & \pminus 0 & \pminus 0 & \pminus 0 & \pminus 0 & \pminus 0 & \pminus 0 & \pminus 0 & \pminus 0 & \pminus 0 & \pminus 0 & \pminus 0 & \pminus 0 & \pminus 0 & \pminus 0 & \pminus 0 & \pminus 0 & \pminus 0 & \pminus 0 & \pminus 0 & \pminus 0 & \pminus 0 & \pminus 0 & \pminus 0 & \pminus 0 & \pminus 0 & \pminus 0 & \pminus 0 & \pminus 0 & \pminus 0 & \pminus 0 & \pminus 0 & \pminus 0 & \pminus 0 & \pminus 0 & \pminus 0 & \pminus 0 & \pminus 0 & \pminus 0 & \pminus 0 \\
\end{array}
\right)

%% file: matrices/EMu_np_12.tex
\left(
\begin{array}{cccccccccccccccccccccccccccccccccccccccccccc}
 \pminus 0 & \pminus 0 & \pminus 0 & \pminus 0 & \pminus 0 & \pminus 0 & \pminus 0 & \pminus 0 & \pminus 0 & \pminus 0 & \pminus 0 & \pminus 0 & \pminus 0 & \pminus 0 & \pminus 0 & \pminus 0 & \pminus 0 & \pminus 0 & \pminus 0 & \pminus 0 & \pminus 0 & \pminus 0 & \pminus 0 & \pminus 0 & \pminus 0 & \pminus 0 & \pminus 0 & \pminus 0 & \pminus 0 & \pminus 0 & \pminus 0 & \pminus 0 & \pminus 0 & \pminus 0 & \pminus 0 & \pminus 0 & \pminus 0 & \pminus 0 & \pminus 0 & \pminus 0 & \pminus 0 & \pminus 0 & \pminus 0 & \pminus 0 \\
 \pminus 0 & \pminus 0 & \pminus 0 & \pminus 0 & \pminus 0 & \pminus 0 & \pminus 0 & \pminus 0 & \pminus 0 & \pminus 0 & \pminus 0 & \pminus 0 & \pminus 0 & \pminus 0 & \pminus 0 & \pminus 0 & \pminus 0 & \pminus 0 & \pminus 0 & \pminus 0 & \pminus 0 & \pminus 0 & \pminus 0 & \pminus 0 & \pminus 0 & \pminus 0 & \pminus 0 & \pminus 0 & \pminus 0 & \pminus 0 & \pminus 0 & \pminus 0 & \pminus 0 & \pminus 0 & \pminus 0 & \pminus 0 & \pminus 0 & \pminus 0 & \pminus 0 & \pminus 0 & \pminus 0 & \pminus 0 & \pminus 0 & \pminus 0 \\
 \pminus 0 & \pminus 0 & \pminus 0 & \pminus 0 & \pminus 0 & \pminus 0 & \pminus 0 & \pminus 0 & \pminus 0 & \pminus 0 & \pminus 0 & \pminus 0 & \pminus 0 & \pminus 0 & \pminus 0 & \pminus 0 & \pminus 0 & \pminus 0 & \pminus 0 & \pminus 0 & \pminus 0 & \pminus 0 & \pminus 0 & \pminus 0 & \pminus 0 & \pminus 0 & \pminus 0 & \pminus 0 & \pminus 0 & \pminus 0 & \pminus 0 & \pminus 0 & \pminus 0 & \pminus 0 & \pminus 0 & \pminus 0 & \pminus 0 & \pminus 0 & \pminus 0 & \pminus 0 & \pminus 0 & \pminus 0 & \pminus 0 & \pminus 0 \\
\end{array}
\right)